\documentclass[12pt,a4paper]{article}
\usepackage[utf8]{inputenc}
\usepackage{eqnarray,amsmath}
\usepackage{amsmath}
\usepackage{amsthm}
\usepackage{amsfonts}
\usepackage{amsmath,bm,bbm}
\usepackage{multirow,array}
\usepackage{amssymb}
\usepackage{graphicx}
\usepackage{quoting}
\quotingsetup{font=small}
\usepackage[left=2cm,right=2cm,top=3cm,bottom=3cm]{geometry}
\usepackage[italian,english]{babel}
\usepackage{setspace}
\usepackage[authoryear,round]{natbib}
\usepackage{lipsum}
\usepackage[nodvipsnames]{color}
\usepackage{verbatim}
\usepackage{natbib}
\usepackage{tikz}
\usetikzlibrary{arrows.meta}
\usetikzlibrary{matrix,arrows}
\usepackage{subfig}
\usepackage{xcolor}
\usepackage{natbib} 
\usepackage{booktabs}  % For better table formatting

\usepackage{enumerate}

\usepackage{url}					%url references -> in captions:    \protect\url{}
\usepackage[pagebackref=false, linktoc=all]{hyperref}
\hypersetup{colorlinks = true, unicode=true,  urlcolor=blue, linkcolor=blue, filecolor=blue, citecolor=blue}    % colorlinks=true, citecolor=red

\usepackage{verbatim}

\usetikzlibrary{calc}

\newtheorem{theorem}{Theorem}

\newtheorem{lemma}[theorem]{Lemma}
\newtheorem{proposition}[theorem]{Proposition}

\newtheorem{example}{Example}

\newtheorem{remark}{Remark}

\begin{document}

\pagenumbering{Alph}
\begin{titlepage}
\title{
How do you know you won’t like it if you’ve (never) tried it? Preference discovery and data design\thanks{
We thank Matteo Aggio, Georgy Artemov, Ivan Balbuzanov, Matteo Bizzarri, Arthur Campbell, Federico Crudu, Alberto Dalmazzo, Domenico Fabrizi, Francesco Feri, Junnan He, Weijia Li, Simon Loertscher, Tiziano Razzolini, Roberto Rozzi, Alex Teytelboym, Chengsi Wang, Simon Weidenholzer, Yves Zenou and the participants to the seminars at ECARES, University of Melbourne, Universities Complutense and Carlos III of Madrid, Monash University, S.~Anna in Pisa, and to the conferences 2024 ASSET in Venice, 2nd DISEI  in Florence, 26th Max Weber Fellows at EUI, 2024 NSE Conference in Twin Cities, 23rd SAET in Santiago de Chile, 35th Stony Brook on Game Theory, WEHIA in Koper. SDL gratefully acknowledges funds from FWO-foundation for the project \textit{``Diffusion of Misinformation in Social Networks"} (id. 42933) and Australian Research Council (ARC) for grant DP200102547 and DP240100158, AM from the F-New Frontiers grant by the University of Siena and from FIN-RIC 2024 grant by Universitas Mercatorum, PP from PRIN grants P20228SXNF and 2022389MRW financed by the Italian Ministry of Research.}}
\author{Sebastiano Della Lena\thanks{Department of Economics, Monash University. Email: sebastiano.dellalena@monash.edu.} 
\and
Alessio Muscillo\thanks{Department of Economics, Statistics and Business, Universitas Mercatorum. Email: alessio.muscillo@unimercatorum.it} 
\and Paolo Pin\thanks{Department of Economics, Universit\`a di Siena \& BIDSA, Universit\`a Bocconi. Email: paolo.pin@unisi.it}}%%(Incomplete and preliminary draft not for circulation)
\date{April 2026 
%\\\smallskip  
%\href{https://www.dropbox.com/scl/fi/jcp3lqhkzjv3k1q8c7rsa/JMP1.pdf?rlkey=gaybo8amoioudricfpvwzk6i3&dl=0}{$[$Updated version here$]$}
}
\maketitle
\abstract{
Consumers discover their preferences through experience, yet the sequence and composition of those experiences are often designed by firms, digital platforms, or policymakers. We introduce a “\textit{data-design}” framework for preference discovery, in which the structure of consumption data shapes learning.
Bundling generates correlated exposure across goods, so utility surprises propagate through the co-consumption network. When estimation errors are known, bias-targeted design can shut down learning and amplify misperceptions. Conversely, robust design uses only the geometry of past co-consumption: \textit{popularity-biased} bundles slow learning, while correlation-breaking bundles accelerate preference discovery. The framework thus explains how dominant platforms can sustain biased demand through exposure design, and why effective regulation may need to intervene on the structure of exposure itself rather than only on prices or market shares.% The framework thus explains how platforms can sustain biased demand, and why effective regulation should intervene on the structure of exposure itself.
%A movie-industry example illustrates the model’s key mechanisms.
 }
 
\vspace{15pt}
{\footnotesize
\noindent {\textit{JEL} classification codes: } D81, D83, D85, L12.
}

\thispagestyle{empty}

\end{titlepage}

\pagenumbering{arabic}

\section{Introduction}

Experience is often necessary for individuals to discover what they value. 
This is true when quality is hard to observe ex ante \citep{nelson1970}, but also when agents do not know what they will like until experience reveals it \citep{slovic1995,delaney2020,grenet2022}. Whether preferences are learned or biases persist therefore depends on how experience is structured, yet in many economic environments that structure is itself designed.

%This issue is economically important because the structure of experience is often designed.
 Firms, digital platforms, and policymakers shape which goods consumers encounter and in which combinations: retail platforms through recommendations and product pairings, digital platforms through playlists and curated feeds, and policymakers through bundled reforms or services. Such practices can help consumers discover overlooked products, but they can also reinforce exposure patterns that slow learning or amplify bias, affecting not only current choices but also which products remain under- or overvalued, and how future demand and market power evolve. This broader concern is consistent with recent policy attention to recommendation, self-preferencing, and bundling practices in digital markets \citep{oecd2023}.

Despite its economic relevance, the mechanism linking designed exposure to preference discovery remains poorly understood. This paper proposes a new \emph{data-design} perspective on preference discovery, in which consumption generates the data from which preferences are inferred.
The key insight is that more experience need not imply better learning: what matters is the informational structure of the data generated by experience. {Our central contribution is to show that preference discovery depends on the covariance structure generated by past exposure.%Our central contribution is to show that preference discovery is an endogenous property of the covariance structure generated by past exposure.
} 
Some consumption patterns reveal preferences, while others sustain or amplify bias, or even prevent learning altogether.

We develop a model in which a consumer (she) is uncertain about her preferences and learns from realized utility, while a provider (he) shapes consumption experiences by choosing which goods or bundles to offer over time. After each experience, the consumer updates her beliefs using past consumption data. Depending on his incentives, the provider may seek to promote or hinder learning.

In this framework, past consumption data define a co-occurrence network of goods (or attributes), in which two goods are linked when they have been experienced together in a bundle. The structure of this network determines how the \emph{consumption surprise} (i.e., the difference between expected and realized utility from a new experience) propagates across goods and shapes the updating of estimated preferences. We interpret these propagation effects as externalities between goods: consuming one good can affect the perceived value of others, including goods that have never appeared in the same bundle but are connected through paths of past co-consumption. As a result, the informativeness of new experiences depends on the correlation structure of exposure. When exposure generates sufficiently rich variation, preferences are learned; when it is concentrated along correlated dimensions, information becomes redundant and learning may fail.
This mechanism reproduces behavioral patterns often associated with constructed preferences \citep{slovic1995}---such as shifting or locally inconsistent evaluations---within a rational learning framework under correlated exposure.\footnote{Here, the ``context" of elicitation can be interpreted as the bundle in which goods are experienced: when goods are evaluated jointly, their effects on perceived utility become statistically interdependent, so that the inferred value of one depends on the others with which it is consumed.}

Our first set of results characterizes how exposure can be designed either to foster discovery or to manipulate belief dynamics when both true and estimated preferences are observed.\footnote{A prominent example is Amazon, which promotes bundling as a way to enhance product discovery (\citealt{amazonbundlingblog}) while simultaneously tightening its ``Consumables Bundling Policy'' to prevent sellers from mixing multiple brands or lower-quality goods under a single SKU (stock--keeping unit, i.e., a unique product listing on the platform) without transparent disclosure (\citealt{amazonpolicy2024}). This dual stance illustrates how bundling can serve both informational and strategic purposes—helping consumers discover new products while also enabling providers to steer perceptions and control demand.} Single-good exposure is the benchmark that maximizes learning by isolating the information generated by each good and minimizing the overall mean squared error of estimated preferences. By contrast, bundles orthogonal to the estimation error provide the benchmark for halting learning: they generate no expected consumption surprise and therefore no informative variation, allowing the provider to continue selling while preserving current estimated preferences. More generally, bundles can be designed to direct consumption surprises toward selected goods. We characterize conditions under which a positive surprise generated by underestimated goods is sufficiently difficult to attribute that it raises not only their estimated value, but also that of goods that are already overestimated. Overall, bundling can be used not only to promote discovery, but also to preserve or amplify misperceptions.

A well-known illustration is Microsoft’s bundling of Internet Explorer with Windows in the late 1990s \citep{usvMicrosoft1999,eumicrosoft2009}. By pre-installing its browser and making it the default option, Microsoft limited exposure to competing products such as Netscape Navigator. In terms of the model, this collapsed variation in consumption data, slowing learning about alternatives and sustaining (possibly) biased demand.  %A well-known illustration of this mechanism is the bundling of Internet Explorer with Microsoft Windows in the late 1990s \citep{usvMicrosoft1999,eumicrosoft2009}. By pre-installing its browser and making it the default option, Microsoft limited consumers’ exposure to competing products such as Netscape Navigator. In terms of the model, this amounted to collapsing variation in consumption data, thereby slowing or halting learning about alternatives and sustaining biased preferences in Microsoft’s favor. 
{A contemporary analogue arises in vertically integrated streaming platforms such as Netflix, which jointly control recommendation, distribution, and increasingly upstream content production. In this environment, exposure design shapes both consumer learning and the future incentives governing content creation.}

Our second set of results concerns robust design: exposure interventions that do not rely on observing true or estimated preferences. Even in this case the provider can control the speed of learning through the geometry of past consumption data. We characterize the most effective robust direction for slowing learning and the most effective robust direction for accelerating it. \emph{Popularity-biased} exposure---formally, bundles aligned with \emph{eigenvector centrality} in the co-occurrence network, which prioritizes goods that are already frequently consumed and strongly connected to other highly connected goods---reinforces the dominant pattern in past consumption and is the most effective robust way to slow learning. By contrast, correlation-breaking exposure---captured by our notion of \emph{correlation centrality}---introduces variation in the least informative direction of past exposure and is the most effective robust way to accelerate learning locally. Thus, even without observing preferences, exposure design can either trap consumers in correlated environments or promote discovery by breaking those correlations.

These results speak directly to recent empirical and policy discussions on recommendation systems. Empirical research documents a persistent \textit{popularity bias} in algorithmic recommendation—an overexposure of already prominent goods that limits novelty and diversity in user experience \cite[as surveyed in][]{klimashevskaia2024}. Our framework provides a microfoundation for this pattern by identifying popularity bias as the robust way to sustain demand while limiting preference discovery. In parallel, recent regulatory initiatives, such as the European Union’s Digital Services Act, call for greater algorithmic choice and for exposure beyond the dominant structure of consumption \citep{busch2023}. This makes the model’s implications directly relevant for platform design and suggests a simple policy application: recommendation algorithms could incorporate correlation-breaking variation to loosen dominant patterns in exposure and enhance informational diversity without major structural changes.

We apply the framework to a monopolistic environment to show how exposure design endogenizes demand formation and thereby becomes a source of dynamic market power.
%We apply our framework to the monopolist’s problem to show how preference discovery changes firm behavior in a simple market environment.
Because exposure shapes how consumers learn what they value, the monopolist no longer simply extracts profits from given preferences, but also chooses how to structure consumption so as to influence future demand. The application shows that, once learning becomes endogenous, product design and pricing become instruments for shaping demand.
We further characterize the welfare implications of misperceived preferences. Both consumer surplus and profits decompose into a price effect and a bundle effect. The price effect is a pure transfer between consumers and producers and cancels in the aggregate, while the bundle effect, driven by allocative distortion, weakly reduces total welfare and can make consumer surplus and profits decrease together.

\paragraph{Related Literature}
%\label{sec:literature}

The paper relates to the literature on taste uncertainty and preference discovery, which studies how consumers learn about their own preferences through experience \citep[e.g.,][]{ok2012,cooke2017,cerreia2023}. We provide a framework in which preference discovery depends on the structure of consumption data. Consumers learn from consumption experiences, and bundled consumption shapes the data they observe; as a result, data structure governs both the direction and the speed of learning. The paper therefore also relates to recent work on platform and algorithmic market environments and their implications for consumer outcomes %on the manipulation of beliefs and exposure by platforms
\citep{calvano2020, acemoglub2024}, by showing that bundling and product design are ways of shaping the data consumers observe and, therefore, preference discovery.

%Taste uncertainty refers to situations in which consumers possess incomplete information about their own preferences and uncover them progressively through experience. Empirical studies, such as those by \cite{grenet2022} and \cite{delaney2020}, provide evidence of this uncertainty, highlighting instances where consumers show incomplete knowledge of their preferences. On the theoretical side, contributions from \cite{cerreia2023}, \cite{cooke2017}, and \cite{ok2012} offer valuable insights into this process. This paper builds on that foundation but shifts the focus from how individuals learn to how the learning environment itself is designed. In contrast to \citet{acemoglub2024}, who study single-product manipulation through presentation and AI-based prediction, we analyze how bundling endogenously structures the data that consumers learn from, influencing both the direction and the speed of preference discovery.

Bundling has long been studied as a tool for price discrimination and surplus extraction \citep{adams1976,mcafee1989}. It has also been analyzed as a source of entry deterrence and market power \citep{nalebuff2004}. More recent mechanism-design work shows that firms may pool or randomize product qualities to screen consumers and manage information rents \citep{loertscher2022,bergemann2022}. 
Our paper identifies an additional informational effect of bundling: it shapes the data through which consumers learn their own preferences. By altering the correlation structure of exposure, bundling affects how consumption experiences translate into belief updating and preference discovery.

%Bundling, the practice of selling multiple products together, is widely recognized as a method of price discrimination. Classic studies such as \citet{adams1976}, \citet{mcafee1989}, and \citet{tirole1988} show how bundling can extract consumer surplus by segmenting buyers according to their reservation values.\footnote{Subsequent research has explored both efficiency and strategic aspects of bundling. \citet{liao2002} shows that bundling can raise consumer surplus under certain conditions, while recent mechanism-design analyses demonstrate that firms may optimally pool or bundle product qualities to manage information rents and increase profits \citep{loertscher2022,bergemann2022}.}We provide a new perspective on bundling, in which it shapes the data through which consumers learn their own preferences. Depending on how goods are combined, bundling can either enhance discovery and welfare---by exposing consumers to new and diverse experiences---or obscure quality and sustain biases, allowing firms to profit from distorted beliefs. 

The paper also relates to the literature on information design and persuasion, which studies how a sender shapes receivers’ beliefs by choosing an information structure \citep{kamenica2011,bergemann2019}. Our framework complements this approach by shifting attention from signal design to data design: the provider does not choose the signal structure directly, but the structure of experiences from which observations are generated, thereby controlling the correlation structure of the data that drive belief updating. %In this sense, the model extends the logic of information design to environments in which beliefs are formed through consumption experiences rather than through designed signals.

%The interconnectedness of estimated preferences as depending on the consumption data The paper also relate to the literatue about nteotk and Io and learnign in networs with We exploit the use network analysis to study the .  \cite{ushchev2018} develop a model of price competition in product-variety networks, showing how network structures influence equilibrium prices through firms' centralities and consumer preferences. Prior research, such as that by \cite{golub2010,golub2012}, has examined social learning in networks where multiple agents learn about a single parameter. In contrast, our study focuses on scenarios where a representative agent, learn about numerous interconnected parameters. This approach allows us to capture the complexity of consumer preference discovery in markets where products are bundled, and it differentiates our work from previous literature on networks and industrial organization \citep{galeotti2009, campbell2013, fainmesser2016,campbell2024}, which emphasizes the role of word-of-mouth and consumer networks. 
%\red{decide what else put about networks}
 %The paper also connects to spectral methods in economics \citep{golub2025} and to intervention analyses based on eigenvector characterizations \citep{galeotti2020,galeotti2021,DasarathaGolubShahMimeo}. In our setting, these eigenvectors come from the information matrix generated by past consumption and characterize the directions along which preference learning can be slowed down or accelerated.
In our paper, past bundled consumption generates a co-consumption network across goods that governs the propagation of utility surprises and the speed of preference identification. This differs from network-based work in industrial organization, where networks capture consumer interactions or product linkages that shape market outcomes \citep{galeotti2009,campbell2013,fainmesser2016,ushchev2018,campbell2024}, and from social-learning models, where many agents learn about a single underlying state through interpersonal interactions \citep{golub2010,golub2012}.

The paper also connects to spectral methods in economics \citep{golub2025}. In particular, the network-intervention literature uses eigenvectors to characterize optimal targeting policies \citep{galeotti2020,galeotti2021,DasarathaGolubShahMimeo}. Here, the relevant eigenvectors arise from the \emph{information matrix} generated by past consumption and identify the directions along which preference learning can be slowed down or accelerated.
%Finally, spectral methods and eigenvector-based analysis appear in several economic applications \citep{golub2025}. In particular, the network-intervention literature uses eigenvectors to characterize optimal targeting policies \citep{galeotti2020,galeotti2021,DasarathaGolubShahMimeo}. In contrast, here the relevant eigenvectors arise from the \emph{information matrix} generated by the consumer's consumption history and characterize the directions along which preference learning can be slowed down or accelerated.
%The methodologies of principal component and spectral analysis, as discussed and employed in different economic applications by \cite{bramoulle2014}, \cite{galeotti2020}, and \cite{galeotti2021}, are instrumental in our analysis, providing insights into the strategies that a provider might employ to manipulate consumer learning.
\\
\\
The following sections cover the formalization of the consumer's learning process and an analysis of the factors affecting consumer learning (Section \ref{sec:model}), strategic implications for the provider (Section \ref{sec:provider}), and an illustrative application where we apply our model to the movie industry (Section \ref{sec:movies}). Section \ref{sec:conclusion} concludes. The Appendix develops several structural extensions and collects the formal proofs. Appendix \ref{app:ext} discusses aggregation and identification, Appendix \ref{subsec:complementarity} extends the model to complementarities and substitution effects, Appendix \ref{app:alpha} allows for an unknown baseline utility, and Appendix \ref{app:proofs} contains the formal proofs of all our results. Finally, Appendix \ref{sec:appendix_movies} provides some additional analysis to the exercise of Section \ref{sec:movies}.

\section{The Model}
\label{sec:model}
We model the repeated interaction between a \emph{consumer} (she) and a \emph{provider} (he) in discrete time. 
At each date $t$, the provider can offer the consumer a \emph{composite good}, that is, a bundle 
$\mathbf{x}_t = (x_{i,t})_{i \in I}$, 
 where $I$ is a finite set of  cardinality $n$. 
 % The component $x_{i,t}$ denotes the quantity of good $i$ consumed at time $t$,  and we allow for the fully general formulation $\mathbf{x}_t \in \mathbb{R}^n$. Most of our theoretical results are derived under this unconstrained specification, which allows entries of $\mathbf{x}_t$ to be continuous or discrete, and positive or negative. When an application requires additional structure---such as non-negativity, dummy bundles, norm constraints such as $\|\mathbf{x}_t\|_2=1$, or nonlinear restrictions induced by complementarities or substitutions among goods (see Appendix~\ref{subsec:complementarity})---we impose it explicitly and indicate which results continue to hold. We denote by $\{\mathbf{e}_i\}_{i=1}^n$ the canonical basis of $\mathbb{R}^n$.
The component $x_{i,t}$ denotes the amount of good (or attribute) $i$ in the bundle consumed at time $t$.
Throughout the paper we adopt a fully general representation in which
$\mathbf{x}_t \in \mathbb{R}^n$, and we denote by $\{\mathbf{e}_i\}_{i\in I}$ the canonical basis of $\mathbb{R}^n$. 
Entries may therefore be continuous or discrete and can take both positive and negative values. For example, they may record quantities or simply whether a good is present in the bundle, and may be measured in levels, relative to a reference bundle, or along characteristics that can go in opposite directions. %Entries may be continuous or discrete. Positive entries represent the presence or intensity of a good or attribute, while negative entries capture deviations from a benchmark or opposite directions of a characteristic. Thus, a bundle may encode product features relative to a reference design, so that increasing one dimension or reducing another is described in the same vector space.
Most of our theoretical results are derived under this unconstrained formulation. Whenever additional feasibility restrictions are imposed, we state them explicitly and clarify which results continue to hold.

We use the term bundle broadly, to refer either to a set of distinct goods consumed jointly or to a product composed of multiple attributes or characteristics. %A bundle can be interpreted either as a set of distinct goods that are consumed jointly, or as a product composed of multiple attributes or characteristics. 
In the latter interpretation, the vector $\mathbf{x}_t$ describes the intensity or presence of the product’s underlying features. A bundle may also represent a team of people working on a common project. Section~\ref{sec:movies} provides an illustrative example along these lines.

%The component $x_{i,t}$ denotes the quantity of good $i$ in the bundle consumed at $t$. 
% Throughout the paper we assume $\mathbf{x}_t \in \mathbb{R}^n$, so that entries may be continuous or discrete, and can take positive or negative values. 
% In some parts of the analysis we impose additional structure, such as restricting to non-negative bundles, $\mathbf{x}_t \in \mathbb{R}^n_+$, or requiring that each bundle is normalized, e.g.~by imposing $\|\mathbf{x}_t\|_2=1$.

% We will state explicitly when such restrictions are in force.
% We denote by $\{\mathbf{e}_i\}_{i\in I}$ the canonical basis of $\mathbb{R}^n$.  

% {\color{red} \bf [I would even be more clear here: our results are for $\mathbb{R}$ -- we will discuss what holds imposing constraints -- anticipate here also the appendix with complementarities -- lo fa Paolo]}

The provider may have different objectives over the consumer’s perceived values
$\hat{\bm{\beta}}_{t}$. For instance, a firm may seek to increase demand, a platform may aim to raise engagement or conversion, and a policymaker may try to build acceptance for a reform. In all cases, the provider acts through the same channel: it designs the consumption environment, thereby shaping the data that govern the evolution of $\hat{\bm{\beta}}_t$.

%In what follows we study the consumer’s dynamic learning problem. Section~\ref{sec:provider} turns to the strategies of the provider.

\subsection{Preferences of the consumer}

The consumer has \emph{linear} preferences over $\mathbb{R}^n$, represented by the vector of marginal utilities $\bm{\beta}:=(\beta_{i})_{i \in I} \in \mathbb{R}^n$, such that, for all $i,j \in I$, $i  \succsim_{\beta} j \ \Leftrightarrow \ \beta_i \geq \beta_j$.  
The profile $\bm{\beta}$ is exogenous and constant over time, and can be interpreted either as the consumer’s subjective (fixed) preferences or as objective qualities of the goods in $I$. %\footnote{Preferences may allow for goods with negative marginal utility, i.e.\ $\beta_i < 0$.} 
Goods in $I$ are assumed to be symmetric and ex-ante exchangeable, that is, relabeling $I$ leaves preferences invariant.

The consumer does not know her preferences $\bm{\beta}$ and faces \emph{taste uncertainty}—that is, incomplete information about her own $\bm{\beta}$. 
We denote with $\hat{\bm{\beta}}_t:=(\hat{\beta}_{i,t})_{i \in I} \in \mathbb{R}^n$ the consumer's estimated preferences at time $t$. 
% We assume that, for each $i \in I$, at time 0  $\hat{\beta}_{i,0}\neq \beta_i $.
The estimated preferences $\hat{\bm{\beta}}_t$ may induce a different ordering of $I$, such that for all $i,j \in I$ it may exist that
$ i  \succsim_{\beta} j$, whereas $j  \succsim_{\hat{\beta}} i$. 

After consuming bundle $\mathbf{x}_t$, the consumer experiences utility
\begin{equation}
\label{eq: utility}
u_t = \alpha + \mathbf{x}_t' \bm{\beta} + \varepsilon_t,
\end{equation}
where $\varepsilon_t \sim \mathcal{N}(0,\sigma^2)$ is i.i.d. across observations, and represents measurement error or idiosyncratic shock. Thus, $u_t$ can be interpreted as a noisy signal about the utility yielded by the bundle $\mathbf{x}_t$. We assume that the consumer knows the baseline utility $\alpha$.\footnote{Appendix \ref{app:alpha} relaxes this assumption by allowing the consumer to $(i)$ hold a misspecified baseline utility and $(ii)$ recognize uncertainty about $\alpha$ and estimate it as well. The main intuitions of the model are robust to these extensions.} 
%As discussed above, the bundle vector $\mathbf{x}_t$ is defined in the general space $\mathbb{R}^n$. Depending on the application, additional restrictions such asnon–negativity, normalization, or dummy bundles may be imposed. When needed, we will state these constraints explicitly.
Finally, note that Equation~\eqref{eq: utility} provides a general representation of the consumer’s utility and in Appendix \ref{subsec:complementarity} we extend the analysis to the case in which the set of goods $I$ may include interaction terms that capture complementarity or substitution among primitive goods. 

Given the realized utility, she makes inference about her preferences and updates $\bm{\hat{\beta}}$:
$$
\mathbf{x}_{t}, \bm{\beta} 
\;\xrightarrow{\ \ \text{consumption} \ \ }\; 
u_t 
\;\xrightarrow{\ \ \text{inference} \ \ }\; 
\bm{\hat{\beta}}_t.
$$

Estimated preferences $\hat{\bm{\beta}}_t$ govern consumption decisions---whether 
to buy and at what price---and broader choices such as voting over policy packages. 
We illustrate a possible example in Section~\ref{sec:monopolist} and abstract from both 
elsewhere to isolate the mapping from bundle design to belief updating.

%\subsection{Consumption Data}
\paragraph{Consumption Data}
To estimate her preferences $\bm{\beta}$, the consumer uses the entire history of past observations up to time $t$. 
Let $\mathbf{X}_t := (\mathbf{x}_s)_{s \leq t}$ denote the $t \times n$ matrix of bundles observed up to $t$.
More explicitly,

$$
%[0,1]^{t \times n} \ni 
\mathbf X_t := 
\underset{n \  \text{goods}}{
\begin{bmatrix}
x_{1,1} &   \dots   & x_{n,1}   \\
\vdots  &   \ddots   & \vdots    \\
x_{1,t} &   \dots   & x_{n,t}   \\
\end{bmatrix}}\text{\scriptsize $t$ observations}%\footnote{\red{RIGHE E COLONNE INVERTITE}}
\qquad
\mathbf u_t
= 
\begin{bmatrix}
    u_1 \\
    \vdots \\
    u_t
\end{bmatrix}
\in \mathbb{R}^{t}.
$$

Thus, at time $t$, the consumer can estimate  her tastes using the entire history of past exposure. A key object is  $\bm{Z}_{t} :=  \mathbf{{X}}_t' \mathbf{{X}}_t \in \mathbb R^{n \times n}$ which summarizes the structure of that exposure.  In the case in which occurrences of each good are dummies, just recording if a good is present or not in the bundle, $\bm{Z}_{t}$ counts in the diagonal how often a good has appeared, and off--diagonally how often two goods have been together. In this case, $\bm{Z}_t$ can be interpreted as a co-occurrence matrix induced by the consumer’s history of observed bundles.\footnote{When the entries of $\mathbf{X}_t$ are  continuous non negative quantities rather than indicators, the interpretation is analogous. Larger diagonal entries correspond to greater exposure to a good, either because it appears more often or in larger quantities, while larger off-diagonal entries correspond to greater joint exposure to pairs of goods.}  In econometric terms, $\bm{Z}_t$ is the \textit{information}, or \textit{precision}, \textit{matrix} associated with the consumer’s past observations.

%In this sense, at every time step $t$, the consumer behaves as an econometrician trying to estimate parameters from a bunch of observations.

%In the econometric literature, given the information available at time $t$,  $\bm{Z}_{t} :=  \mathbf{{X}}_t' \mathbf{{X}}_t \in \mathbb R^{n \times n}$ is  called the \emph{precision matrix} or the \emph{concentration matrix}.  In the case in which occurrences of each good are dummies, just recording if a good is present or not in the bundle,  $\bm{Z}_{t}$ counts in the diagonal how often a good has appeared, and off--diagonally how often two goods have been together. Thus,  in this case, $\bm{Z}_{t}$ represents (a version of) the \textit{co-occurrence} adjacency matrix of the goods in $I$, given the stream of observation available at $t$.\footnote{%   When, more generally, the elements of $\mathbf{X}_t$ represents continuous quantities, the interpretation is qualitative analogous, in the sense that if $\bm{Z}_{ii,t}>\bm{Z}_{jj,t}$, the mean square quantity of observation is larger for $i$ than for $j$,  and we can say that good $i$ has been observed \emph{more often} and/or in \emph{larger quantities} than good $j$. A similar argument applies for the off--diagonal elements, implying that larger values indicate more joint observations for a couple of goods.}

Before analyzing preference updating%in Section~\ref{subsec:update}
, note that throughout the main text we assume that $\bm{Z}_t$ is full rank. Appendix~\ref{subsec:multicoll} considers the case in which some observations are redundant, so multicollinearity reduces the effective dimension of the problem.

%\subsection{Updating of preferences}\label{subsec:update}
\paragraph{Updating of preferences}
At each $t$, the consumer estimates $\bm{\beta}$ from the previous $t$ observations. 
Formally, the estimator $\hat{\bm{\beta}}_t$ maximizes the likelihood function
\begin{equation}
\hat{\bm{\beta}}_t \in \arg \max_{\bm\beta} \; \mathcal{L}(\bm{\beta}\,|\,\mathbf{u}_t,\mathbf{X}_t)
:= \prod_{s=1}^t f(u_s\,|\,\bm{\beta}, \mathbf{x}_s),
\label{eq:likelihood}
\end{equation}
where  $f\left(u_s|\bm{\beta}, \mathbf{x}_s\right)$ is the density function of $u_s$ given the preferences $\bm{\beta}$ and the bundles of consumption $\mathbf{x}_s $, forming the whole history of consumption $\mathbf{X}_t $.

Since utility is linear and errors are normally distributed with mean zero and constant variance, 
maximizing \eqref{eq:likelihood} is equivalent to ordinary least squares (OLS). 
%Thus, we can state following proposition describing the estimated preferences of the consumer at $t$.

\begin{lemma}
\label{prop: lemma_1}
The estimator of ${\boldsymbol \beta}$ after $t$ observations is
\begin{equation}
    \hat{\bm{\beta}}_t = \bm{W}_{t}\,\mathbf{X}_t' \left(\mathbf{u}_t-\bm{1}\alpha\right),
\label{eq: beta}
\end{equation}
where $\bm{W}_{t}:=\bm{Z}^{-1}_{t} = (\mathbf{X}_t' \mathbf{X}_t)^{-1}$. 
For each $i \in I$, this expands to
\begin{equation}
   \hat{\beta}_{i,t} = \sum_{s=1}^t \left(\sum_{j=1}^n w_{ji,t} x_{j,s}\right)
u_s %\big(\mathbf{x}_s'\bm{\beta} + \varepsilon_s + (\alpha-\hat\alpha)\big),
    \label{eq: betai}
\end{equation}
with ex-post uncertainty
\begin{equation}
  \operatorname{Var}\!\big[\hat{\bm \beta}_t\big] = \sigma^2 \bm{W}_{t}.
  \label{eq:variance}
\end{equation}
\end{lemma}

%Equations \eqref{eq: beta} and \eqref{eq: betai} in
Lemma \ref{prop: lemma_1} characterizes the vector of estimated preferences at $t$, given the whole history of consumption and the statistical structure generated by the history of bundles.

The matrix $\bm{W}_{t} $   
is the (scaled) \textit{variance-covariance matrix} of estimators' vector and captures the degree of uncertainty associated with the estimated preference.\footnote{{Note that the actual variance-covariance matrix is $\sigma^2\bm{W}_{t}$. Throughout the paper, we refer to $\bm{W}_{t}$ simply as the variance-covariance matrix to simplify the exposition.}}
%As well known from OLS estimation, 
Diagonal entries $w_{ii,t}$ are always non-negative and measure the variance of the estimate of
$\beta_i$. Economically, they capture how precisely the consumer has learned
the value of good $i$ after observing $t$ bundles.
Off-diagonal entries $w_{ij,t}$ are symmetric and measure the covariance between the estimates of $\beta_i$ and $\beta_j$. They capture the extent to which past joint consumption makes learning about one good informative about the other.
%Entries in the diagonal are always non negative and represent the variance of the regression coefficients. They estimate the uncertainty that the consumer has ex--post (i.e.~once observed $t$ bundles) about the estimation of the corresponding parameter.
%Off--diagonal entries are symmetric and represent covariance between the corresponding regression coefficients, that is how much an error in the estimation of one coefficient is expected to be correlated with an error in the other.
We can rearrange equation \eqref{eq: beta} as  $\hat{\bm{\beta}}_t =\bm{\beta}
+ \bm{W}_{t}(\mathbf{X}_t)' \bm{\varepsilon}_t =\bm{\beta}
+\frac{Var\left[\hat{\bm{\beta}}_t\right]   }{\sigma^2}(\mathbf{X}_t)' \bm{\varepsilon}_t$. Therefore, since the errors have zero conditional mean, all the expected estimated preferences are unbiased.%\footnote{ For completion, we remind that in standard OLS, for good $i$, the quantity $\sigma \sqrt{w_{ii}}$ is the standard deviation in the uncertainty of  the estimation of $\hat{\beta}_i$.  In standard OLS the econometrician estimates also $\hat{\sigma}$ from the data.  Even if we do not need this distinction in our paper, because both the consumer and the provider will need only $\bm{W}_{t}$, what an agent in our setup would need to compute depends on whether she/he knows the true $\sigma$ or she/he has to estimate $\hat{\sigma}$ from the data.}

\subsection{Dynamic updating and correlated externalities} \label{sec:dynup}

We can relate the estimates at each $t$ to the estimated preferences at the previous period.
When at time $t$ the consumer gets a new bundle $\mathbf{x}_{t}$,  she expects that the utility she will get is $\alpha+\mathbf{x}'_{t}\bm{\hat{\beta}}_{t-1}$.
However, because of errors in her estimation and of the added noise $\varepsilon_t$, she will likely experience something different.
We call the \emph{consumption surprise}, which can be positive or negative, $\Delta u_t :=u_t-\mathbb{E}_{\hat{\bm{\beta}}_{t-1}}[u_t]=\mathbf{x}'_{t}\bm{\beta} + \varepsilon_t -\mathbf{x}'_{t}\bm{\hat{\beta}}_{t-1}$, that is, the distance between the actually experienced utility and the expected utility.
The consumption surprise will affect the new estimation according to a simple linear updating rule. 

\begin{proposition}
    \label{prop:1}
    The estimated preferences at time $t$ in equation \eqref{eq: beta}
    can be written as 
    \begin{equation}
    \hat{\bm{\beta}}_t =\hat{\bm{\beta}}_{t-1}+ \bm{\omega}_t \Delta u_t
    \label{eq: betap1}
    \end{equation}
    with $\bm{\omega}_t:=\frac{1}{1+ \mathbf{x}_{t}'\bm{W}_{t-1} \mathbf{x}_{t}} \bm{W}_{t-1}  \mathbf{x}_t$. In particular, for each $i \in I$
    \begin{equation}
    \hat{\beta}_{i,t} = \hat{\beta}_{i,t-1} + \frac{\sum_{j=1 }^{n} w_{ij,t-1} x_{j,t} }{1+ Var[\hat{u}_t] /\sigma^2}\Delta u_{t}.
    \label{eq:linear_updating}
    \end{equation}
    where ${Var}[\hat{u}_t]:=\sigma^2(\mathbf{x}_{t})' \bm{W}_{t-1}  \mathbf{x}_{t}$ is the variance of the predicted utility at time $t$.
\end{proposition}

Proposition \ref{prop:1} shows that preference updating is driven by consumption surprises, and that each surprise is distributed across goods according to the covariance structure generated by past consumption.\footnote{Note that equation~\eqref{eq: betap1} can also be interpreted in Bayesian terms. 
Because of the normality assumption, the recursive least-squares process is mathematically equivalent to sequential Bayesian updating in a linear--Gaussian model in which the consumer holds a prior
$
\bm{\beta}\sim\mathcal{N}(\hat{\bm{\beta}}_{t-1},\sigma^2\bm{W}_{t-1})
$
and observes
$
u_t=\mathbf{x}_t'\bm{\beta}+\varepsilon_t$,
with $
\varepsilon_t\sim\mathcal{N}(0,\sigma^2)$.
At each date, the posterior mean and covariance coincide with the recursive least-squares update.
To initialize the recursion, one must specify an initial Gaussian prior 
$
\bm{\beta}\sim\mathcal{N}(\hat{\bm{\beta}}_{0},\sigma^2\bm{W}_{0})
$.
With finite $\bm{W}_0$, the posterior mean corresponds to a mixed (ridge-type) estimator. The classical OLS recursion is recovered under diffuse (flat) initialization, obtained as the limit $\bm{W}_0=\rho \bm{I}$ with $\rho\to\infty$, i.e.\ an improper Gaussian prior with infinite variance (which is intuitively, for us, the limit of a null history $\bm{Z}_0$). See \citet[Section~4]{Pollock2003Recursive} for an explicit derivation and discussion of diffuse initial conditions in recursive regression.} 

The intuition is as follows. When the realized utility of bundle $\mathbf{x}_t$ exceeds its predicted utility, that is, when $\Delta u_t>0$, the consumer revises upward the estimated value of the bundle; when $\Delta u_t<0$, she revises it downward. This revision is then propagated to other goods through $\bm{W}_{t-1}$. %In particular, the update of good $i$ depends on the weighted exposure of $i$ to the consumed bundle, measured by $\sum_{j=1}^n w_{ij,t-1}x_{j,t}$. If this term is zero, good $i$ is unaffected. If it is positive, a positive surprise raises $\hat{\beta}_{i,t}$ and a negative surprise lowers it. If it is negative, the opposite occurs. We refer to these two cases as correlated and anti-correlated externalities.
For instance, if for a given good $i$ we have $w_{ij,t-1}=0$ for every $j$ such that $x_{j,t}>0$, then $\hat{\beta}_{i,t}=\hat{\beta}_{i,t-1}$ and good $i$ is unaffected. Conversely, when $i$ is connected to at least one consumed good, the update depends on the sign and magnitude of the entries $w_{ij,t-1}$. If $w_{ij,t-1}>0$ and $x_{j,t}>0$, a positive surprise increases $\hat{\beta}_{i,t}$, while a negative surprise decreases it; if $w_{ij,t-1}<0$, the effect is reversed. We call these two cases \textbf{correlated} and \textbf{anti-correlated externalities} of $j$ on $i$.
Throughout, these ``externalities'' are informational spillovers in belief updating, induced by the covariance structure of the estimator, rather than technological or payoff externalities.
The absolute value $|w_{ij,t-1}|$ measures the strength of the effect, i.e.\ how much consumption of good $j$ influences the updating of good $i$. In aggregate, the total effect of the bundle $\mathbf{x}_t$ on $\hat{\beta}_{i,t}$ is positive (negative) if and only if $\sum_{j=1 }^{n} w_{ij,t-1} x_{j,t} \cdot \Delta u_t
>0$ $(<0)$.

Some general properties of these effects are immediate. Since $\bm{W}_{t}$ is symmetric, they are symmetric as well. Because the diagonal entries of $\bm{W}_{t}$ are positive, each good has a positive effect on itself, and the magnitude is larger the higher the uncertainty about that good.

The structure of $\bm{W}_{t}$ can be understood through its inverse $\bm{Z}_{t}$, which can be interpreted as the adjacency matrix of a co-occurrence network where goods are nodes and two nodes $i,j$ are directly linked if and only if they have appeared together in at least one bundle. In this network, $w_{ij,t}=0$ whenever there is no path between $i$ and $j$. Hence, the consumer updates her estimated preferences only for goods that belong to the same connected component of the co-occurrence network.

Two additional properties are worth noting. First, for each good $i$ that has been consumed together with some other good, there exists at least one $j$ with $w_{ij,t}\leq 0$. Thus, every good is subject to at least one non-strictly positive correlation externality. Second, if the network described by $\bm{Z}_{t}$ is bipartite, then $w_{ij,t}\geq 0$ when $i$ and $j$ are on the same side, and $w_{ij,t}\leq 0$ when they are on opposite sides.\footnote{To see the first claim, pick any \(j\) with \(z_{ij,t}>0\). Since \(\bm{W}_t \bm{Z}_t = \bm{I}\), the \((i,j)\)-entry satisfies \(\sum_k w_{ik,t} z_{kj,t}=0\). But \(z_{ij,t}>0\), while all \(z_{kj,t}\ge 0\), so the sum can be zero only if at least one coefficient \(w_{ik,t}\) is negative. For the second claim, if the graph associated with $\mathbf Z_t$ is bipartite, its nodes can be partitioned into two sets $A$ and $B$. Let $\mathbf D$ be the diagonal matrix with $D_{ii}=1$ for $i\in A$ and $D_{ii}=-1$ for $i\in B$, and consider the auxiliary matrix $\mathbf D \mathbf Z_t \mathbf D$, introduced only for the proof. Because every nonzero off-diagonal entry of $\mathbf Z_t$ links nodes on opposite sides of the bipartition, premultiplying and postmultiplying by $\mathbf D$ changes the sign of those entries, while preserving symmetry and positive definiteness. Thus $\mathbf D \mathbf Z_t \mathbf D$ is a Stieltjes matrix, so its inverse is entrywise nonnegative. Since $(\mathbf D \mathbf Z_t \mathbf D)^{-1}=\mathbf D \mathbf Z_t^{-1} \mathbf D=\mathbf D \mathbf W_t \mathbf D$, this implies $D_{ii} w_{ij,t} D_{jj}\geq 0$ for every $i,j$. Now $D_{ii}D_{jj}=1$ when $i$ and $j$ lie on the same side of the bipartition, and $D_{ii}D_{jj}=-1$ otherwise. It follows that $w_{ij,t}\geq 0$ in the former case and $w_{ij,t}\leq 0$ in the latter. For the properties of Stieltjes matrices see, e.g., \cite{berman1994}.} 
{For example, if bundles are electronic devices and goods are the hardware and the software components (considered in aggregate for each device), the resulting network is bipartite. In this case, two software components (or two hardware components) always exert positive correlation externalities on each other, while hardware and software components exert negative correlation externalities. The intuition is that discovering that a hardware component performs well leads the consumer to reassess all the software experienced with it, which in turn feeds back on related hardware, and so on through the bipartite structure.}

\begin{example}[Line]
\label{ex:line}
Let us consider a path of consumption that creates a line network among goods as shown in Figure \ref{fig:line} and described by the following co-occurrences matrix and the implied variance-covariance matrix.\footnote{We assume all the goods are tasted one time in isolation to avoid multicollinearity.}   
\begin{figure}[h!]
  \centering
  \begin{tikzpicture}
    % Nodes on the line
    \foreach \x in {0, 1, 2, 3} {
        \node[circle, draw] (node\x) at (\x, 0) {};
    }
    % Lines connecting the nodes
    \draw (node0) -- (node1);
    \draw (node1) to[out=0, in=180] (node2);
    \draw (node2) to[out=0, in=180] (node3);
    \node[below=1.5cm] at (1.1,0) {$(a)$ Co-occurrences network $\bm{Z}_t$};
\end{tikzpicture}
\ \ \ \ \ \ \ \ \ \ \ \ \ \ \ \ 
\begin{tikzpicture}
    % Nodes on the line
    \foreach \x in {0, 1, 2, 3} {
        \node[circle, draw] (node\x) at (\x, 0) {};
    }
 \node[below=1.5cm] at (1.1,0) {$(b)$ Covariance network $\bm{W}_t$};
    % Lines connecting the nodes
    \draw[red] (node0) -- (node1) -- (node2) -- (node3) -- cycle;

    % Circular edges above the line
    \foreach \start/\end in {0/2, 1/3, 2/0, 3/1} {
        \draw[blue] (node\start) to[out=90, in=90] (node\end);
    }

    % Circular edge below the line
    \draw[red] (node0) to[out=-90, in=-90] (node3);
\end{tikzpicture}
    \caption{\footnotesize  Line}
    \label{fig:line}
\end{figure}

$$\bm{Z}_{t-1}=
\begin{bmatrix}
2 & 1 & 0 & 0 \\
1 & 3 & 1 & 0 \\
0 & 1 & 3 & 1 \\
0 & 0 & 1 & 2 \\
\end{bmatrix}
%\ \ \ \ \lambda(\bm{Z}_{t})=3+\sqrt{2}\ \ \ \ \mathbf{v}(\bm{Z}_{t-1}) = \begin{bmatrix}
 %   1 \\
  %  1 + \sqrt{2} \\
   % 1 + \sqrt{2} \\
   % 1
%\end{bmatrix}
\ \ \ \ \ \ \ \ \ \ \ \
\bm{W}_{t-1}=\frac{1}{21} \cdot \begin{bmatrix}
13 & -5 & 2 & -1 \\
-5 & 10 & -4 & 2 \\
2 & -4 & 10 & -5 \\
-1 & 2 & -5 & 13 \\
\end{bmatrix}
%\ \ \ \ \lambda(\bm{W}_{t})=1\ \ \ \ \mathbf{v}(\bm{W}_{t-1}) = 
%\begin{bmatrix}
 %   -1 \\
  %  1 \\
   % -1 \\
   % 1
%\end{bmatrix}
$$
\\
Given this path of consumption, if the consumer tastes again good 1 alone and experiences a unitary positive surprise, $\Delta u_{t} =+ 1$, the updating of the estimated preferences of the directly connected good and the good at distance three decrease, whereas those of the good at distance two increases. See Figure \ref{fig:line2}
 below.
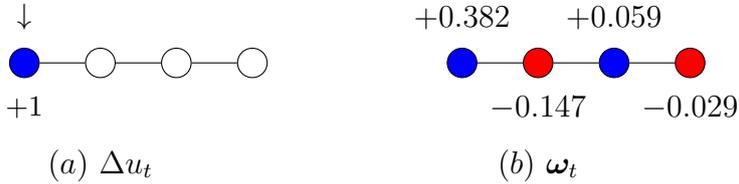
\begin{figure}[h!]
\label{ex:line2}
  \centering
\begin{tikzpicture}
    % Nodes on the line
    \foreach \x/\col in {0/blue, 1/white, 2/white, 3/white} {
        \node[circle, draw, fill=\col] (node\x) at (\x, 0) {};
    }

\node[above=0.3cm] at (node0) {\small $\downarrow$};
\node[below=0.3cm] at (node0) {\small $+1$};
    % Lines connecting the nodes
    \draw (node0) -- (node1);
    \draw (node1) to[out=0, in=180] (node2);
    \draw (node2) to[out=0, in=180] (node3);

    \node[below=1cm] at (node1) {$(a)$ $\Delta u_t$};
\end{tikzpicture}
\ \ \ \ \ \ \ \ \ \ \ \ 
\begin{tikzpicture}
    % Nodes on the line
    \foreach \x/\col in {0/blue, 1/red, 2/blue, 3/red} {
        \node[circle, draw, fill=\col] (node\x) at (\x, 0) {};
    }
\node[above=0.3cm] at (node0) {$+0.382 $};% $.51$};
\node[below=0.3cm] at (node1) {$-0.147 $};%-0.26$};
\node[above=0.3cm] at (node2) {$+0.059$};%+0.04$};
\node[below=0.3cm] at (node3) {$-0.029$};%- 0.009$};

    % Lines connecting the nodes
    \draw (node0) -- (node1);
    \draw (node1) to[out=0, in=180] (node2);
    \draw (node2) to[out=0, in=180] (node3);
       \node[below=1cm] at (node1) {$(b)$ $\bm{\omega}_t$ %$\Delta \hat{\bm{\beta}}_t$
       };
\end{tikzpicture}
    \caption{\footnotesize  Updating after tasting good 1 and experiencing a unitary positive surprise.}
    \label{fig:line2}
\end{figure}

\end{example}

\subsection{Data Design and Learning}\label{sec:nolearn}

The previous section and examples show how consumption surprises propagate across connected goods, leading to preference revisions that depend on the structure of the network of goods that summarizes the consumer’s consumption history. 
We now build on that mechanism and discuss how, at each point in time, bundles can be designed to target the consumer’s current estimation error, thereby shaping the updating of estimated preferences.

Let us define $\Delta\bm{\beta}_{t}:=\hat{\bm{\beta}}_{t}-\bm{\beta}$ the  estimation error at $t$.
Then, the expected surprise after consuming a bundle $\mathbf{x}_{t+1}$ is 
$\mathbb{E}\left[\Delta u_{t+1}\right]= -\mathbf{x}'_{t+1}\Delta {\bm{\beta}}_t=-\sum_{i \in I }x_{i,t+1}\Delta{\beta}_{i,t}
$.

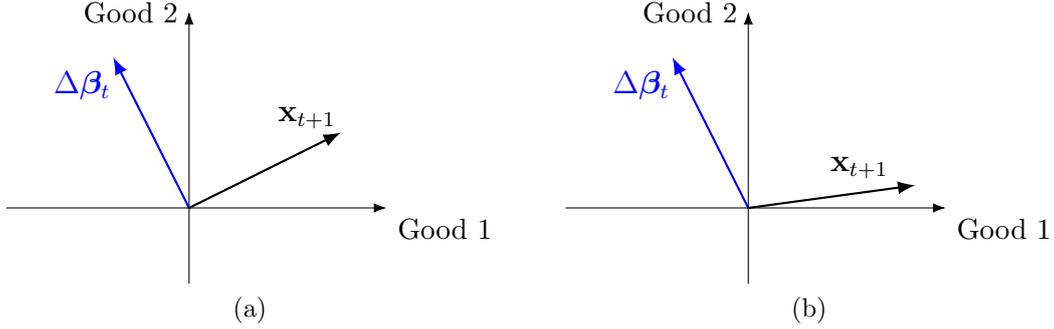
\begin{figure}[h!]
\centering

%---------------------------------
% (a) Orthogonal case
%---------------------------------
\subfloat[\small ]{
\begin{tikzpicture}[
    line cap=round,
    line join=round,
    scale=1.0,
    baseline={(O)}
]

  % Origin
  \coordinate (O) at (0,0);

  % Axes
  \draw[-{Latex}] (-2.4,0) -- (2.6,0) node[below right] {\small Good 1};
  \draw[-{Latex}] (0,-1.0) -- (0,2.6) node[left] {\small Good 2};

  % Vectors
  \coordinate (Db) at (-1,2);
  \coordinate (xv) at (2,1);

  \draw[-{Latex}, thick, blue] (O) -- (Db)
    node[pos=0.6, xshift=-23pt, yshift=12pt] {$\Delta\bm{\beta}_t$};

  \draw[-{Latex}, thick] (O) -- (xv)
    node[pos=0.6, xshift=10pt, yshift=17pt] {$\mathbf{x}_{t+1}$};

\end{tikzpicture}
}
\hspace{1em}
%---------------------------------
% (b) Biased case
%---------------------------------
\subfloat[\small ]{
\begin{tikzpicture}[
    line cap=round,
    line join=round,
    scale=1.0,
    baseline={(O)}
]

  % Origin
  \coordinate (O) at (0,0);

  % Axes
  \draw[-{Latex}] (-2.4,0) -- (2.6,0) node[below right] {\small Good 1};
  \draw[-{Latex}] (0,-1.0) -- (0,2.6) node[left] {\small Good 2};

  % Vectors
  \coordinate (Db) at (-1,2);
  \coordinate (xv) at (2.2,0.3);

  \draw[-{Latex}, thick, blue] (O) -- (Db)
    node[pos=0.6, xshift=-23pt, yshift=12pt] {$\Delta\bm{\beta}_t$};

  \draw[-{Latex}, thick] (O) -- (xv)
    node[pos=0.6, xshift=4pt, yshift=10pt] {$\mathbf{x}_{t+1}$};

\end{tikzpicture}
}

\caption{ \footnotesize Example with $\Delta \beta_1<0$ and $\Delta \beta_2>0 $.
(a) Orthogonal bundles $\mathbf x_{t+1} \perp \Delta \bm{\beta}_t$ do not generate expected surprise and thus prevent learning;
(b) Bundles such that $\mathbf x_{t+1}' \Delta\bm{\beta}_t < 0$ will generate positive expected surprise.
}
\label{fig:geometry}

\end{figure}

Let us first notice that not every consumption experience leads the consumer to revise her estimated preferences in expectation. Learning requires that realized utility systematically deviates from what the consumer’s current beliefs predict. When the chosen bundle combines goods in such a way that overestimation and underestimation errors offset each other, expected realized utility coincides with predicted utility. Formally, this occurs whenever
$\mathbb{E}[\Delta u_{t+1}] = -\mathbf{x}_{t+1}' \Delta \bm{\beta}_t = 0$.
In this case, in expectation, the experience generates  \textit{no consumption surprise} and therefore conveys no information about the direction of the estimation error. As a consequence, beliefs remain unchanged in expectation, $ \mathbb{E}[\hat{\bm{\beta}}_{t+1}] = \hat{\bm{\beta}}_t$.
Geometrically, the set of such bundles forms a hyperplane orthogonal to the current estimation error vector $\Delta \bm{\beta}_t$ (see Figure~\ref{fig:geometry}a).

We now turn to consumption experiences that \emph{do} generate learning. When a bundle places relatively more weight on goods whose values are currently \emph{underestimated}, predicted utility is biased downward and the bundle generates, in expectation, \emph{positive consumption surprise}. Formally, $\mathbb{E}[\Delta u_{t+1}]>0$ if and only if $\mathbf{x}'_{t+1}\Delta\bm{\beta}_t<0$ (see Figure~\ref{fig:geometry}b). Similarly, bundles that place relatively more weight on goods whose values are currently \emph{overestimated} generate, in expectation, \emph{negative consumption surprise}.

Proposition~\ref{prop:1} shows that a positive (or negative) surprise generated by the bundle $\mathbf{x}_{t+1}$ is transmitted across goods through the covariance structure encoded in $\bm{W}_t$. In particular, the expected revision of $\hat\beta_{i,t}$ is proportional to $\sum_{j=1}^n w_{ij,t} x_{j,t+1}$, scaled by the magnitude of the surprise. Thus, the bundle determines both the sign of expected surprise,  through the inner product $\mathbf{x}'_{t+1}\Delta\bm{\beta}_t$, and how the resulting revision is distributed across goods, through the exposure vector $\bm{W}_t\mathbf{x}_{t+1}$.
As a consequence, learning need not always move estimated preferences toward their true values for every good. Following a consumption surprise, the induced revision can also increase the estimated value of goods that are already overestimated, or decrease the estimated value of goods that are already underestimated. The presence and direction of such spillovers depend on the sign pattern implied by $\bm{W}_t$.

To make this mechanism transparent, the following result characterizes expected updating after consuming bundles composed by two goods and focusing on the case with opposite-signed estimation errors---one good overestimated and the other underestimated.

%To make the mechanism transparent, we focus on bundles composed by   two-good  case with a nonnegative bundle and opposite-signed estimation errors---one good overestimated and the other underestimated---and characterize how the bundle composition shapes expected updating.

\begin{proposition}
Consider $i, j$ such that   $\Delta\beta_{i,t}>0$  and
$\Delta\beta_{j,t}<0$ and a bundle $\mathbf{x}_{t+1}=x_{i,t+1} \cdot \mathbf{e}_i +x_{j,t+1} \cdot \mathbf{e}_j $ with $x_{i,t+1}>0$ and $x_{j,t+1}>0$. Then, in expectation:

\begin{enumerate}
    \item If $\dfrac{x_{j,t+1}}{x_{i,t+1}}=-\dfrac{\Delta\beta_{i,t}}{\Delta\beta_{j,t}}$, then there is no preference updating, i.e. $\mathbb{E}[\hat{\bm{\beta}}_{t+1}]=\hat{\bm{\beta}}_t$.
    
    \item If $\dfrac{x_{j,t+1}}{x_{i,t+1}}>-\dfrac{\Delta\beta_{i,t}}{\Delta\beta_{j,t}}$, then the estimated  preferences of both goods increase, i.e. $\mathbb{E}[\hat{{\beta}}_{i,t+1}]>\hat{{\beta}}_{i,t}$ and $\mathbb{E}[\hat{{\beta}}_{j,t+1}]>\hat{{\beta}}_{j,t}$, if and only if either
    \begin{itemize}
    \item[(a)] the goods are positively correlated---i.e., $w_{ij,t}>0$;
    \item[(b)] or, the bundle is sufficiently balanced---i.e., 
    $\dfrac{|w_{ij,t}|}{w_{jj,t}} < \dfrac{x_{j,t+1}}{x_{i,t+1}} < \dfrac{w_{ii,t}}{|w_{ij,t}|}$.
    \end{itemize}
\end{enumerate}
\label{cor:1}
\end{proposition}

Part 1 of Proposition \ref{cor:1} refers to bundles orthogonal to the estimation error where the relative weights on the two goods are chosen so that their estimation errors offset each other in the aggregate. 
If the two estimation errors have opposite signs, %With nonnegative bundles, 
such offsetting is possible only when elements of $\mathbf{x}_{t+1}$ are non-negative, otherwise 
 every non-negative and non-degenerate bundle produces a surprise, and the consumer keeps updating in that direction until the bias is corrected. 
 Orthogonal bundles are economically interesting  because they preserve the provider’s ability to monetize current demand while preventing systematic correction of consumer misperceptions.

Part 2 of Proposition \ref{cor:1}  considers bundles that put more weight on the underestimated good, so that expected surprise is positive and updating occurs in expectation.  The sign of the updating for each good is determined by the interaction between bundle composition and the covariance structure. If the two goods are positively correlated (2.a), a positive surprise  always increases both estimated coefficients, including that of the good that is initially overestimated. If the goods are negatively correlated (2.b), a joint increase requires the bundle to be sufficiently balanced: when one good dominates the bundle, the source of the surprise is more identifiable and updating corrects the estimation error of that good; when the bundle is more balanced, attribution is weaker and the estimated preferences of both goods can move in the same direction. By the same logic, negative surprise can further reduce the estimated coefficient of a good that is already underestimated. %Figure~\ref{fig:directbias} illustrates these cases.

\begin{remark}
    Data design through bias-targeted bundling can both shut down learning and amplify misperceptions, making overvalued goods even more overvalued and undervalued goods even more undervalued.
\end{remark}

\begin{figure}[h!]
\centering
\includegraphics[scale=0.45]{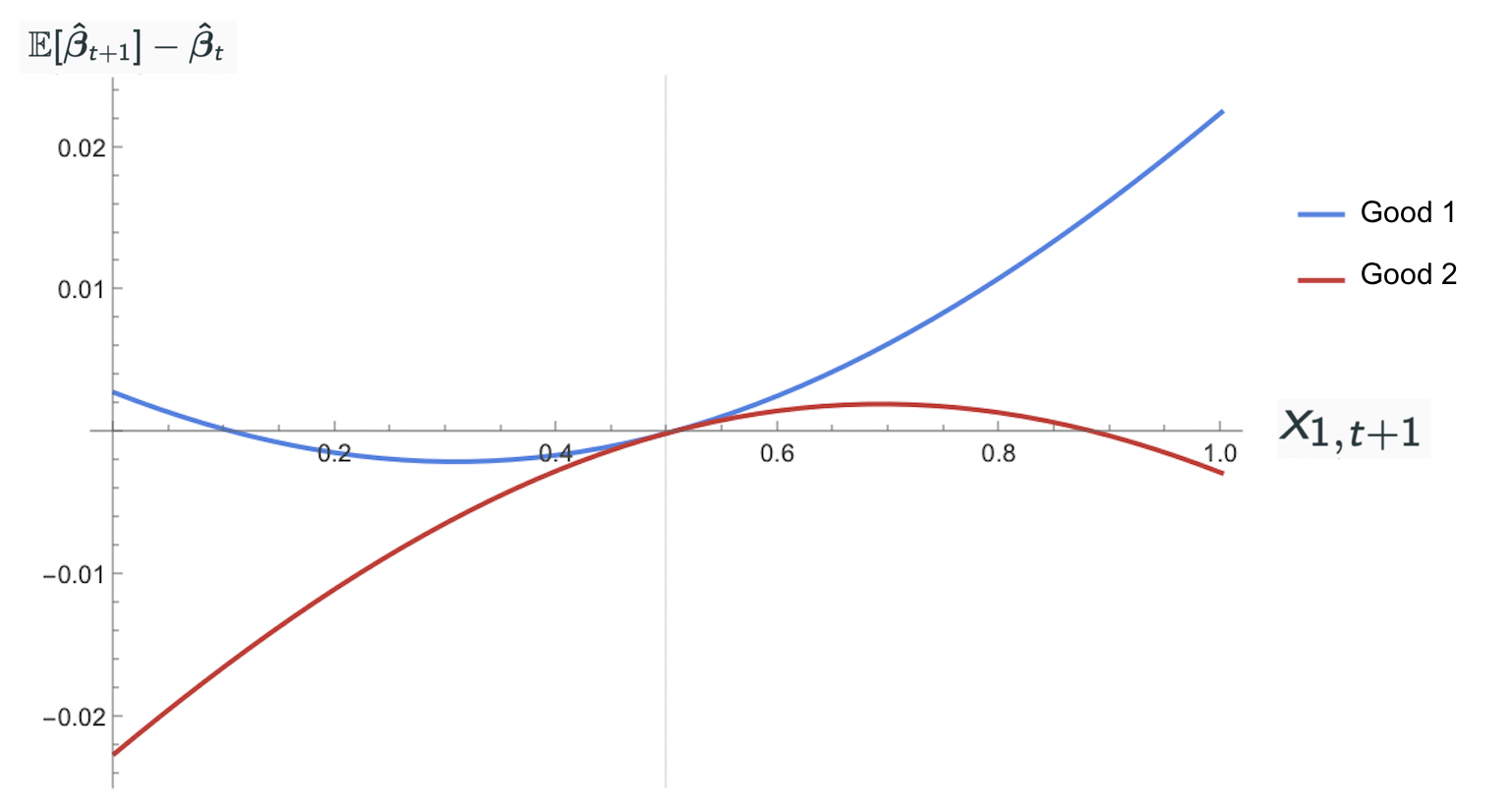} 
\caption{\footnotesize{
Updates $\mathbb{E}[\bm{\hat{\beta}}_{t+1}] - \bm{\hat{\beta}}_{t}$ after consuming the bundle $\mathbf{x}_{t+1} = (x_{1,t+1}, 1 - x_{1,t+1})$, when  $w_{12,t}<0$, and the vector of
biases in estimated preferences is
$\Delta\bm{\beta}_{t} = (-0.2, 0.2)$.
If $x_{1,t+1} = \frac{1}{2}$, $\mathbf{x}_{t+1} \perp \Delta\bm{\beta}_{t}$.}}
\label{fig:directbias}
\end{figure}

Figure~\ref{fig:directbias} illustrates the mechanism in a two-good environment, $I=\{1,2\}$, where the goods are negatively correlated given past consumption, $w_{12,t}<0$. The figure plots the expected change in estimated preferences, $\mathbb{E}[\hat{\bm{\beta}}_{t+1}]-\hat{\bm{\beta}}_{t}$, as the bundle $\mathbf{x}_{t+1}=(x_{1,t+1},1-x_{1,t+1})$ varies.

 At $x_{1,t+1}=1/2$, the bundle is orthogonal to the bias vector $\Delta\bm{\beta}_t$, so $\mathbb{E}[\Delta u_{t+1}]=0$ and there is no expected updating. Moving away from this orthogonal point generates systematic
updates in beliefs. In a neighborhood of $x_{1,t+1}=1/2$ increasing $x_{1,t+1}$---that is, placing more weight on the underestimated good---locally increases the estimated preferences of both goods, while decreasing it lowers both.
As the bundle moves toward the extremes and one good becomes dominant, the source of the consumption surprise becomes identifiable and the updating of estimated preferences reduces the biases%estimated preferences move toward the true ones
: the estimated preference of the underestimated good increases, while that of the overestimated good decreases, regardless of whether expected surprise is positive or negative.

These local results characterize how bundle composition shapes the direction of belief updating at a given history. We now turn to the dynamic counterpart of the same question: how repeated bundle design affects asymptotic learning and the speed at which estimation errors vanish.

%Here, we start with the analysis of what are the conditions that lead to have correlated or anti-correlated externalities, and on how they depend on the network obtained from the co--occurrence matrix $\bm{Z}_{t}$.
%In particular, there is a relationship between the length and weightings of the path between $i$ and $j$ and the sign of $w_{ij,t}$. As we have seen form the examples in Section \ref{sec:dynup}, if the network is a bi-partited graph, then, the sign of $w_{ji,t}$ is  positive if there is a path of odd length between $i$ and $j$, and negative if there is a path of even length between $i$ and $j$ (recall the smartphone example in the introduction, with hardware and software components). Note also that a particular case of bipartite network is a network without loop, that is a \emph{tree}.
%This observation, and more results, follows from the spectral analysis of those matrices $\bm{Z}_{t}$ and $\bm{W}_{t}$.

%\subsubsection{Long-run}\label{subsec:longrun}
\paragraph{Long-run}
Having characterized the local geometry of updating, 
we now analyze long-run learning, identifying the 
conditions under which $\hat{\bm{\beta}}_t$ converges 
to $\bm{\beta}$ and the consumption patterns that make
learning faster. We measure learning quality by the MSE, $\mathbb{E}\left[\|\Delta\bm{\beta}_t\|_2^2\right]$, and ask which bundle sequences minimize it at every $t$ simultaneously. For this analysis, we normalize $\|\mathbf{x}_t\|_2 = 1$, so that the results depend solely on the relative composition of the bundle. We also define $\lambda^{max}_t:=\lambda^{max}(\bm{Z}_t)$ the maximal eigenvalue of matrix $\bm{Z}_t$.

\begin{proposition}
Consider the estimated preferences dynamics in equation \eqref{eq: betap1}.
\begin{itemize}
    \item 
Then, $\mathrm{plim}_{t\rightarrow \infty} \Delta \bm{\beta}_{t}=\bm{0}$ if and only if the bundle sequence $(\mathbf{x}_t)_{t\geq 1}$ generates information along every direction of $\mathbb{R}^n$, that is, $\lambda^{\min}_t \to \infty$ as $t \to \infty$.

\item Moreover, under $||\mathbf{x}_t||_2 = 1$, $\mathbb{E}\big[||\Delta\boldsymbol{\beta}_t||_2^2\big] 
\geq \frac{\sigma^2 n^2}{t}$, with equality if
each observation contains a single good and all singletons 
appear with equal frequency. \end{itemize}
\label{prop:LT}
\end{proposition}

The first part of Proposition~\ref{prop:LT} characterizes when long-run learning of true preferences succeeds: the bundle sequence must keep generating informational variation in every direction of the space of goods. This condition can fail in several ways. A simple case is when some goods are eventually dropped from all bundles: their coefficients are then never identified, and the corresponding directions of $\mathbb{R}^n$ accumulate no further information. A more subtle failure occurs when the provider offers bundles orthogonal to the current estimation error $\Delta\bm{\beta}_t$. From Proposition~\ref{cor:1}, such bundles generate zero expected consumption surprise: the estimator not only fails to converge in the limit, but exhibits no expected drift toward the truth at any period, so the directed component of belief revision is frozen entirely. Both pathologies share the same root cause---the bundle sequence collapses into a region of $\mathbb{R}^n$ that does not span the parameter space.

%The first part of Proposition~\ref{prop:LT} identifies the two conditions that make long-run learning of true preferences fail. Condition $(i)$ implies that, for learning to occur, each good must continue to be consumed. It reflects the consistency of the OLS estimator: by the law of large numbers, as the number of observations grows to infinity, the estimator converges to the true parameter.Condition $(ii)$ implies that, for learning to occur, bundles must generate sufficient informational variation. It follows from Proposition \ref{cor:1}: when a bundle is orthogonal to $\Delta\bm{\beta}_t$, expected consumption surprise is zero, so the estimates follow a random walk with zero expected drift and fail to converge to the true parameters.
%Notice that if we restrict attention to nonnegative bundles, $\mathbf{x}_t \ge 0$, condition (ii) can hold only if at least one good is overvalued and another is undervalued.

The second part of Proposition~\ref{prop:LT} concerns 
the speed of learning. The total mean squared error 
is minimized at every $t$ when each good is consumed in isolation
at a time. Intuitively, with only single-good bundles 
the co-occurrence matrix $\mathbf{Z}_t$ is gradually 
driven toward a diagonal form, as cross-covariances 
between distinct goods vanish asymptotically, and the 
consumer learns about each good without confounding. 
Any bundling inflates the total mean squared error above 
its minimum, slowing learning or even stopping it 
altogether.

\begin{remark}
Single--goods exposure provides the benchmark for fastest preference discovery; bundles orthogonal to the estimation error provide the benchmark for halting it altogether.
\end{remark}
\begin{comment}
\begin{figure}[htbp]
\centering
%\subfloat[\textit{Popularity-biased} bundle, $\mathbf{v}^N$]{
%\includegraphics[scale=0.34]{vnintervection}} \ \ \ \ \ \ \ \ \ \
%\subfloat[\textit{Correlation-breaking} bundle, $\mathbf{v}^C$]{%
%\includegraphics[scale=0.34]{vcintervection}}
%\\
\subfloat[Orthogonal bundle, $\perp \Delta \beta$]{
\includegraphics[scale=0.35]{xortintervention}} \ \ \ \ \ \ 
\subfloat[Single-goods interventions ]{
\includegraphics[scale=0.35]{singleintervention}} 
\caption{Learning dynamics with two goods ($|I|=2$) under different bundle designs. With $\bm{\beta}=(1,1)$ and $\hat{\bm{\beta}}_0=(0.9,1.2)$.}
\label{fig:interventions}
\end{figure}
\end{comment}
%,The second part of Proposition~\ref{prop:LT} concerns the speed of learning. Learning is fastest when only one good is consumed at a time, because consuming multiple goods simultaneously introduces covariance noise across goods. Intuitively, with only single-good bundles the co-occurrence matrix $\mathbf{Z}_t$ is gradually driven toward a diagonal form, as cross-covariances between distinct goods vanish asymptotically. This implies that a provider who aims to favor consumers’ preference discovery would tend to provide single goods, to induce cleaner inference and avoid confounding. Conversely, bundling goods together can slow learning down, or even stop it.

\subsection{Robust Design}
\label{sec:spectral}

In the previous section we showed that, when the estimation error $\Delta\bm{\beta}_t$ is observed, bundles can be designed to target specific biases. In many applied contexts, however, true and estimated preferences are private information, so $\Delta\bm{\beta}_t$ is unknown and such interventions are infeasible. This section studies bundle designs that are robust to (that is, independent of) the consumer’s true preferences and current estimation errors. We show how bundle composition shapes the geometry of exposure and how the resulting spectral structure of past consumption determines the speed of learning.

\begin{comment}

Interventions discussed in the previous section rely on knowledge of the current estimation error $\Delta\bm{\beta}_t$, so bundles can be designed to target it, either to stop learning or to manipulate specific biases.
In many applied settings, however, true and estimated preferences are private information, so $\Delta\bm{\beta}_t$ is not observed and and such interventions are infeasible. 

Nevertheless, learning speed depends on the correlation structure of consumption data. In this section, we show how bundle composition controls the geometry of exposure and the resulting spectral structure of past consumption, thereby affecting learning without requiring knowledge of true preferences or current estimation errors.

By controlling the geometry of exposure, data design can promote or hinder learning using only  correlation structure of consumption data. Thus, by exploiting the spectral
structure of past consumption data is therefore robust to the consumer’s true preferences and current estimation errors.
\end{comment}

\subsubsection{Spectral Analysis}
Let us first notice that since the co-occurrence matrix $\bm{Z}_{t}$  is positive definite and symmetric, it has all positive real eigenvalues.
%\footnote{Recall that bundles are defined in the general space $\mathbb{R}^n$.In some parts of this section we restrict attention to feasible bundleswith non–negative components, $\mathbf{x}_t\in\mathbb{R}^n_+$.Under this restriction some eigenvector directions cannot be implemented exactly as feasible bundles.}
Thus, we can decompose it as $$\bm{Z}_{t}=\bm{V}_t\bm{ \Lambda}_t\bm{V}_t',$$ where $\bm{ \Lambda}_t$ is a diagonal matrix, whose diagonal entries $(\lambda_{i,t})_{i \in I}$ are the eigenvalues of $\bm{Z}_{t}$ ordered in descending order from greatest to smallest, and $\bm{V}_t:=(\mathbf{v}_{1,t},...,\mathbf{v}_{n,t})$ is an orthonormal  $(n \times n)$-matrix where the $i$-th column, $\mathbf{v}_{i,t}$, is the real eigenvector of $\bm{Z}_{t}$ associated to the eigenvalue $\lambda_{i,t}$. 
We define $\lambda^{max}_t:=\lambda_{1,t}$ as the leading eigenvalue of the co-occurrences matrix $\bm{Z}_{t}$ and ${\mathbf{v}}^{N}_t:=\mathbf{v}_{1,t}$ its associated eigenvector, which corresponds to the largest principal component or the \textbf{eigenvector centralities} of the goods in the co-occurrences network.
Therefore, ${\mathbf{v}}^{N}_t$ identifies the dominant co-consumption pattern in the data observed up to time $t$. Goods with larger entries in ${\mathbf{v}}^{N}_t$ contribute more strongly to the most recurrent direction of joint exposure in the consumer's history. Accordingly, bundles aligned with ${\mathbf{v}}^{N}_t$ place relatively greater weight on goods that have been most frequently experienced together.

Since $\bm{W}_{t}=\bm{Z}^{-1}_{t}$, the same spectral decomposition implies $\bm{W}_{t}=\bm{V}_t\bm{\Lambda}^{-1}_t\bm{V}'_t$. Let $\lambda^{\min}_t:=\lambda_{n,t}$, so that $1/\lambda^{\min}_t$ is the largest eigenvalue of $\bm{W}_{t}$, and denote by ${\mathbf{v}}^{C}_t:=\mathbf{v}_{n,t}$ the associated eigenvector. Then ${\mathbf{v}}^{C}_t$ summarizes the dominant sign structure of correlations across goods: its entries split goods into two groups, with correlated externalities within each group and anti-correlated externalities across groups.\footnote{This sign-based split is the analogue of the standard spectral bipartition used in network analysis: goods with positive entries lie on one side and goods with negative entries on the other, reflecting a dependence structure in which cross-group links are relatively more important than within-group ones (on this, see e.g., \citealt{bramoulle2014, galeotti2020}).} For this reason, we refer to the entries of ${\mathbf{v}}^{C}_t$ as the \textbf{correlation centralities}.

A useful spectral measure for us is the \emph{condition number}. For symmetric square matrices the condition number is the ratio between the maximal and minimal eigenvalues. In our case,  ${\bf Z}_t$ and ${\bf W}_t$ have the same condition number because, as they are inverse matrices, eigenvalues are reciprocal of each other, so the maximum and minimum are swapped. 
Thus,
\begin{equation}
\kappa_t :=\kappa (\bm{Z}_t) \equiv \kappa (\bm{W}_t)= \frac{\lambda^{max}_t}{\lambda^{min}_t} \in [1, \infty ) .
\label{eq:conditionnumber}
\end{equation}
%The condition number is always well defined because, as discussed above, all eigenvalues are strictly positive.
% The condition number measures how unstable a linear operator is. In particular, if $\kappa_t \rightarrow 1$ the system in equation \eqref{eq: beta} is well-conditioned and the solutions are reliable and stable. The larger $\kappa_t $ the more the system is ill-conditioned and highly sensitive to perturbations.  When $\kappa_t \rightarrow \infty$, the matrix   ${\bf Z}_t$ (and ${\bf W}_t$) becomes close to being singular---i.e., there is almost multicollinearity among goods and the system approaches the situation in which there are infinite solutions.
The condition number measures how unevenly information is distributed across directions in the parameter space. %In particular, 
If $\kappa_t$ is close to 1 the system in equation \eqref{eq: beta} is well-conditioned and learning proceeds at similar rates across dimensions. The larger $\kappa_t$, the more asymmetric the learning process becomes, as some directions become much more precisely identified than others.

The next proposition shows how the spectral structure of $\bm{Z}_t$ can be used to control the speed of
learning{, measured by one step changes in the MSE,} in a robust way. Here, ``robust" means that the effect does not depend on the {knowledge and} direction of the current estimation error $\Delta\bm{\beta}_t$.
Specifically, we identify bundle choices that increase or decrease the condition number, and thereby slow down or speed up learning.

\begin{proposition}
\label{prop:robust}
Restrict attention to feasible bundles with $\|\mathbf{x}_{t+1}\|_2=1$. Then, at each time $t$:
\begin{itemize}
    \item The bundle $\mathbf{x}_{t+1} \equiv \mathbf{v}^N_t$ maximally increases the condition number, $\kappa_{t+1}>\kappa_{t}$, and is the most effective robust way to \textit{slow down} learning, both locally and in the long run, since this direction remains optimal if chosen repeatedly, implying $\kappa_t\to\infty$.
    \smallskip
    \item The bundle $x_{t+1}\equiv v_t^C$ weakly decreases the condition number, $\kappa_{t+1}\le \kappa_t$, and is the
most effective robust way to speed up learning locally. The decrease is strict whenever the smallest
eigenvalue of $\mathbf Z_t$ is simple and $\kappa_t>1$.
\end{itemize}
\end{proposition}

% \red{new proposal above}
%     \begin{proposition}
% \label{prop:robust}
% Restrict attention to feasible bundles with $\|x_{t+1}\|_2=1$.
% Then, at each date $t$:

% \begin{enumerate}
%     \item The bundle $x_{t+1}=v_t^N$, where $v_t^N$ is the eigenvector associated with $\lambda_t^{\max}$, yields the largest one-step increase in the condition number:
%     \[
%     \kappa_{t+1}>\kappa_t.
%     \]
%     Moreover, if this direction is chosen repeatedly over time, it remains optimal at every subsequent date and implies $\kappa_t\to\infty$. Hence, $v_t^N$ is the most effective robust direction to slow down learning both locally and in the long run.

%     \item The bundle $x_{t+1}=v_t^C$, where $v_t^C$ is the eigenvector associated with $\lambda_t^{\min}$, yields the largest one-step decrease in the condition number:
%     \[
%     \kappa_{t+1}<\kappa_t.
%     \]
%     Hence, $v_t^C$ is the most effective robust direction to speed up learning \emph{locally}. In general, however, this improvement is only local. 
% \end{enumerate}
% \end{proposition}

%{\bf [should we add in the formal statement the result that $\mathbf{v}^N_t$ remains the same?] @  \red{I would keep this only for the appendix and the discussion}}

%{\bf [add in the disxcussion that  $\mathbf{v}^N_t$ is always all positive -- $\mathbf{v}^N_t$ is never]  @  \red{If there are negative elements in x and in Z I think that also the elemetns of vn can have negative entris } }

\begin{figure}[htbp]
\centering
\subfloat[\textit{Popularity-biased} bundle, $\mathbf{v}^N$]{
\includegraphics[scale=0.35]{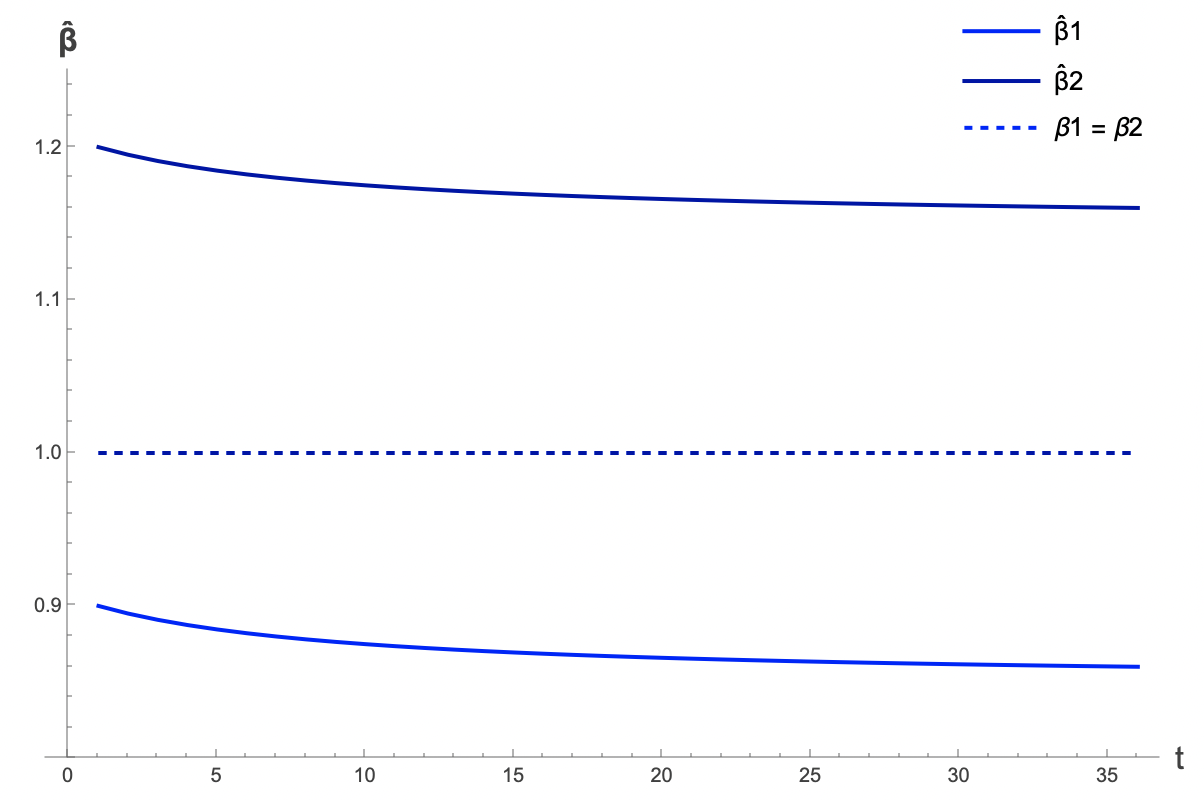}} \ \ \ \ \ \ 
\subfloat[\textit{Correlation-breaking} bundle, $\mathbf{v}^C$]{%
\includegraphics[scale=0.35]{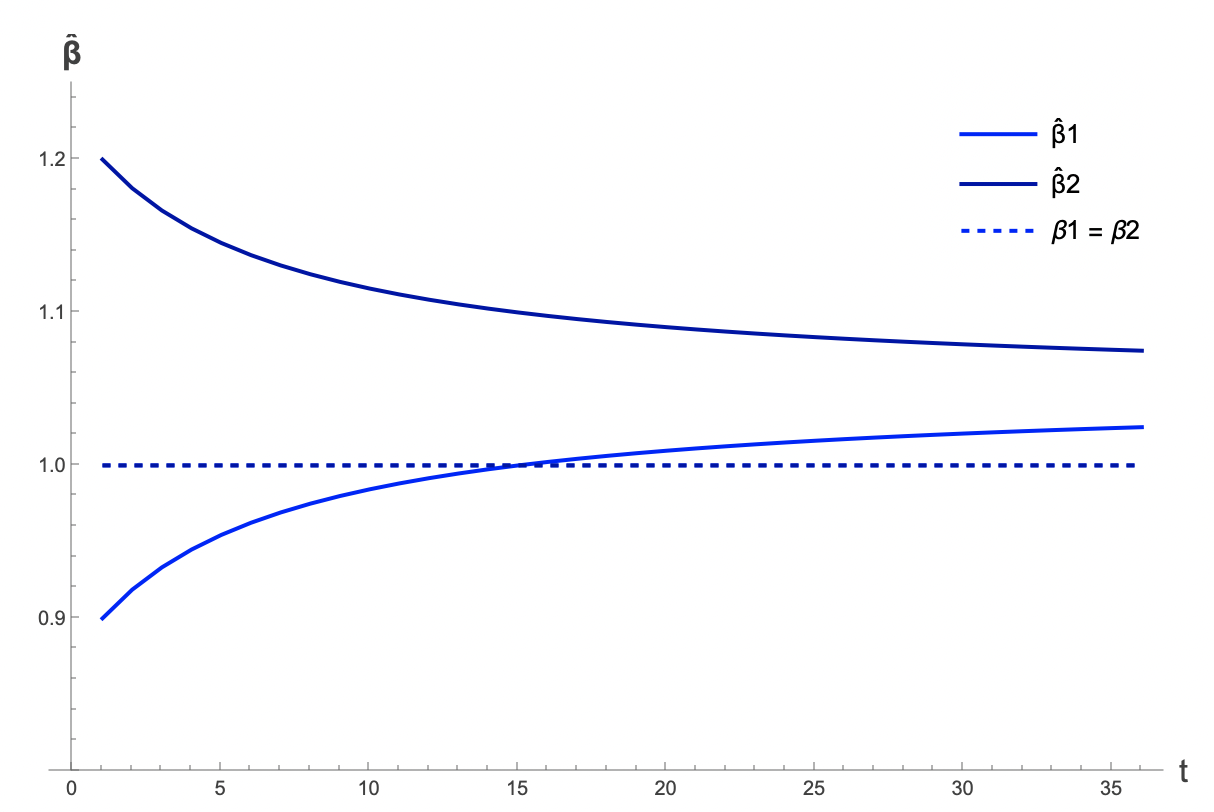}}
\\
\subfloat[Orthogonal bundle, $\perp \Delta \beta$]{
\includegraphics[scale=0.34]{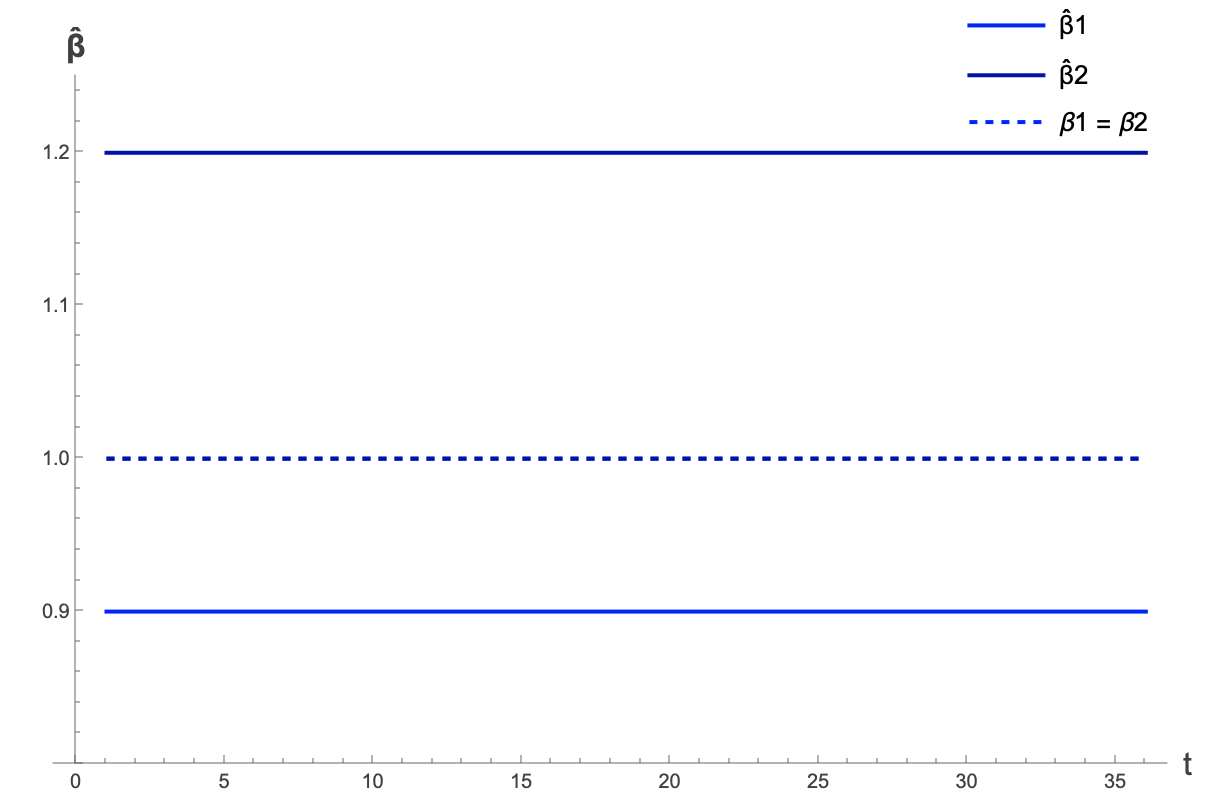}} \ \ \ \ \ \ 
\subfloat[Single-goods interventions ]{
\includegraphics[scale=0.34]{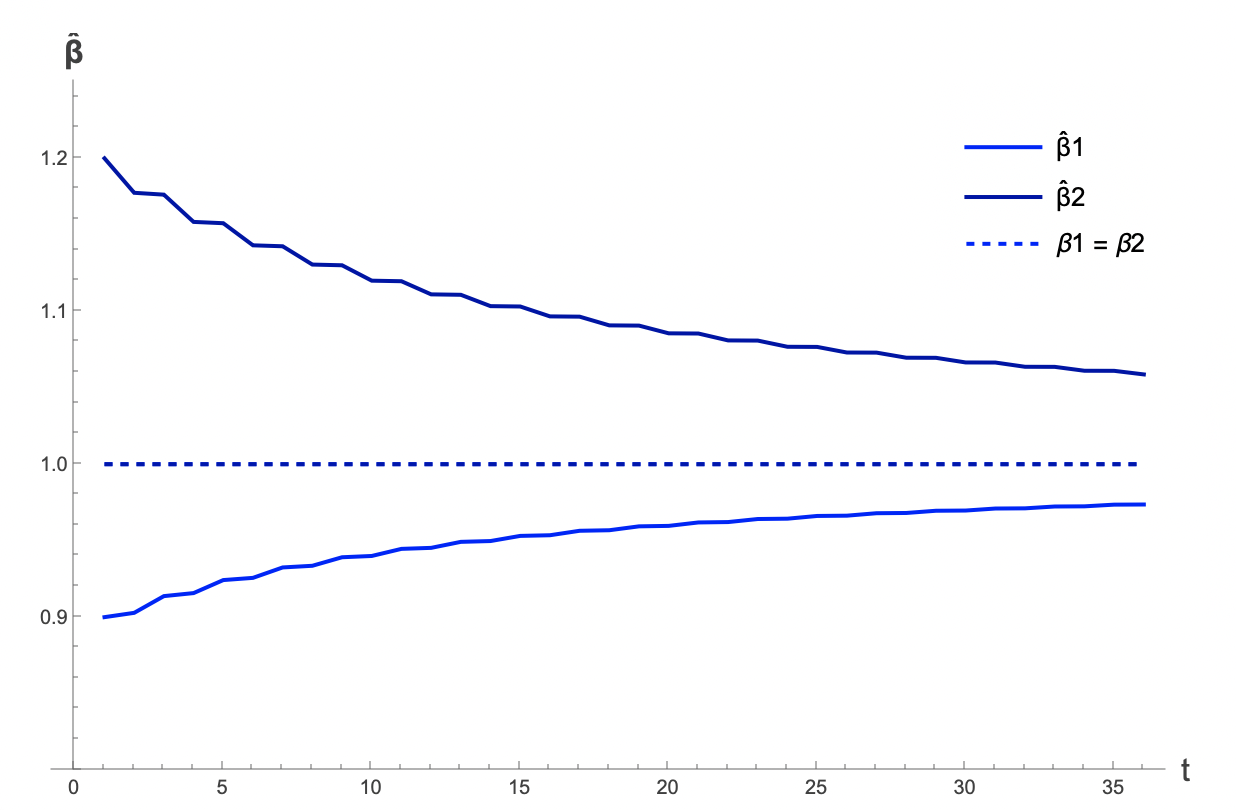}} 
%\caption{ \footnotesize  Learning dynamics with two goods ($|I|=2$) under different bundle designs, starting from $\hat{\bm{\beta}}_0=(0.9,1.2)$, with $\bm{\beta}=(1,1)$ and  $\sigma^2 = 0$.}
\caption{\footnotesize  Learning dynamics with two goods ($|I|=2$) under different bundle designs, starting from $\hat{\bm{\beta}}_0=(0.9,1.2)$, with $\bm{\beta}=(1,1)$ and  $\sigma^2 = 0$.}
\label{fig:interventions}
\end{figure}

Proposition \ref{prop:robust} provides a robust characterization of how bundling affects the speed of learning through the condition number. Consuming $\mathbf{v}^N_t$ raises $\kappa_t$ and is the most effective robust way to slow down learning. 
Technically, adding $\mathbf{x}_{t+1}\equiv \mathbf{v}^N_t$ leaves the eigenvectors of $\bm{Z}_t$ unchanged and only raises the leading eigenvalue, so that $\lambda^{\max}_{t+1}=\lambda^{\max}_t+1$, and $\kappa_t$ diverges asymptotically.
Intuitively, $\mathbf{v}^N_t$ consists of the most recurrent co-consumption pattern in the consumer’s history, so the resulting observation is largely redundant: it reinforces information already embedded in $\bm{Z}_t$ rather than adding new variation. Economically, this corresponds to a form of \emph{popularity bias}: over-exposing the consumer to goods that are already very frequently consumed together reduces the informativeness of new data and makes inference increasingly ill-conditioned. 
This is illustrated in Figure~\ref{fig:interventions}(a)  (a two-good specialization of the line-network setting of Example~2, with parameters given in the figure caption), where the popularity-biased bundle, $\mathbf{v}^N$, generates only limited updating and preferences' updating remains close to the no-learning benchmark with orthogonal bundles in Figure~\ref{fig:interventions}(c).

Conversely, consuming $\mathbf{v}^C_t$ lowers $\kappa_t$ and is the most effective robust way to speed up learning \emph{locally} at a given time $t$. The bundle targets the weakest direction of $\bm{Z}_t$ and acts as a \emph{correlation-breaking} bundle: it counteracts the dominant dependence structure in the existing data and improves identification across multiple goods in a single step. As above, adding \(\mathbf{x}_{t+1}\equiv \mathbf{v}_t^C\) leaves the eigenvectors of \(\mathbf Z_t\) unchanged and raises only its smallest eigenvalue by one. As a result, that eigen-direction need not remain the least-explored one at \(t+1\): if the second-smallest eigenvalue is below \(\lambda_t^{\min}+1\), it becomes the new minimum. The most effective correlation-breaking bundle must therefore, in general, be reconsidered at each step, since the least-explored direction may change after each update.

Finally, note that the single-good exposures discussed in Proposition~\ref{prop:LT} are asymptotically optimal for maximizing the speed of learning and are also robust, since they do not depend on the current estimation error $\Delta\bm{\beta}_t$. However, they add information along one dimension at a time. By contrast, $\mathbf{v}^C_t$ delivers the most effective robust one-step improvement, as it adds information across multiple goods simultaneously at each $t$.
This is illustrated in Figure~\ref{fig:interventions}(b): exposure to the correlation-breaking bundle, $\mathbf{v}^C$, generates fastest learning and moves beliefs toward the truth, but the dynamics stop before reaching unbiased estimators. At the same time, Figure~\ref{fig:interventions}(d) shows that single-good exposure alternated across goods delivers sharper and unbiased long-run identification, precisely because alternation spans the parameter space across periods rather than preserving residual joint dependence within a fixed bundle.

\begin{remark}
    Robust data design uses only the geometry of past co-consumption. Popularity-biased bundles reinforce correlated exposure and slow learning; correlation-breaking bundles reduce it and accelerate learning. 
\end{remark}

The following example reports the associated eigenvector and correlation centralities, illustrating how the two bundles $\mathbf{v}^N_t$ and $\mathbf{v}^C_t$ are interpreted in this setting.

\begin{example}[Line network extension]\label{ex:line_ext}
Consider the line co-occurrence network from Example~\ref{ex:line}. The eigenvalues and the associated eigenvector and correlation centralities are
\[
\lambda^{\max}_t = 3 + \sqrt{2}, \qquad \lambda^{\min}_t = 1, \qquad 
\mathbf{v}^N_t
= \begin{bmatrix}
    1 \\
    1 + \sqrt{2} \\
    1 + \sqrt{2} \\
    1
\end{bmatrix}, \qquad 
\mathbf{v}^C_t =
\begin{bmatrix}
    -1 \\
    1 \\
    -1 \\
    1
\end{bmatrix}.
\]

The bundle $\mathbf{v}^N_t$ places more weight on the two central goods, which are the most frequently co-consumed in the line. This corresponds to a \emph{popularity-biased} exposure: it loads on the dominant co-consumption pattern and tends to add redundant information.
The bundle $\mathbf{v}^C_t$ instead alternates signs and partitions goods into two sides. This sign pattern captures the dominant correlation structure encoded in $\bm{W}_t$: goods on the same side exhibit correlated externalities, whereas goods on opposite sides exhibit anti-correlated externalities. In this sense, $\mathbf{v}^C_t$ acts as a \emph{correlation-breaking} exposure, as it loads directly on the two-sided dependence structure.
\end{example}

Finally, note that the negative entries of $\mathbf{v}^C$ are a natural feature of many of the applied settings the framework is designed to study. In contexts where bundles describe attributes or characteristics, negative entries simply represent variations relative to a reference design, with one direction of an attribute traded off against another---a pullover composed of cotton, wool, color warmth, and softness, or a smartphone configured with screen size, camera resolution, processing speed, and battery capacity. In policy design, where bundles describe combinations of public goods or reform components, negative entries correspond to scaling back specific provisions: a fiscal package may increase healthcare spending while reducing defense outlays, and the political analogue of $\mathbf{v}^C$ is a reform composition that breaks the dominant pattern of past legislative bundling. In recommendation systems, where the platform may face non-negativity constraints on the items it can offer, the strategy is implemented by approximating $\mathbf{v}^C$ within the feasible space---for instance, by over-weighting goods that are weakly connected in the co-consumption network rather than literally offering negative recommendations.

\section{The provider}
\label{sec:provider}

The previous sections characterized how bundle composition shapes preference learning. We now turn to the provider, who can strategically design exposure by choosing which bundles to provide. 
Since each bundle affects both current perceived utility and the future path of beliefs, the provider can influence the direction, persistence, and speed of learning. The discussion below identifies three canonical strategies, and Section~\ref{sec:monopolist} presents an application of the framework to the case of a monopolist.
\\
\\
\textbf{Strategy 1} \textit{Give single goods.}
When the provider observes the current estimation error and wants to affect the perceived value of a specific good with maximal precision, single-good exposure provides the sharpest intervention. If the goal is to raise the estimated value of good $i$, the optimal companion good $j$ solves
$$
j^* \in \arg\min_j \frac{w_{ji,t}}{1+w_{jj,t}}\Delta\beta_{j,t},
$$
so that $j^*$ is either an undervalued good with a positive correlation externality on $i$, or an overvalued good with a negative externality. If the objective is instead to reduce the perceived value of $i$, the logic reverses:
$$
j^* \in \arg\max_j \frac{w_{ji,t}}{1+w_{jj,t}}\Delta\beta_{j,t}.
$$
When good $i$ is itself sufficiently misvalued, offering $i$ alone is optimal. In this sense, single-good exposure provides the benchmark case in which the provider targets one dimension of learning as directly as possible. 
Single-good exposure may also be optimal under alternative objectives, such as maximizing surprise, as in \citet{ely2015}, since the most undervalued goods generate the largest positive gap between expected and realized utility. 
\\
\\
\textbf{Strategy 2} \textit{Bias-targeted bundles.}
Beyond single-good exposure, the provider can design bundles that target the bias vector $\Delta\bm{\beta}_t$.
The geometry of learning in Section \ref{sec:nolearn} shows that the alignment of~$\mathbf{x}_{t+1}$ with $\Delta\bm{\beta}_t$ determines how estimated preferences evolve.
Any bundle orthogonal to $\Delta\bm{\beta}_t$ yields zero expected surprise and leaves estimated preferences unchanged.
If~$\mathbf{x}_{t+1}'\Delta\bm{\beta}_t<0$, the bundle places greater weight on underestimated goods, generating a positive surprise that raises their estimated coefficients and, through correlation effects, may also increase the perceived value of already overvalued goods.
If $\mathbf{x}_{t+1}'\Delta\bm{\beta}_t>0$, the opposite occurs: the bundle emphasizes overestimated goods, producing a negative surprise and lowering their estimated values.
The provider can thus control both the direction and the intensity of preference updating through the orientation of $\mathbf{x}_{t+1}$ relative to $\Delta\bm{\beta}_t$. 

%In particular, 
The same logic allows the provider to use bundles either to \emph{exploit} current misperceptions (e.g., sustain the valuation of an already overvalued good) or to \emph{discover} undervalued goods by inducing positive surprise along dimensions where the consumer underestimates value.
Overall, the provider shapes the dynamics of preference estimation through the structure of exposure itself.
By adjusting the alignment of bundles with current estimation errors, he can preserve, attenuate, or amplify estimation biases.

A useful illustration is Microsoft’s bundling of Internet Explorer with the Windows operating
system. By pairing a potentially overvalued product with a component that consumers were
going to use anyway, the provider made evaluation occur at the level of the bundle rather than
at the level of the browser alone. In our framework, this weakens attribution: realized utility
from the joint experience is not cleanly traced back to a single component, so the adjustment
of estimated preferences for the browser may be slowed and partly displaced onto the bundled
environment. In this sense, bundling can help preserve the perceived value of an already favored
good, even without shutting down learning exactly.

A contrasting illustration is the bundling of applications within the Microsoft Office suite.
Many users initially purchased Office mainly for Word, while undervaluing applications such
as Excel or PowerPoint. By bundling these products together, Microsoft ensured repeated
exposure to the initially undervalued applications. As users interacted with them, realized
utility could exceed expectations, generating positive surprise and upward revision of their
estimated value. In this sense, bundling can also act as a discovery device, accelerating learning
about goods that were initially undervalued.

More generally, the mechanism applies in the case of composite goods, bundled services, or team production, where overall performance reflects the joint contribution of multiple elements. In such environments, bundling a new undervalued product (or team member) with an established one can accelerate its discovery if the established component is correctly valued. Conversely, bundling can also preserve the perceived value of an overvalued component by embedding it in an aggregate whose overall value is underestimated.
\\
\\
\textbf{Strategy 3} \textit{Spectral bundles.}
This strategy follows directly from Proposition~\ref{prop:robust} and the related discussion.
Even without information about the estimation error $\Delta \bm{\beta}_t$, the provider can exploit the spectral structure of the data matrix $\mathbf{Z}_t$ to regulate the speed of learning.
Offering a bundle proportional to the leading eigenvector $\mathbf{v}_t^N$   aligns consumption with the direction of maximal redundancy, thereby increasing multicollinearity and slowing convergence.
Conversely, offering a bundle proportional to the correlation-centrality vector $\mathbf{v}_t^C$ introduces orthogonal variation, reduces redundancy, and accelerates learning.
These spectral strategies act purely through the geometry of exposure: they determine how quickly information accumulates, independently of any knowledge of preferences or estimated coefficients.
%As discussed in previous sections, the presence of strong complementarities or substitutions constrains the feasibility of these designs, but when such interactions are small the provider can still implement the eigenvector strategy over the primitive goods, preserving the same informational effects.

A related phenomenon arises in digital recommendation systems, where exposure tends to concentrate on already popular goods---known as \textit{popularity bias}---thereby reinforcing existing consumption patterns and limiting informational diversity \citep[see][]{klimashevskaia2024}.
In terms of the model, such algorithms effectively reproduce $\mathbf{v}_t^{N}$-type bundles, which allocate goods in proportion to their eigenvector centrality in the network of past co-consumption data, thus slowing learning and trapping consumers in correlated environments.
Introducing correlation-central variation, analogous to $\mathbf{v}_t^{C}$, would instead diversify exposure, improve identification, and accelerate convergence in estimated preferences.

The following Table \ref{tab:csmall} summarizes the provider's best strategies depending on available information and on the provider's objective (favoring or hindering consumer's learning).
{In competition-policy terms, these strategies capture how a dominant provider may either preserve demand distortions that protect incumbent products or accelerate discovery of products with higher latent profitability.}

\begin{table}[!ht]
\centering
\renewcommand{\arraystretch}{1.15} % Adjust the spacing inside the cells
\setlength{\extrarowheight}{2pt} % Increase the space inside each cell
\begin{tabular}{|c||c|c|} 
\hline
 & \textbf{Favor learning} & \textbf{Hinder learning}
 \\ 
%\hline
%& Sensitive to Initial Conditions & Independent of Initial Conditions  
%\\ 
\hline
\hline
%\textbf{Complete info} 
\textbf{Known} %$\Delta \bm{\beta}_t$ 
& Single goods   &  Bias-targeted %$\mathbf{x} \perp \Delta \bm{\beta}_{t_0} $
\\ 
%about 
$\Delta \bm{\beta}_t$  
&  &  bundles  \\
\hline 
%\textbf{Incomplete info}
\textbf{Unknown} %$\Delta \bm{\beta}_t$
& $\mathbf{x} \equiv \mathbf{v}^C$  &  $\mathbf{x} \equiv \mathbf{v}^N$ 
\\ 
%about 
$\Delta \bm{\beta}_t$  
& Break Correlation & Popularity bias\\ 
\hline
\end{tabular}
\caption{\footnotesize {Provider's best strategies depending on information and if he wants to favor consumer's learning or not.}}
\label{tab:csmall}
\end{table}

\subsection{A Monopolistic Provider}
\label{sec:monopolist}

% We now present an application in which the strategies previously discussed are adopted by a monopolist. The goal is to provide a unified example that illustrates when the monopolist optimally chooses \emph{single goods} (Strategy 1), \emph{bias-targeted} bundles (Strategy 2) to either foster discovery or manipulate estimated preferences, or  \emph{spectral} bundles (Strategy 3) that offer a \emph{robust} exposure rule driven only by the geometry of the observation design.

%We present the monopolist as a stylized application of the three strategies discussed above, rather than as a general industrial-organization model. The purpose of this section is to show how single-good exposure, bias-targeted bundling, and spectral bundles emerge under different informational environments within one simple pricing problem.

We apply our framework to a monopolistic market. A monopolist chooses which products or bundles to offer over time to a consumer who learns her preferences through consumption. Because exposure shapes how the consumer learns what she values, the monopolist faces a tradeoff between extracting current profits and shaping future demand. Product design and bundling thus become a dynamic strategic instruments. We show how the three canonical strategies---single-good exposure, bias-targeted bundling, and spectral bundles---emerge under different informational environments.

%We apply our framework to a monopolistic market. A monopolist chooses which bundles to offer over time to a consumer who learns her preferences through consumption. Because exposure shapes how the consumer learns what she values, the monopolist faces a tradeoff between extracting current profits and shaping future demand. We show how the three canonical strategies---single-good exposure, bias-targeted bundling, and spectral bundles---emerge under different informational environments.

Let us consider a monopolist who, at each $t\in \{1,2\}$, chooses an exposure bundle $\mathbf{x}_t$.\footnote{
The two-period horizon is purely expositional. Similar qualitative trade-offs arise in the infinite-horizon problem.} Marginal linear costs are heterogeneous and denoted by $\bm{\gamma}=(\gamma_i)_{i\in I}$. Let $\delta>0$ denote the relative importance of period-2 profits in the monopolist’s objective. The consumer purchases whenever the price does not exceed her perceived value, so period-$t$ profit is
\[
\Pi_t=(p_t-\mathbf{x}_t'\bm{\gamma})
\cdot
\mathbbm{1}_{\;p_t\le \mathbf{x}_t'\hat{\bm{\beta}}_{t-1}}.
\]
Optimal pricing implies $p_t=\mathbf{x}_t'\hat{\bm{\beta}}_{t-1}$. The monopolist therefore solves
\begin{equation}
\label{eq:optimal_mono}
\max_{\mathbf{x}_1, \mathbf{x}_2} \ \ \mathbf{x}_1'(\hat{\bm{\beta}}_{0}-\bm{\gamma}) + \delta \ \mathbb{E}\left[ \mathbf{x}_2'(\hat{\bm{\beta}}_{1}-\bm{\gamma})\right].    
\end{equation}

\paragraph{Complete information}
We now consider the case in which the monopolist has complete information about both true and estimated preferences. We assume that for each $t\in\{1,2\}$ the monopolist chooses $\mathbf{x}_t\in \mathbb{R}^n$ such that $\|\mathbf{x}\|_1 = 1$, where the entries represent either quantities of the goods composing the bundle or attribute configurations of a single product offered to the consumer. %$\sum_i x_{i,t}=1$ and $x_{i,t}\ge 0$,so that the consumer's experience depends on the share of each good in the bundle. 
Under complete information, the monopolist can design bundles to target estimation errors.
At $t=2$ profits are linear in $\mathbf{x}_2$, so it is generically optimal to sell a single good in the second period. Therefore, the monopolist solves 
\[
\max_{ {||\mathbf{x}_1||=1 }}\;
\mathbf{x}_1'(\hat{\bm{\beta}}_0-\bm{\gamma})
+
\delta \  \mathbb{E}\left[\max_{i\in I}\{\hat{\beta}_{i,1}-\gamma_i\}\right].
\]

The following proposition characterizes optimal first-period exposure when $\delta$ is sufficiently large---so future profits matter---and the signal is sufficiently precise---so exposure produces substantial updates in $\hat{\bm{\beta}}_1$. Under these conditions, the period-1 choice is primarily driven by its effect on period-2 perceived profitability.

\begin{proposition}
\label{prop:monCI}
 Assume $\delta$ is sufficiently large and $\sigma^2$ is sufficiently small. Suppose the monopolist observes $\bm{\beta}$ and $\hat{\bm{\beta}}_0$, and let
$i \in \arg\max_\ell \big\{\hat{\beta}_{\ell,0} - \gamma_\ell\big\}$,  $j \in \arg\max_\ell \big\{\beta_\ell - \gamma_\ell\big\}$.
Then:
\begin{itemize}
    \item If $i = j$, the monopolist sells good $i$ in period 2. The optimal period-1 bundle either sells $i$ directly or uses a manipulation bundle that preserves $\mathbb{E}[\hat{\beta}_{i,1}]$, depending on whether $i$ is overestimated.
    \item If $i \neq j$ and $\hat{\beta}_{i,0} - \gamma_i > \beta_j - \gamma_j$, the monopolist uses a period-1 manipulation bundle that preserves or increases $\mathbb{E}[\hat{\beta}_{i,1}]$, and sells $i$ in period 2.
    \item If $i \neq j$ and $\beta_j - \gamma_j > \hat{\beta}_{i,0} - \gamma_i$, the monopolist uses a period-1 discovery bundle that increases $\mathbb{E}[\hat{\beta}_{j,1}]$, and sells $j$ in period 2.
\end{itemize}
\end{proposition}

\begin{comment}
\begin{proposition}
\label{prop:monCI}
Assume $\delta$ is sufficiently large and $\sigma^2$ is sufficiently small. Suppose the monopolist observes $\bm{\beta}$ and $\hat{\bm{\beta}}_t$ and let
$
i \in \arg\max_\ell\{\hat\beta_{\ell0}-\gamma_\ell\}$,
$j \in \arg\max_\ell\{\beta_\ell-\gamma_\ell\}$.
Optimal first-period bundles involve at most two goods. Moreover:

\begin{itemize}
\item If $i=j$, it is optimal to sell the single good $i$ in both periods.

\item If $i\neq j$ and $\hat\beta_{i0}-\gamma_i>\beta_j-\gamma_j$, it is optimal to use a period-1 (manipulation) bundle that increases $\mathbb{E}[\hat\beta_{i1}]$ and sell $i$ in period 2.

\item If $i\neq j$ and $\beta_j-\gamma_j>\hat\beta_{i0}-\gamma_i$, it is optimal to use a period-1 (discovery) bundle that increases $\mathbb{E}[\hat\beta_{j1}]$ and sell $j$ in period 2.
\end{itemize}
\end{proposition}
\end{comment}
When the current consumer's beliefs already identify the good with the highest true margin, the monopolist provider has no incentive to distort exposure: he sells that good in both periods. When beliefs mis-rank goods, first-period exposure is chosen strategically to shape second-period demand. In particular, exposure can be used either to preserve the perceived advantage of the currently favored good (manipulation) or to accelerate the upward revision of an undervalued but truly more profitable good (discovery).

This mechanism applies in environments in which a firm can allocate short-run exposure across goods, while second-period monetization concentrates on a single component. For example, in printer--cartridge aftermarkets, firms often sell the printer with starter cartridges or related services, while profits later concentrate on a high-margin component such as replacement cartridges. Similarly, in software ecosystems, firms may initially sell a suite that jointly exposes users to multiple applications, while subsequent revenue concentrates on the component with the highest standalone margin (for instance, an enterprise add-on that becomes the main driver of renewal or upgrade decisions).

\paragraph{Incomplete information}

We now consider the case in which the monopolist has incomplete information about the true and estimated preferences. 
We assume that the monopolist holds a neutral prior over the perceived margins, $\mathbb{E}[\hat{\beta}_{i,0}-\gamma_i]=0$ for every $i\in I$. Economically, this captures a monopolist who, lacking information on consumer preferences and estimation errors, treats goods as ex ante equivalent and does not expect any one of them to be systematically more profitable than another. 
We distinguish two regimes according to the monopolist's prior on the consumer's consumption surprise $\Delta u_1$: priors are \emph{pessimistic} when the monopolist expects the consumer to be disappointed on average, so $\mathbb{E}[\Delta u_1]<0$, and \emph{optimistic} when she expects the consumer to be positively surprised, so $\mathbb{E}[\Delta u_1]>0$.\footnote{The notion of priors used here differs from the one discussed in Section~\ref{sec:dynup}. There, priors refer to the consumer's beliefs about $\bm{\beta}$ in the Bayesian interpretation of the %recursive least-squares 
updating rule. Here instead the priors are those of the monopolist, who does not observe $\bm{\beta}$ or $\hat{\bm{\beta}}_0$ and forms beliefs over the margins $\hat{\beta}_{i,0}-\gamma_i$ and the future consumption surprise $\Delta u_1$.% Moreover, we assume that, under the monopolist's prior, the sign of the expected surprise is bundle-invariant.
}
 {Moreover, we assume that, under the monopolist's prior, the expected surprise is bundle-invariant.}

To capture platforms that recommend or bundle goods through a stable algorithmic policy, we focus on stationary strategies $\mathbf{x}_1=\mathbf{x}_2=\mathbf{x}$ with $\|\mathbf{x}\|_2=1$. This constraint captures a reduced-form limit on how strongly the platform can tilt exposure toward a narrow set of goods, due to variety/engagement considerations or regulatory/reputation constraints. Thus, the monopolist solves
\begin{align*}
\max_{\|\mathbf{x}\|_2=1} \ \ &
\mathbb{E}\left[
\mathbf{x}'(\hat{\bm{\beta}}_0-\bm{\gamma})
+
\delta \mathbf{x}'(\hat{\bm{\beta}}_1-\bm{\gamma})
\right]
\\
=
\max_{\|\mathbf{x}\|_2=1} \ \ &
\mathbb{E}\left[
\mathbf{x}'(\hat{\bm{\beta}}_0-\bm{\gamma})
+
\delta \left(
\mathbf{x}'(\hat{\bm{\beta}}_0-\bm{\gamma})
+
\frac{\mathbf{x}'\bm{W}_0\mathbf{x}}{1+\mathbf{x}'\bm{W}_0\mathbf{x}} \Delta u_1
\right)
\right].
\end{align*}

\begin{proposition}
\label{prop:MonII}
Suppose the monopolist does not observe $\bm{\beta}$ and $\hat{\bm{\beta}}_0$ and holds priors with $\mathbb{E}[\hat{\beta}_{i,0}-\gamma_i]=0$. Under stationary strategies $\mathbf{x}_1=\mathbf{x}_2=\mathbf{x}$ with $\|\mathbf{x}\|_2=1$:
\begin{itemize}
\item if priors are pessimistic, the optimal bundle is $\mathbf{x}^*=\mathbf{v}_0^N$ (popularity-bias);
\item if priors are optimistic, the optimal bundle is $\mathbf{x}^*=\mathbf{v}_0^C$ (correlation-breaking).
\end{itemize}
\end{proposition}

The result distinguishes between two benchmark exposure regimes.
A popularity--biased rule
directs exposure toward goods that are already widely consumed. For example, Netflix’s
``Top 10'' lists, YouTube’s ``Trending'' page, or Amazon’s ``Best Sellers'' rankings systemati-
cally amplify items that are already central in the consumption network. In our framework,
this corresponds to exposure tilting toward the dominant direction of past consumption. Because new experiences remain concentrated along already well-explored dimensions, belief
revision is limited and dominant products---such as established artists, blockbuster titles, or
well-known brands---remain overexposed even when alternatives yield higher margins or quality.

By contrast, correlation-breaking exposure resembles diversification policies observed in recommender systems that deliberately inject variety, such as Spotify’s ``Discover Weekly'' or
category-balancing rules in online marketplaces. In our framework, this corresponds to exposure tilting toward directions that are weakly represented in past consumption data. These
mechanisms rotate attention toward less connected goods, generate more informative consump-
tion data, and increase the probability that previously underexposed products are discovered.
In this case, exposure tends to accelerate belief updating and reallocate demand toward goods
whose profitability was initially underestimated.
%The characterization in terms of $\mathbf{v}_0^N$ and $\mathbf{v}_0^C$ is derived in the unconstrained ambient space. Under additional feasibility restrictions, the interpretation of these spectral directions is unchanged, but their exact implementation may no longer be possible. If bundle components are required to be non-negative, $\mathbf{v}_0^N$ remains feasible by Perron--Frobenius, whereas $\mathbf{v}_0^C$ may fail to be implementable because of its mixed sign pattern. If bundles are restricted to dummy vectors, both $\mathbf{v}_0^N$ and $\mathbf{v}_0^C$ should instead be read as benchmark directions, since their entries need not be binary. We return to this point in Appendix~\ref{subsec:complementarity}, where complementarities further restrict the feasible set by tying interaction terms to primitive quantities.

\subsection{Welfare implications}

To understand the economic relevance of misperceived preferences, we study their welfare implications in the monopolistic environment described above within a single period (hence we drop the time index). When the estimated preferences $\hat{\bm{\beta}}$ differ from the true vector $\bm{\beta}$, both pricing and bundle selection respond to the resulting estimation error. As we show below, the induced changes in both consumer surplus and profits can be decomposed into: $(i)$ a \emph{price effect}, holding the implemented bundle fixed, and $(ii)$ a \emph{bundle (allocation) effect},  driven by changes in the implemented bundle.

Fix a generic perceived preference vector $\mathbf{b}$ and let
$\mathbf{x}(\mathbf{b})$ denote a unique bundle selection under beliefs
$\mathbf{b}$.\footnote{Formally, $\mathbf{x}(\cdot)$ is a deterministic mapping
from perceived preference vectors to bundles. If the underlying problem admits
multiple optimal bundles, the mapping selects one of them arbitrarily. One may
think, for instance, of the bundle chosen by the monopolist or by the consumer
according to the optimal choices characterized in the previous section.}
In what follows, we adopt the optimal pricing rule associated with the maximization problem in
\eqref{eq:optimal_mono}, so that the price charged under beliefs $\mathbf{b}$ is
$
p(\mathbf{b})=\mathbf{x}(\mathbf{b})'\mathbf{b}
$.
Finally, define the belief-induced change in the implemented bundle as
$
\Delta \mathbf{x}(\mathbf{b}) := \mathbf{x}(\mathbf{b})-\mathbf{x}(\bm{\beta})$,
that is, the difference between the bundle implemented under beliefs
$\mathbf{b}$ and the bundle that would be implemented under the true preference
vector $\bm{\beta}$.

\paragraph{Consumer surplus.} 
Consumer (expected) surplus is $CS(\mathbf{b}) := \alpha +\mathbf{x}(\mathbf{b})'\bm{\beta} -\mathbf{x}(\mathbf{b})'\mathbf{b} $. The variation of consumer surplus induced by misperceived preferences is 
\begin{align*}
\Delta CS := CS(\hat{\bm{\beta}})-CS(\bm{\beta})  & = -\mathbf{x}(\hat{\bm{\beta}})'\Delta\bm{\beta}  \\ & 
= 
-
\underbrace{\mathbf{x}(\bm{\beta})'\Delta\bm{\beta}}_{\text{price  effect}}
-
\underbrace{\Delta\mathbf{x}(\hat{\bm{\beta}})'\Delta\bm{\beta}}_{\text{bundle  effect}}.
\end{align*}

The decomposition isolates two forces. The \emph{price effect} captures how misperceived marginal valuations translate into mispricing of the benchmark bundle $\mathbf{x}(\bm{\beta})$: if estimated valuations are higher than true ones on average (so $\mathbf{x}(\bm{\beta})'\Delta\bm{\beta}>0$), the monopolist sets a higher price and consumer surplus falls. Conversely, underestimation ($\mathbf{x}(\bm{\beta})'\Delta\bm{\beta}<0$) lowers the price and mechanically raises consumer surplus.

The \emph{bundle effect} reflects the additional distortion coming from the belief-induced change in the implemented bundle. By revealed preference applied to the monopolist's optimization at $\bm{\beta}$ and $\hat{\bm{\beta}}$, the bundle shift always satisfies $\Delta\mathbf{x}(\hat{\bm{\beta}})'\Delta\bm{\beta} \geq 0$, so the bundle effect on consumer surplus is weakly negative: misperception steers consumers toward products that are too expensive relative to their true value, and the consumer is hurt by the resulting allocative distortion in addition to the transfer captured by the price effect.

%The \emph{bundle effect} reflects the additional distortion coming from the belief-induced change in the implemented bundle. Its sign depends on the alignment between the bundle shift $\Delta \mathbf{x}(\hat{\bm{\beta}})$ and the estimation error $\Delta\bm{\beta}$. When the bundle tilts toward goods whose marginal values are overestimated (so $\Delta\mathbf{x}(\hat{\bm{\beta}})'\Delta\bm{\beta}>0$), consumers are steered toward products that are “too expensive relative to their true value,” and surplus falls. In contrast, if misperception shifts consumption toward dimensions that are underestimated (so $\Delta\mathbf{x}(\hat{\bm{\beta}})'\Delta\bm{\beta}<0$), the bundle effect is positive and can partially offset—and in principle dominate—the price effect.

\paragraph{Profits.}
Profits are $\Pi(\mathbf{b}) =\mathbf{x}(\mathbf{b})'(\mathbf{b}-\bm{\gamma})$.  The variation of profits induced by misperceived preferences is 
\[
\Delta\Pi := \Pi(\hat{\bm{\beta}})-\Pi(\bm{\beta})
=
\underbrace{\mathbf{x}(\bm{\beta})'\Delta\bm{\beta}}_{\text{price effect}}
+
\underbrace{\Delta\mathbf{x}(\hat{\bm{\beta}})'\big(\hat{\bm{\beta}}-\bm{\gamma}\big)}_{\text{bundle  effect}}.
\]

The price effect is the exact counterpart of the one in consumer surplus: it is a pure transfer. When $\mathbf{x}(\bm{\beta})'\Delta\bm{\beta}>0$, the benchmark bundle is priced higher, profits rise, and consumer surplus falls by the same amount; when $\mathbf{x}(\bm{\beta})'\Delta\bm{\beta}<0$, the transfer goes the other way.

The bundle effect reflects how the belief-induced change in the implemented bundle affects profits through perceived unit margins $(\hat{\bm{\beta}} - \bm{\gamma})$. Because the monopolist optimizes the bundle under her own beliefs, the bundle effect on profits is weakly positive: $\Delta\mathbf{x}(\hat{\bm{\beta}})'(\hat{\bm{\beta}} - \bm{\gamma}) \geq 0$ by revealed preference, since $\mathbf{x}(\hat{\bm{\beta}})$ maximizes $\mathbf{x}'(\hat{\bm{\beta}} - \bm{\gamma})$ over the feasible bundle space.

%The bundle effect reflects how the belief-induced change in the implemented bundle affects profits through perceived unit margins $(\hat{\bm{\beta}}-\bm{\gamma})$. Profits increase if $\Delta\mathbf{x}(\hat{\bm{\beta}})$ shifts consumption toward goods with higher perceived margins, and decrease if it shifts weight toward goods with lower (or negative) perceived margins.

\paragraph{Total welfare.}
Total welfare is $SW(\mathbf{b}) =CS(\mathbf{b}) +\Pi(\mathbf{b})$.  The variation of total welfare induced by misperceived preferences is 
\[
\Delta SW := SW(\hat{\bm{\beta}})-SW(\bm{\beta}) =
\Delta\mathbf{x}(\hat{\bm{\beta}})'(\bm{\beta}-\bm{\gamma})\leq 0.
\]

Pricing effects cancel in the aggregate, since they constitute a pure transfer between consumers and the monopolist. Misperceived preferences therefore affect total welfare only through the bundle effect, which is always weakly negative: $\mathbf{x}(\bm{\beta})$ is the welfare-maximizing allocation under the optimal pricing rule, so any belief-induced shift moves consumption away from it. When the resulting allocative distortion outweighs the price-effect transfer, consumer surplus and profits can both fall below their benchmark levels. {This shows that data-design distortions are not merely redistributive: by altering the informational basis of future demand, they can generate persistent allocative inefficiencies even when prices are privately optimal.}

\section{An illustrative example: The Movie Industry}
\label{sec:movies}
%\red{Maybe we should start with a brief discussion to recall that the model can be applied to preference discovery also in a setting where a investor should decide in which ``asset" invest: actors, soccer players, etc. In such a case Y is the outcome  of a joint effort (e.g., team work productivity) and the consumer is actually a economic agents who should evaluate the individual quality. In such an example the provider can be a agent (of actors or sportsman) that try to push for his clients. }

As a final exercise, we apply our model to a real dataset. In doing so, we do not aim to provide theoretical support for our model nor to find empirical evidence of causality in the data. Our goal is to clarify the mechanism by which we model the updating of estimated preferences, using an enjoyable and easy-to-remember dataset. Here, we treat movies as bundles and actors as goods.
This allows us to use the \emph{co--starring} network, where nodes represent actors, and two actors are linked if they have acted together in at least one movie, as in \cite{watts1999small}, one of the first to apply complex networks to the social sciences in a chapter of his book. Specifically, we include actors as dummy variables in the bundle related to a movie.

We used the \href{https://www.imdb.com/}{IMDb movie database}. This dataset contains comprehensive information on more than 140,000 movies, dating back to the early 20th century (and on more than 97,000 actors). However, for only around 1,000 of these movies, it also includes \href{https://www.boxofficemojo.com/chart/top\_lifetime_gross_adjusted/?adjust\_gross\_to=2022}{\emph{box office} earnings}, which are adjusted for inflation to 2022 US dollars.\footnote{As explained on the
\href{https://help.imdb.com/article/imdbpro/industry-research/box-office-mojo-by-imdbpro-faq/GCWTV4MQKGWRAUAP?ref_=mojo\_cso\_md\#inflation}{IMDbPro website}, ``[a]djusting for ticket price inflation is not an exact science and should be used for a general idea of what a movie might have made if released in a different year, assuming it sold the same number of tickets. [...] Since these figures are based on average ticket prices they cannot take into effect other factors that may affect a movie's overall popularity and success.''}
We assume, with the necessary caveats set aside for this illustrative exercise, that the global audience represents our representative consumer and that the box office earnings of each movie reflect how much people liked it. In fact, we use this information as a proxy for the \emph{utility} perceived by the representative consumer.

In this smaller set of 986 movies with box office data, there are 1,836 actors. Each movie features between 1 and 4 actors, with an average of 3.9 actors per movie. Conversely, each actor appears in an average of 2.1 movies (with a minimum of 1 and a maximum of 21).\footnote{The minimum number of movies per actor is 1, the median 1, the maximum is 21 and the top 5-th percentile is 6.}
Only 121 of these actors appear in at least 6 movies with box office data (the top 5th percentile). We restrict our analysis to these 121 actors to facilitate the clear display of the co--occurrence network.\footnote{There are 123 actors, but if two actors happen to appear in exactly the same movies they are indistinguishable so we remove them to avoid multicollinearity (see Appendix \ref{subsec:multicoll}). This is the case of Daniel Radcliffe and Rupert Grint from the \emph{Harry Potter} saga, for example.}
In this way, we are left with 680 movies, released between 1964 and 2020. We order these movies chronologically: for every $t \in \{ 1, \dots , 680 \}$, vector $\mathbf x_t \in \mathbb R^{121}$ represents a movie where component $i$ is 1 if actor $i$ starred that movie and 0 otherwise, and $u_t$ its box office earnings (expressed in millions of dollars). In this simplified interpretation of the exercise, $\beta_i$ can be seen as the additional earnings that a particular actor $i$ contributes to a movie's total earnings, and $\hat{\beta}_{i,t}$ is an estimate of this value based on the history up to time $t$. Here, the history refers to the sequential release of movies over time, $t \in \{ 1, \dots , 680 \}$.

\subsection{Networks and centrality measures}

\begin{figure}[h]
    % \centering
    \vspace{-30mm}
    \hspace{-20mm}
    \setlength{\abovecaptionskip}{-20mm}
    % \hspace*{-30mm}
    \includegraphics[width=1.24\textwidth]{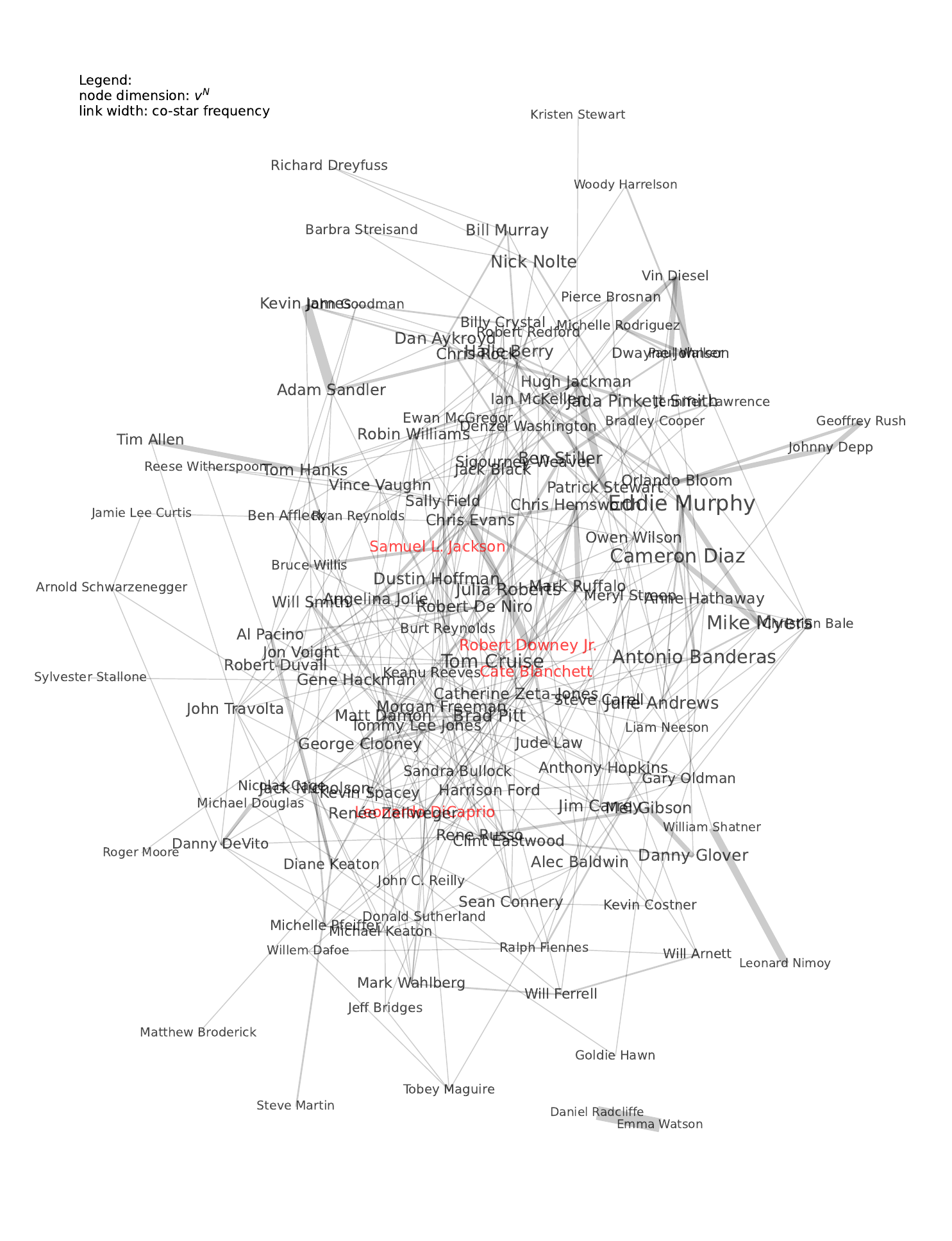}
    \caption{\footnotesize Co--starring network of the 121 actors in our database, with larger font size for more central nodes according to network centrality $\mathbf v^N$. The link width is proportional to the number of movies two actors co-star in. Samuel L. Jackson, Robert Downey Jr., Cate Blanchett and Leonardo DiCaprio are in red for readability, as we focus on them in Section \ref{subsec:moviebeta}.}
    \label{fig:net1}
\end{figure}

Figure \ref{fig:net1} illustrates the co--starring network represented by the adjacency matrix $\bm{Z}_{680}$, based on the 680 movies we selected. 
In the figure, the font size of the actor names is proportional to their network centrality $\mathbf v^N$ while the width of the links is proportional to the number of movies in which the two actors have appeared together. 

Clusters of strongly connected actors often correspond to high-grossing movies from well-known sagas. For instance, Hugh Jackman, Ian McKellen, and Patrick Stewart from the \emph{X-Men} saga; Johnny Depp, Geoffrey Rush, and Orlando Bloom from the \emph{Pirates of the Caribbean} saga; Dwayne Johnson, Michelle Rodriguez, and Vin Diesel from the \emph{Fast and Furious} saga; and William Shatner and Leonard Nimoy from the \emph{Star Trek} saga.
Table \ref{table:network_centrality} reports the first 10 actors, ranked by network centrality.

\begin{table}[h!]
    \centering
    \begin{tabular}{|c|c|c|}
    \hline
    \textbf{Rank} & \textbf{Actor/Actress} & \textbf{Network Centrality} \\ 
    \hline
    1 & Eddie Murphy & 0.766 \\ 
    \hline
    2 & Cameron Diaz & 0.360 \\ 
    \hline
    3 & Mike Myers & 0.302 \\ 
    \hline
    4 & Tom Cruise & 0.260 \\ 
    \hline
    5 & Antonio Banderas & 0.208 \\ 
    \hline
    6 & Brad Pitt & 0.086 \\ 
    \hline
    7 & Nick Nolte & 0.085 \\ 
    \hline
    8 & Julie Andrews & 0.084 \\ 
    \hline
    9 & Julia Roberts & 0.082 \\ 
    \hline
    10 & Jada Pinkett Smith	& 0.072 \\ 
    \hline
    \end{tabular}
    \caption{\footnotesize Top 10 actors for network centrality $\mathbf v^N$.}
    \label{table:network_centrality}
\end{table}

If we consider the same network but we focus on correlation centrality $\mathbf v^C$, this measure generates a partition of the nodes into two groups, where each group tends to have a positive externality on those in the same group, and a negative one on those in the other group (see Section \ref{sec:spectral}).
Table \ref{table:correlation_centrality} reports the first $5$ actors for each of the two groups, according to this measure. 
Interestingly, the two top scorers are William Shatner and Leonard Nimoy who almost always starred together in the \emph{Star Trek} saga. Moreover, in the same saga they starred also together with Patrick Stewart who, in turn, is also part of the \emph{X-men} saga where he acts very often together with Hugh Jackman, Ian McKellen and Halle Berry. In line with our model, actors who are very often together will tend to have opposite effects on one another, as signaled by the opposite signs of their correlation centrality.\footnote{See Figure \ref{fig:net2} in Appendix \ref{sec:appendix_movies}.}

% Now we consider the same network, but we focus on correlation centrality $\mathbf v^C$.
% Figure \ref{fig:net2} reports the same network, but highlighting this second measure.
% As discussed in Section \ref{sec:spectral}, this measure generates a partition of the nodes into two groups, where each group tends to have a positive externality on those in the same group, and a negative one on those in the other group.
% In this figure, the size of the font in the names of the actors is bigger for more central nodes, according to correlation centrality.
% Table \ref{table:correlation_centrality} reports the first $5$ actors for each of the two groups, according to this measure. Interestingly, the two top scorer are William Shatner and Leonard Nimoy who almost always starred together in the \emph{Star Trek} saga. Moreover, in the same saga they starred also together with Patrick Stewart who, in turn, is also part of the \emph{X-men} saga where he acts very often together with Hugh Jackman, Ian McKellen and Halle Berry. As predicted by our model, actors who are very often together will tend to have opposite effects on one another, as signaled by the opposite signs of their correlation centrality.

\begin{table}[h!]
\centering
\begin{tabular}{|c|c|c|c|}
\hline
\textbf{Group} & \textbf{Rank} & \textbf{Actor/Actress} & \textbf{Correlation Centrality} \\ 
\hline
\multirow{5}{*}{\textbf{\textcolor{blue}{Positive}}} & 1 & {\textcolor{blue}{William Shatner}} & {\textcolor{blue}{0.701}} \\ 
\cline{2-4}
& 2 & Hugh Jackman & 0.074 \\ 
\cline{2-4}
& 3 & Ian McKellen & 0.041 \\ 
\cline{2-4}
& 4 & Halle Berry & 0.034 \\ 
\cline{2-4}
& 5 & Mel Gibson & 0.016 \\ 
\hline
\hline
\multirow{5}{*}{\textbf{\textcolor{orange}{Negative}}} & 1 & {\textcolor{orange}{Leonard Nimoy}} & {\textcolor{orange}{$-0.683$}} \\ 
\cline{2-4}
& 2 & Patrick Stewart & $-0.181$ \\ 
\cline{2-4}
& 3 & Orlando Bloom & $-0.026$ \\ 
\cline{2-4}
& 4 & Ryan Reynolds & $-0.013$ \\ 
\cline{2-4}
& 5 & Danny Glover & $-0.010$ \\ 
\hline
\end{tabular}
\caption{\footnotesize Top 5 actors for correlation centrality $\mathbf v^C$ in two groups: positive and negative.}
\label{table:correlation_centrality}
\end{table}

\subsection{Updating of $\hat\beta$'s}
\label{subsec:moviebeta}

\begin{figure}[htb]
    \hspace{-19mm}
\includegraphics[width=1.2\textwidth]{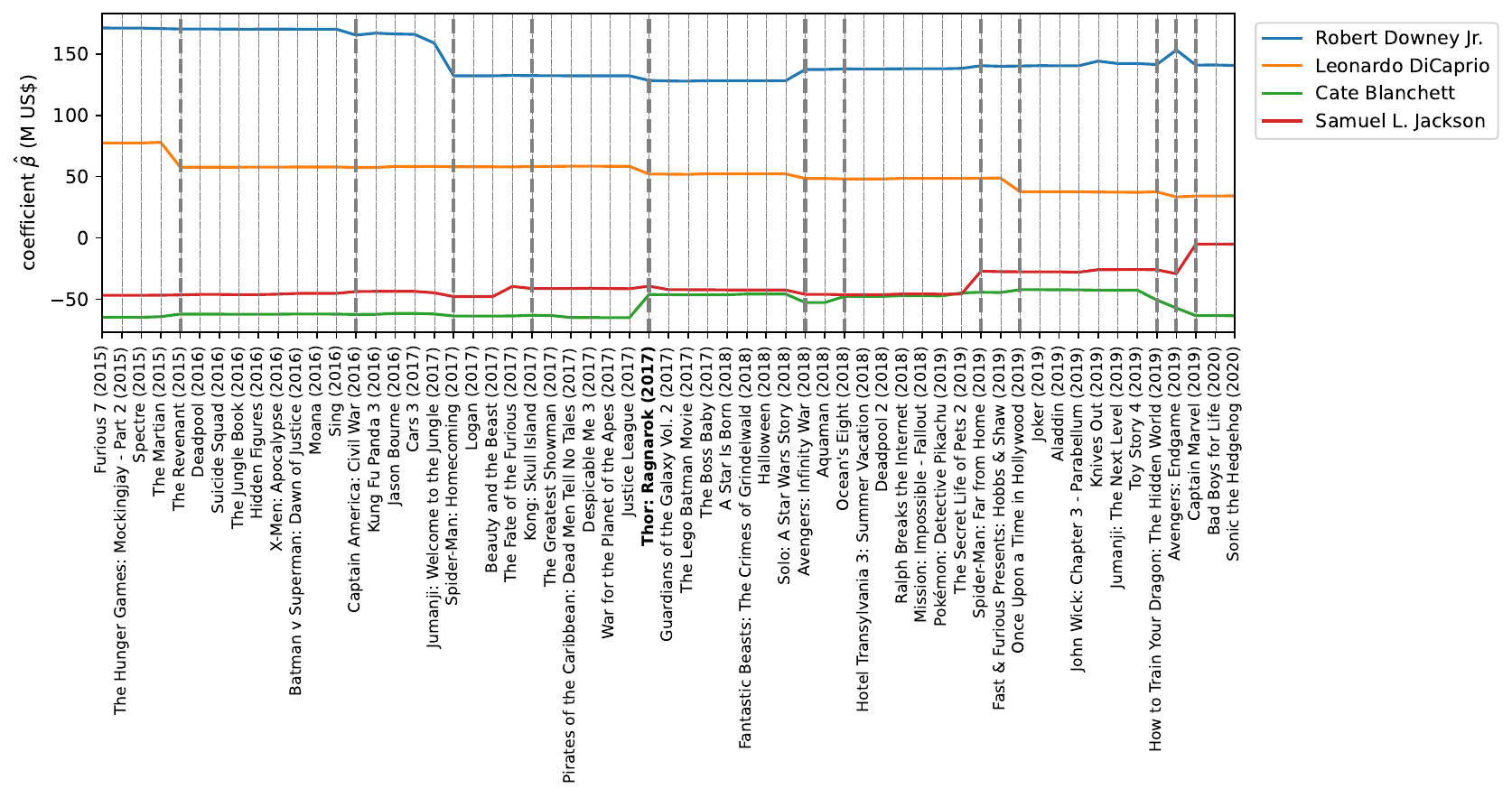}
        \caption{\footnotesize  History of the $\hat\beta$'s of four actors (in million US\$): Robert Downey Jr., Leonardo DiCaprio, Cate Blanchett and Samuel L. Jackson. Dashed lines denote movies where at least one of them acted, for readability. Cate Blanchett stars in \emph{Thor: Ragnarok (2017)} which explains the jump in her $\hat\beta$. Notably, however, even if the other actors do not appear in such a movie, their $\hat\beta$'s are also updated because they are linked to her via the co-starring network. Their correlation coefficients with her determine the direction and intensity of these updates. Since $\bm Z_t$ first reaches full rank at $t=554$ (2011), all coefficient paths displayed here lie in the post-full-rank region of the sample.}
    \label{fig:DDBJ_zoom}
\end{figure}

The previous descriptive statistics came from an ex--post analysis of the full history of \emph{consumption}, that is, considering all the movies up to the last time/movie $t=680$.
Instead, here we provide an intuition of how, according to our model, actors' estimated preferences $\hat\beta$'s evolve over time.
For this purpose, we focus on four actors: Robert Downey Jr., Leonardo DiCaprio, Cate Blanchett and Samuel L. Jackson.
Figure \ref{fig:DDBJ_zoom} reports how their estimated $\hat\beta$'s change as new movies are released over time. On the horizontal axis we have the chronological order of movies in our database, by release year. \footnote{%
Since $\mathbf x_t$ has 121 dimensions and many actors debut and enter only later in the chronological sample, the recursive OLS estimator initially faces a rank-deficient phase in which $\bm Z_t$ is singular. To reduce identification concerns related to this initial phase, we restrict attention to actors who appear in at least 4 movies both before and after $t=554$, the first observation at which $\bm Z_t$ reaches full rank (corresponding to \emph{Puss in Boots}, 2011). This ensures that the actors retained for the analysis are observed sufficiently often on both sides of the full-rank threshold, and rules out cases in which the post-2011 dynamics are driven by actors who enter only very late in the sample or whose estimated $\hat\beta$ is very noisy.
The four actors highlighted in Figure \ref{fig:DDBJ_zoom} --- Robert Downey Jr., Leonardo DiCaprio, Cate Blanchett, and Samuel L. Jackson --- belong to this restricted set.}

To make sense of the evolution of the $\hat\beta$'s, let us focus on an example.
Cate Blanchett stars in \emph{Thor: Ragnarok (2017)}, so her estimated $\hat\beta$ is updated, as is visible in the corresponding jump of the green line in Figure \ref{fig:DDBJ_zoom}. 
Crucially, and coherently with our model, that same movie updates also the $\hat\beta$'s of the other three actors even if they were not part of that movie. This is because they are linked to Cate Blanchett in the co-starring network and, therefore, affected by her update: the estimate of Robert Downey Jr. and Leonardo DiCaprio decrease while Samuel L. Jackson's increases. The direction and strength of their updates, i.e. sign and intensity, are driven by their correlation coefficient with Cate Blanchett.

\clearpage

\section{Conclusion}
\label{sec:conclusion}

This paper develops a framework to study how preferences are discovered through experience when the consumption pattern is designed. We show how the structure of consumption data determines the propagation of information across goods. Bundling creates a co-occurrence network that induces correlation externalities among goods, so that a surprise about one affects the estimated preferences of the others. %, and goods that co-occur often tend to exhibit negatively correlated updates. 
These network effects shape the speed, direction, and eventual biases of the preference discovery process.

From this foundation, we characterize the strategies available to a provider who designs the consumption experiences. When the provider knows both true and perceived preferences, he can manipulate belief formation directly---either accelerating discovery or sustaining bias---by aligning bundles with the consumer’s current estimation errors.   Even without such information, however, he can still regulate learning through the spectral properties of the data. Bundles aligned with the leading eigenvector of the co-occurrence network (\textit{popularity-bias}) reinforce existing correlations and slow preference discovery, while bundles aligned with the correlation-centrality vector reduce correlation in the data and accelerate identification. These spectral strategies act purely through the geometry of past experiences.

The analysis provides a formal basis for understanding how designed consumption experiences shape the formation of preferences. It highlights a fundamental tension between discovery and control: the same mechanisms that can help consumers learn their true preferences and increase welfare can also be used to slow down learning and preserve biased estimated preferences. More broadly, the mechanisms we characterize do not depend on real-world providers explicitly implementing the strategies identified by our model, nor on the designers of complex recommendation or AI-based systems fully understanding the logic by which exposure patterns affect preference discovery. Our framework provides the tools to assess these effects regardless. %But this does not reduce the relevance of the mechanism.Our framework identifies
We identify a novel rationale for bundling: beyond price discrimination or product design, bundling can shape the data through which consumers discover their preferences, thereby affecting profits and other objectives related to market steering and control. Precisely because these effects may arise even through opaque or partially automated systems, they should be taken into account by policymakers across a wide range of contexts.

More generally, the paper connects consumer learning, data design, and network structure in a unified framework, and opens several directions for future work. Natural extensions include competition among providers, consumer experimentation and search, and social learning among consumers. Studying these environments would help clarify how preference discovery interacts with broader questions of welfare, market power, and regulation. A particularly promising extension concerns vertically related digital markets in which platforms jointly control downstream exposure and upstream production. In such settings, data-design may influence both final demand and the information available to competing producers.

% \paragraph{Future research}
% Competition. Social learning. Consumer experimentation.  

%\clearpage

\bibliographystyle{apalike}
\bibliography{biblio}

\newpage
 
\appendix
\setcounter{equation}{0}

\begin{center}
	{\LARGE \textbf{Appendix}}%: Proofs of all results in the main text}\label{App: Proofs}} 
\end{center}

\section{Aggregation and Identification}
\label{app:ext}

This appendix addresses two structural properties underlying the baseline model. 
First, we show that the learning dynamics are preserved when the consumer is interpreted as a representative agent obtained by linear aggregation of heterogeneous individuals with independent idiosyncratic shocks. The recursive least–squares structure remains unchanged, with aggregation affecting only the variance of the error term.
Second, we discuss identification and rank conditions. When some goods always appear in fixed proportions, the precision matrix may fail to be full rank, generating multicollinearity. We show that the problem can be reduced to a lower–dimensional representation without altering the updating mechanism or the qualitative properties of learning.

Together, these results clarify that the core mechanism does not rely on a literal single-agent interpretation or on full-rank consumption data, but only on the informational geometry encoded in the precision matrix.

\subsection{Representative consumer}
\label{subsec:repcon}

Note that, having assumed linear preferences in \eqref{eq: utility}, the decision-maker ($DM$) can actually be representative of many consumers.
Indeed, suppose that we have $K$ consumers and we adopt a linear combination $\boldsymbol{\phi} \in {[0,1]}^K$ of them, with $\sum_{k=1}^{K} \phi_k = 1$.
Imagine that they all had the same history of consumption, but different $\alpha$'s, $\boldsymbol{\beta}$'s, and i.i.d. errors $\varepsilon_{k,t}$.
Call $\mathbf{B} \in \mathbb{R}^{I \times K}$ the rectangular matrix where $\mathbf{B}_{ik}= \beta_{i,k}$ is the real coefficient of consumer $k$ for good $i$, and similarly $\hat{\mathbf{B}}_t$ their estimated preferences at time $t$.
$\mathbf{u}_t \in \mathbb{R}^K$ will be the vector of all utilities at time $t$, and ${\mathbf{U}}_t \in \mathbb{R}^{K \times t}$ the rectangular matrix recording all past utilities up to time $t$, so that $\mathbf{U}_{kt}= u_{k,t}$.
All their experiences at time $t$ can be aggregated as
\[ \sum_{k=1}^{K} \phi_k u_{k,t} = \mathbf{u}'_t \boldsymbol{\phi} = \boldsymbol{\alpha}' \boldsymbol{\phi} + \mathbf{x}_t' \mathbf{B} \boldsymbol{\phi} + \boldsymbol{\varepsilon}'_t \boldsymbol{\phi} . \]

From equation \eqref{eq: beta}, their estimated preferences can be aggregated as
\[
\hat{\mathbf{B}}_t \boldsymbol{\phi}
= \sum_{k=1}^{K} \phi_k \hat{\boldsymbol{\beta}}_{k}
= \bm{W}_{t} (\mathbf{X}_t)' \bigl(\mathbf{U}_t \boldsymbol{\phi}\bigr),
\]
where $\bm{W}_{t} (\mathbf{X}_t)'$ is common to all consumers.

This means that everything can be aggregated linearly, considering them as a single representative consumer who has real preferences $\boldsymbol{\alpha}' \boldsymbol{\phi}$ and $\mathbf{B} \boldsymbol{\phi}$, and estimated preferences $\hat{\mathbf{B}}_t \boldsymbol{\phi}$ at time $t$.
What only changes is the error term $\boldsymbol{\varepsilon}'_t \boldsymbol{\phi}$ that washes away idiosyncratic fluctuations and has a reduced variance of $\boldsymbol{\phi}' \boldsymbol{\phi} ~\sigma^2 \leq \sigma^2$ (since all errors are independent, we do not need to take into account covariances of the random variables).%{\color{red} \bf [check! - wisdom of the croud? but different preferences]}
\footnote{%
Note that in principle, from an econometric point of view, the sum of $\boldsymbol{\alpha}' \boldsymbol{\phi} + \boldsymbol{\varepsilon}'_t \boldsymbol{\phi}$ can also include the effect of missing variables (missing goods, in our setup), as long as their effect is not correlated with the observed variables (the goods we include in the analysis, in our setup).}

So, the interpretation of our consumer as a representative consumer does not change any result, but makes it plausible to model an external provider who knows the true preferences of the representative consumer, while she does not.
Indeed, the provider can run surveys and market experiments to get this information, while each individual consumer cannot, as she does not know the other consumers and may not even know the weights $\boldsymbol{\phi}$.\footnote{It would be interesting, and we leave this for future research, to model a social network of consumers, where they share and aggregate information over time.}

\subsection{Multicollinearity}
\label{subsec:multicoll}

Before characterizing the learning process, we note that some of 
the goods in $I$ may always appear together in fixed proportions 
across bundles. In such cases, their individual contributions 
cannot be identified, and the consumer effectively treats them as 
a single good. Formally, this corresponds to some columns of 
$\mathbf{X}_t$ being linear combinations of others, so that the 
precision matrix $\bm{Z}_t=\mathbf{X}_t'\mathbf{X}_t$ is not 
full rank.

Let $m(\mathbf{X}_t)$ denote the rank of $\mathbf{X}_t$. When 
$m(\mathbf{X}_t) < n$, the consumer can reduce the dimensionality 
of the problem by projecting $\mathbf{X}_t$ onto an 
$m(\mathbf{X}_t)$-dimensional subspace, obtaining a reduced matrix 
$\underline{\mathbf{X}}_t \in \mathbb{R}^{t \times m(\mathbf{X}_t)}$ 
on which the precision matrix is full rank and preference updating 
remains well defined. In principle, any basis of the column space 
of $\mathbf{X}_t$ yields a valid reduction. Different choices, 
however, may differ in economic interpretability. A natural and 
symmetric approach is to proceed class by class: for each set of 
perfectly collinear goods, replace that set with a single 
representative composite, for example its equally weighted average, 
while leaving all other goods unchanged. If multiple disjoint 
collinear classes are present, the same construction is applied 
separately to each class. Equal weighting is simply a normalization 
that treats goods symmetrically within a collinear class. The 
associated restricted preference vector and estimate are denoted 
$\bm{\beta}^m_t$ and $\hat{\bm{\beta}}^m_t$, and the restricted 
precision matrix is 
$\bm{Z}^m_t = \underline{\mathbf{X}}_t'\underline{\mathbf{X}}_t$.

In the main body of the paper, we assume that the observed data do not exhibit 
multicollinearity, with the implicit understanding that, if 
detected, the consumer reduces the dimensionality of the problem 
using the method described above. Note that, as $\mathbf{X}_t$ 
grows by acquiring new rows over time, the rank $m(\mathbf{X}_t)$ 
cannot decrease: once two goods are observed in sufficiently 
varied proportions, their contributions become separately 
identified and collinearity is resolved. This implies that 
multicollinearity cannot spontaneously arise during the learning 
process. Nevertheless, as we discuss in 
Section~\ref{sec:provider}, the provider can approach the 
boundary of multicollinearity asymptotically through strategic 
bundling choices that gradually reduce the information content 
of new observations.

\section{Complementarities and substitution effects}
\label{subsec:complementarity}

In this appendix we extend the baseline model to allow for complementarities across components and revisit the main results under this more general specification.
As anticipated in the main text, complementarities can be interpreted as
restrictions on the feasible bundle space: interaction regressors are
mechanically determined by primitive quantities and therefore cannot be
freely chosen by the provider.

\subsection{Augmented linear representation}

The linear specification in equation~\eqref{eq: utility} allows us to analyze also cases in which there are complementarities or substitution between goods. We do so by treating $I$ as composed by primitive goods and the interaction terms between them. Considering all pairwise interactions, if there are $m$ primitive goods, then the cardinality of $I$ is
$n = m + \frac{m(m-1)}{2}$.

Formally, let $\mathbf{x}^p_t$ denote the $m$-dimensional vector of primitive goods consumed at time $t$, with associated coefficients $\bm{\beta}^p$. For each pair $(i,j)$ with $i<j$, we define the interaction regressor $x_{i,t}x_{j,t}$ with associated coefficient $\beta_{ij}=\beta_{ji}$. The full regressor vector and coefficient vector can thus be written as
$\mathbf{x}_t = \big(\,\mathbf{x}^p_t,\; (x_{i,t}x_{j,t})_{\,1 \le j < i \le m}\,\big)'$ and
$\bm{\beta} = \big(\,\bm{\beta}^p,\; (\beta_{ij})_{\,1 \le j < i \le m}\,\big)'$.

When some elements of $I$ represent interaction terms, 
we can rewrite the utility in~\eqref{eq: utility} as
\begin{align}
u_t =& \alpha + \mathbf{x}_t' \bm{\beta}  + \varepsilon_t \nonumber \\
=& \alpha + \mathbf{x}^{p \prime}_t \bm{\beta}^p + \sum_{i< j}\beta_{ij}\,x_{i,t}x_{j,t} + \varepsilon_t,
\label{eq:u_complementarities}
\end{align}
where $\beta_{ij}>0$ represents complementarity  and $\beta_{ij}<0$ represents substitution of goods $i$ and $j$.

Allowing the set $I$ to include interaction terms does not qualitatively affect the consumer’s learning process given her consumption history. 
It does, however, restrict the provider’s capacity to manipulate bundles. 
As shown in~\eqref{eq:u_complementarities}, the consumer’s preferences may include coefficients $\beta_{ij}$ capturing complementarities and substitutions, but the provider’s choice set remains $m$-dimensional. 
The remaining $n-m$ components of $\mathbf{x}_t$ are mechanically determined by the interactions $x_{i,t}x_{j,t}$ and cannot be freely chosen.

%It is important to stress that allowing the set $I$ to include interactions between primitive goods does not qualitatively affect the consumer’s preference updating process, given her history of consumption (as discussed in Section~\ref{subsec:update}). It does, however, constrain the provider’s capacity to manipulate bundles. As shown in equation~\eqref{eq:u_complementarities}, the consumer’s preferences may include coefficients $\beta_{ij}$ that capture complementarities and substitutions, but the provider’s choice set remains $m$-dimensional. The remaining $n-m$ components of $\mathbf{x}_t$ are mechanically determined as interaction terms $x_{i,t}x_{j,t}$ and cannot be chosen independently.

In what follows, we discuss how each of our main results adapts when interactions are present, clarifying which propositions remain valid and under what assumptions.

As previously discussed, Lemma \ref{prop: lemma_1} and Proposition \ref{prop:1} hold independently of whether complementarities or substitutions are present.  
Since the updating rule operates on the full regressor vector $\mathbf{x}_t$, interaction terms can be treated as additional ``goods'' with corresponding coefficients $\beta_{ij}$.  
Hence, all results concerning the convergence of estimated preferences, the decomposition of biases, and the characterization of the learning dynamics remain valid under the extended specification.

\subsection{Orthogonal bundles under complementarities}

For Proposition \ref{cor:1}, we can show that even in the presence of complementarities and substitution effects it is always possible to construct a bundle orthogonal to $\Delta\bm{\beta}_t$. 
% Hence, the proposition remains valid when $\alpha=\hat{\alpha}$.
% However, when $\alpha \neq \hat{\alpha}$, a bundle exactly proportional to the bias vector $\Delta\bm{\beta}_t$ may no longer be feasible, since the presence of interaction terms restricts the set of attainable combinations of goods.
% When these effects are sufficiently small, the proportional bundle can be locally approximated, preserving the interpretation of $\mathbf{x}_t \propto \Delta\bm{\beta}_t$ as the direction of maximal belief reinforcement.
This is done in the following result.

\begin{proposition}
    \label{app:ortho_compl}
Consider $i, j$ such that $\Delta\beta_{i,t}>0$, $\Delta\beta_{j,t}<0$ and $\Delta\beta_{ij,t}\neq 0$, and a bundle
$\mathbf{x}_{t+1}=x_{i,t+1} \cdot \mathbf{e}_i +x_{j,t+1} \cdot \mathbf{e}_j $ with $x_{i,t+1}>0$ and $x_{j,t+1}>0$. Then, in expectation, if $x_{i,t+1}$ is equal to the only outcome of the formula
\[
\frac{\Delta\beta_{i,t}+\Delta\beta_{ij,t}-\Delta\beta_{j,t}
\pm
\sqrt{\Delta\beta_{i,t}^2
+2\Delta\beta_{i,t}(\Delta\beta_{ij,t}-\Delta\beta_{j,t})
+(\Delta\beta_{ij,t}+\Delta\beta_{j,t})^2}}
{2\Delta\beta_{ij,t}}
\]
strictly between $0$ and $1$, 
and $x_{j,t+1}=1-x_{i,t+1}$,
there is no preference updating, $\mathbb{E}[\hat{\bm{\beta}}_{t+1}]=\hat{\bm{\beta}}_t$.
\end{proposition}

\subsection{Recursive learning with interaction terms}

Proposition~\ref{prop:LT} extends to the case with complementarities and substitution only in its long-run identification logic. Since the augmented parameter vector includes both primitive and interaction coefficients, convergence requires variation not only in primitive goods but also in their pairwise interaction regressors.

A learning path consisting exclusively of single-good bundles cannot identify interaction coefficients, since interaction regressors are observed only when the corresponding primitives co-vary. A minimal learning path must therefore include both singleton exposures (one primitive at a time) and pair exposures (two primitives jointly), so as to generate variation in $x_i x_j$.

From now on, we consider dummy bundles and normalize primitive exposures by imposing
\[
\|\mathbf{x}_t\|_\infty = 1 \quad \text{for all } t,
\]
which fixes the scale of primitive quantities and keeps interaction terms uniformly bounded.

Even with histories restricted to singletons and pairs, pair exposures mechanically couple primitive and interaction regressors, so $\mathbf{Z}_t$ is no longer diagonal in the augmented parameter space. The next result shows, however, that the inverse information matrix still retains a sparse local structure.

\begin{proposition}
\label{prop:Wzeros_interactions}
Consider the augmented specification with $m$ primitive goods and all $\frac{m(m-1)}{2}$ pairwise interaction terms, under dummy consumption.
Assume the consumer's history consists only of singleton bundles and pair bundles.
Let $s_i$ be the number of singleton observations of good $i$, and $c_{ij}$ the number of pair observations of $(i,j)$, and let $\mathbf{Z}$ be the corresponding information matrix, with $\mathbf{W}=\mathbf{Z}^{-1}$.

Then, for any interaction term $(i,j)$ and any primitive good $h\notin\{i,j\}$,
\[
\mathbf{W}_{ij,h}=0 .
\]
Equivalently, in the primitive--interaction off-diagonal blocks of $\mathbf{W}$, an interaction regressor $(i,j)$ has nonzero entries only in the columns corresponding to $i$ and $j$.
\end{proposition}

Proposition \ref{prop:Wzeros_interactions} shows that, under singleton and pair exposures, the inverse information matrix retains a remarkably sparse structure despite the presence of interaction terms. Although pair bundles mechanically couple primitive and interaction regressor in $\mathbf{Z}$, this dependence does not propagate arbitrarily in $\mathbf{W}=\mathbf{Z}^{-1}$.

In particular, an interaction regressor $(i,j)$ is conditionally orthogonal to every primitive $h\notin\{i,j\}$. Thus, learning about complementarities between $i$ and $j$ does not directly affect the estimated marginal utility of unrelated goods. The only primitive coefficients directly linked to $\beta_{ij}$ through the updating rule are $\beta_i$ and $\beta_j$.

This structural property has two implications. First, it preserves a form of local learning: even with complementarities, inference remains confined to the minimal subnetwork generated by the bundle. Second, it limits the scope of correlation externalities induced by interaction terms, preventing higher-order spillovers across unrelated primitives.

Hence, while complementarities enlarge the parameter space, they do not generate full cross-contamination of beliefs. The geometry of exposure still governs the propagation of updates, but the interaction structure preserves a localized pattern of conditional dependence.\footnote{This property may be relevant for experimental economists when designing sessions. If more than one treatment is varied, restricting attention to at most two treatments at a time preserves conditional orthogonality between unrelated treatments in the inverse information matrix, limiting cross-contamination in estimated coefficients and improving interpretability.}

% \subsection{Spectral properties and implications for bundle design}

% Finally, regarding Proposition \ref{prop:robust} when complementarities or substitutions are present, the structure of the feasible set may prevent the provider from implementing bundles that correspond exactly to the eigenvectors of $\mathbf{Z}_t$.  
% The interaction terms impose nonlinear constraints that tie the consumption levels of primitive goods to those of their products.  
% Nevertheless, when these effects are moderate, eigenbundles can be locally approximated by feasible combinations of goods that replicate the same principal directions of variation in consumption.  
% Hence, the interpretation of eigenbundles as the optimal informational directions remains approximately valid.

% {\bf \color{red} [here we may say something about the provider section]}
\subsection{Spectral properties and implications for bundle design}

% Finally, Proposition \ref{prop:robust} does not extend unchanged once complementarities or
% substitutions are introduced. With interaction terms, the provider chooses primitive consumption
% levels, while the augmented regressors are mechanically generated by those choices. The feasible
% set in the augmented space is therefore constrained, generally nonlinear, and need not contain the
% eigenvectors of $\mathbf Z_t$.

% For this reason, the exact spectral characterization in Proposition \ref{prop:robust} is specific to
% the baseline model. When complementarities are present, the eigenvectors of $\mathbf Z_t$ still
% identify directions of greatest redundancy and weakest identification, but they should be interpreted
% as benchmark directions rather than implementable bundles.
% The provider's problem then becomes a constrained design problem over primitive bundles. In this
% sense, complementarities turn the spectral prescription of Proposition \ref{prop:robust} into a
% second-best implementation problem. Deriving the exact optimal policy in this environment requires
% solving a constrained nonlinear design problem, which we leave to future work.

Finally, Propositions \ref{prop:robust} and \ref{prop:MonII} do not extend unchanged once feasibility constraints are imposed. %This issue already arises under the simple restriction $\mathbf{x}_t\in\mathbb{R}^n_{+}$: while the leading eigenvector $\mathbf{v}^N_t$ remains feasible by Perron--Frobenius, the correlation-breaking direction $\mathbf{v}^C_t$ may fail to be implementable because of its mixed sign pattern.
With complementarities or substitutions %, the problem becomes stronger. T
the provider chooses only primitive consumption levels, while the augmented regressors are mechanically induced by those choices. The feasible set in the augmented space is therefore constrained, generally nonlinear, and need not contain the eigenvectors of $\mathbf Z_t$.

Nevertheless, the core message of Propositions \ref{prop:robust} and 
\ref{prop:MonII} survives. The first-best spectral bundles are $\mathbf{v}^N(\mathbf{Z}_t)$ and $\mathbf{v}^C(\mathbf{Z}_t)$ computed on the full augmented matrix, but as discussed above these are generically not implementable since the feasible set does not contain arbitrary vectors in the augmented space. A natural second--best benchmark consists of $\mathbf{v}^N(\mathbf{Z}^p)$ 
and $\mathbf{v}^C(\mathbf{Z}^p)$, the eigenvectors of the primitive co-occurrence 
matrix, which are always feasible. The loss in efficiency from using these second-best bundles rather than the first-best
depends on the importance of complementarities in the consumer’s utility. While the
spectral directions themselves are determined by the consumption history and therefore
do not depend on the utility parameters, the welfare implications of deviating from the
first-best do. In particular, when interaction coefficients $\beta_{ij}$ are small relative to
primitive coefficients $\beta_i$, the efficiency loss is limited, whereas stronger complementarities or substitutabilities amplify the cost of such deviations.
By contrast, the logic of Proposition~\ref{prop:monCI} is more robust: since complete-information 
manipulation and discovery rely on bias-targeted bundles rather than on exact 
spectral implementation, and Proposition~\ref{app:ortho_compl} shows that 
orthogonal two-good bundles can still be constructed in the presence of 
interaction terms, the underlying mechanism survives under complementarities. 
Deriving the exact constrained optimal policy requires solving a nonlinear 
design problem, which we leave to future work.

% \red{the proble with complemetarity becae to choose the feasible bunfle that minimize the vector similartity with vn and vc}
% For this reason, the exact spectral characterizations in Propositions \ref{prop:robust} and \ref{prop:MonII} are specific to the baseline model. When complementarities are present, the eigenvectors of $\mathbf Z_t$ still identify directions of greatest redundancy and weakest identification, but they should be interpreted as benchmark directions rather than implementable bundles. By contrast, the logic of Proposition \ref{prop:monCI} is more robust: since complete-information manipulation and discovery rely on bias-targeted bundles rather than on exact spectral implementation, and Proposition \ref{app:ortho_compl} shows that orthogonal two-good bundles can still be constructed in the presence of interaction terms, the underlying mechanism survives under complementarities. In this sense, complementarities turn the spectral prescriptions of Propositions \ref{prop:robust} and \ref{prop:MonII} into second-best implementation problems. Deriving the exact optimal policy in this richer environment requires solving a constrained nonlinear design problem, which we leave to future work.

\section{Unknown Baseline Utility}
\label{app:alpha}

In the baseline model, we assume that the consumer knows the 
baseline utility $\alpha$. We now relax this assumption and 
consider two cases. In the first, we consider the case of  
\emph{misspecified intercept}, in which the consumer holds a fixed 
incorrect value $\hat{\alpha} \neq \alpha$ and estimates 
$\bm{\beta}$ alone. In the second, we consider the case of \emph{estimated intercept}, in which the consumer recognizes her 
uncertainty about $\alpha$ and estimates the intercept jointly with 
$\bm{\beta}$.

\paragraph{Misspecified intercept.}
The main results of the model can be retrieved under this 
specification, with the following modifications.
The estimator in Lemma~\ref{prop: lemma_1} becomes
$$
\hat{\bm{\beta}}_t 
= \bm{W}_{t}\,\mathbf{X}_t'(\mathbf{u}_t - \hat{\alpha}\bm{1}),
$$
which expands to
$$
\hat{\bm{\beta}}_t 
= \bm{\beta} 
+ (\alpha - \hat{\alpha})\bm{W}_{t}(\mathbf{X}_t)'\bm{1} 
+ \bm{W}_{t}(\mathbf{X}_t)'\bm{\varepsilon}_t.$$
The second term on the right-hand side is a permanent bias 
induced by the intercept misspecification, which does not 
vanish as long as $\hat{\alpha} \neq \alpha$; unlike the noise 
term, which vanishes asymptotically since $\bm{\varepsilon}_t$ 
has zero mean. 
Hence, perfect asymptotic learning of $\bm{\beta}$ never occurs 
when the intercept is misspecified. In particular, 
Proposition~\ref{prop:LT} no longer holds: even if 
conditions~$(i)$--$(ii)$ are satisfied, the permanent bias 
$(\alpha-\hat{\alpha})\bm{W}_t(\mathbf{X}_t)'\bm{1}$ 
prevents $\Delta\bm{\beta}_t$ from converging to zero.

The updating rule in Proposition~\ref{prop:1} is 
unchanged in form, but the consumption surprise at time $t$ 
becomes
$\Delta u_t 
= \alpha + \mathbf{x}'_t\bm{\beta} + \varepsilon_t 
- \hat{\alpha} - \mathbf{x}'_t\hat{\bm{\beta}}_{t-1}$,
so that even in the absence of estimation error 
($\hat{\bm{\beta}}_{t-1} = \bm{\beta}$), the consumer 
experiences a non-zero expected surprise 
$\mathbb{E}[\Delta u_t] = \alpha - \hat{\alpha}$ at every 
period. This permanent surprise does not vanish with experience 
and continuously distorts the updating of $\hat{\bm{\beta}}_t$ 
away from $\bm{\beta}$.

\begin{figure}[h!]
\label{fig:nolearn}
\centering

%---------------------------------
% (a) Orthogonal case (α̂ = α)
%---------------------------------
\subfloat[\small $\hat{\alpha}=\alpha$]{
\begin{tikzpicture}[line cap=round, line join=round, scale=1.0, baseline={(0,-0.2)}]
  % Axes (same vertical span for both figures)
  \draw[-{Latex}] (-2.4,0) -- (2.6,0) node[below right] {\small Good 1};
  \draw[-{Latex}] (0,-1.0) -- (0,2.6) node[left] {\small Good 2 \ };

  % Vectors (steeper Δβ, orthogonal x)
  \coordinate (O)  at (0,0);
  \coordinate (Db) at (-1,2);
  \coordinate (xv) at ( 2,1);

  % Draw vectors
  \draw[-{Latex},thick, blue] (O) -- (Db)
    node[pos=0.6, xshift=-23pt, yshift=12pt] {$\Delta\bm{\beta}_t$};
  \draw[-{Latex},thick] (O) -- (xv)
    node[pos=0.6, xshift= 10pt, yshift=17pt] {$\mathbf{x}_{t+1}$};
\end{tikzpicture}
\label{fig:orthogonal_case}
}
\hspace{0.75cm}
%---------------------------------
% (b) Biased case (α̂ ≠ α)
%---------------------------------
\subfloat[\small $\hat{\alpha}\neq\alpha$]{
\begin{tikzpicture}[line cap=round, line join=round, scale=1.0, baseline={(0,-0.2)}]
  % Axes (identical bounds to left panel)
  \draw[-{Latex}] (-2.4,0) -- (2.6,0) node[below right] {\small Good 1};
  \draw[-{Latex}] (0,-1.0) -- (0,2.6) node[left] {\small Good 2 \ };

  % Δβ (in blue)
  \coordinate (O)  at (0,0);
  \coordinate (Db) at (-1,2);
  \draw[-{Latex},thick,blue] (O) -- (Db)
    node[pos=0.6, xshift=-23pt, yshift=12pt, text=blue] {$\Delta\bm{\beta}_t$};

  % Two x vectors colinear with Δβ
  \coordinate (xQII) at (-0.55, 1.1);
  \coordinate (xQIV) at ( 0.55,-1.1);
  \draw[-{Latex},thick] (O) -- (xQII)
    node[pos=0.55, xshift=-20pt, yshift= 6pt] {$\mathbf{x}_{t+1}$};
  \draw[-{Latex},thick] (O) -- (xQIV)
    node[pos=0.55, xshift= 20pt, yshift=-6pt] {$\mathbf{x}_{t+1}$};

  % Dashed projections
  \draw[densely dashed] (xQII) -- (0, 1.1);
  \draw[densely dashed] (xQIV) -- (0,-1.1);

  % Labels for sign of (α−α̂)
  \node[right]  at (0, 1) {\footnotesize  \ $\alpha-\hat{\alpha}>0$};
  \node[left]   at (0,-1) {\footnotesize   $\alpha-\hat{\alpha}<0$ \ };
\end{tikzpicture}
}

\caption{\footnotesize  Bundles that do not generate expected surprise and thus prevent learning.  
(a) If $\hat{\alpha}=\alpha$, any bundle orthogonal to $\Delta\bm{\beta}_t$;  
(b) if $\hat{\alpha}\neq\alpha$, bundles lying on a shifted hyperplane orthogonal to $\Delta\bm{\beta}_t$, with the shift determined by the sign of $(\hat{\alpha}-\alpha)$.}
\label{fig:learning_vectors}
\end{figure}
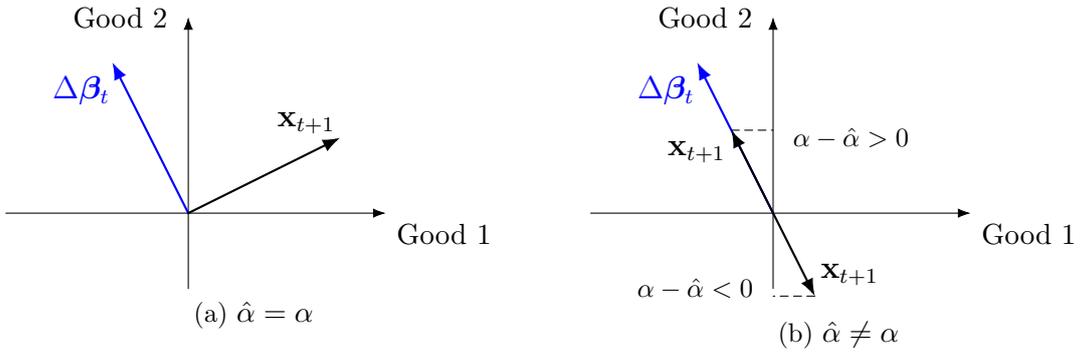

If the consumer holds a misspecified intercept, it is still possible for the provider to halt 
learning. However, the structure of the zero-surprise condition 
changes relative to the baseline model. When the intercept is 
correctly specified, any bundle orthogonal to $\Delta\bm{\beta}_t$ 
generates zero expected surprise. When the intercept is 
misspecified, orthogonal bundles no longer suffice: the provider 
must instead offer bundles that lie on a shifted hyperplane, 
as characterized by Proposition~\ref{prop:nolearn} below 
and illustrated in Figure~\ref{fig:learning_vectors} for both cases.

\begin{proposition}\label{prop:nolearn}
Let $\Delta\bm{\beta}_t := \hat{\bm{\beta}}_t - \bm{\beta}$ 
denote the bias in estimated preferences at time $t$, and let 
$\mathbf{x}_{t+1} \neq \bm{0}$ be the bundle consumed at time 
$t{+}1$. Then, in expectation, the consumer experiences no 
surprise, $\mathbb{E}[\Delta u_{t+1}] = 0$, and therefore does 
not update her estimated preferences, 
$\mathbb{E}[\hat{\bm{\beta}}_{t+1}] = \hat{\bm{\beta}}_t$, 
if and only if
\begin{equation}
\mathbf{x}_{t+1}
= 
-\frac{\hat{\alpha}-\alpha}{\|\Delta\bm{\beta}_t\|_2^{2}} 
\Delta\bm{\beta}_t
+ \mathbf{z},
\label{eq:xnolearn}
\end{equation}
where $\mathbf{z}$ is any vector orthogonal to 
$\Delta\bm{\beta}_t$.
\end{proposition}

Finally, note that the permanent bias in $\hat{\bm{\beta}}_t$ induced 
by intercept misspecification has direct implications for the 
provider's profits. When the consumer underestimates the 
baseline utility, that is $\hat{\alpha} < \alpha$, her estimated 
marginal utilities $\hat{\bm{\beta}}_t$ are systematically 
inflated relative to their true values. A provider aware of 
this can price above true marginal utility and extract larger 
profits than in the baseline model, independently of any 
active manipulation of bundles.

\paragraph{Estimated intercept.}
If instead the consumer recognizes her uncertainty about 
$\alpha$ and estimates it jointly with $\bm{\beta}$, the OLS 
problem uses the augmented design 
$\mathbf{X}^*_t = (\bm{1}_t, \mathbf{X}_t)$ with parameter 
vector $(\alpha,\bm{\beta})'$. By the Frisch--Waugh--Lovell 
theorem, one can partial out the intercept via the demeaning 
matrix $\bm{M}_t := \bm{I}_t - \tfrac{1}{t}\bm{1}_t \bm{1}_t'$, 
where $\bm{I}_t$ is the $t\times t$ identity and $\bm{1}_t$ is 
the $t$-dimensional vector of ones. Letting 
$\bar{u}_t := \tfrac{1}{t}\bm{1}_t'\mathbf{u}_t$ and 
$\bar{\mathbf{x}}_t := \tfrac{1}{t}\mathbf{X}_t'\bm{1}_t$, 
the slope and intercept estimators are
\[
\hat{\bm{\beta}}_t 
= (\mathbf{X}_t' \bm{M}_t \mathbf{X}_t)^{-1}
\mathbf{X}_t' \bm{M}_t \mathbf{u}_t,
\qquad
\hat{\alpha}_t = \bar{u}_t - 
\bar{\mathbf{x}}_t'\hat{\bm{\beta}}_t.
\]
Since $\bm{M}_t \bm{1}_t = \bm{0}$, the intercept drops out 
of the slope regression and is recovered from sample means. 
When regressors are mean-centered ($\bar{\mathbf{x}}_t = 
\bm{0}$), $\bm{M}_t\mathbf{X}_t = \mathbf{X}_t$ and the slope 
estimator reduces to 
$(\mathbf{X}_t'\mathbf{X}_t)^{-1}\mathbf{X}_t'\mathbf{u}_t$, 
exactly as in the baseline model.

The qualitative logic of the model remains the same, but the 
relevant geometry is now defined in terms of demeaned 
regressor. %In particular, 
The precision matrix becomes
\[
\bm{Z}^M_t := \mathbf{X}_t'\bm{M}_t\mathbf{X}_t,
\]
and the effect of a new observation on $\hat{\bm{\beta}}_t$ is 
governed by the demeaned component of the new bundle, rather 
than by the raw bundle itself. For this reason, the results of 
the main text do not extend by a simple notational replacement 
of $\bm{Z}_t = \mathbf{X}_t'\mathbf{X}_t$ with 
$\bm{Z}^M_t$. Rather, the same intuition survives after 
residualizing bundles with respect to the intercept: the 
updating rule, the no-surprise conditions, and the spectral 
arguments must all be reformulated using demeaned bundles. 
Accordingly, the co-occurrence structure and the associated 
eigen-directions should here be interpreted as those induced by 
the residualized design matrix $\bm{M}_t\mathbf{X}_t$.

\section{Proofs}\label{app:proofs}
\subsection*{Proof of Lemma~\ref{prop: lemma_1}}
The proof can be derived by adapting standard OLS estimation allowing for a misspecified intercept. \hfill$\blacksquare$

\subsection*{Proof of Proposition \ref{prop:1}}

We begin by considering the vector containing all $\hat{\beta}_{k,t}$ given a chosen $\mathbf{x}_{t}$. Utilizing the  \cite{sherman1950} formula in conjunction with equation \eqref{eq: beta}, we can express the update rule for $\bm{\hat{\beta}}_{t}$ as:
\begin{equation}
\bm{\hat{\beta}}_{t} = \bm{\hat{\beta}}_{t-1} + \frac{1}{1+ \mathbf{x}'_{t}\left(\mathbf{X}_{t-1}'\mathbf{X}_{t-1} \right)^{-1} \mathbf{x}_{t}} \left(\mathbf{X}_{t-1}'\mathbf{X}_{t-1} \right)^{-1} \mathbf{x}_{t} \left(u_{t}- \mathbf{x}'_{t}\bm{\hat{\beta}}_{t-1}\right).
\end{equation}
Recognizing that $\bm{W}_{t-1} = \left(\mathbf{X}_{t-1}'\mathbf{X}_{t-1} \right)^{-1}$, this simplifies to:
\begin{equation}
\bm{\hat{\beta}}_{t} = \bm{\hat{\beta}}_{t-1} + \frac{1}{1+ \mathbf{x}'_{t}\bm{W}_{t-1} \mathbf{x}_{t}} \bm{W}_{t-1} \mathbf{x}_{t} \left(u_{t}- \mathbf{x}'_{t}\bm{\hat{\beta}}_{t-1}\right).
\end{equation}
Substituting $u_{t}= \mathbf{x}'_{t}\bm{\beta} + \varepsilon_t$, we obtain:
\begin{equation}
\label{eq: consequence_sherman_morrison}
\bm{\hat{\beta}}_{t} = \bm{\hat{\beta}}_{t-1} + \frac{1}{1+ \mathbf{x}'_{t}\bm{W}_{t-1} \mathbf{x}_{t}} \bm{W}_{t-1} \mathbf{x}_{t} \left( \mathbf{x}'_{t}\bm{\beta} + \varepsilon_t- \mathbf{x}'_{t}\bm{\hat{\beta}}_{t-1}\right).
\end{equation}
This can be further simplified as:
\begin{equation}
\bm{\hat{\beta}}_{t} = \bm{\hat{\beta}}_{t-1} + \bm{\omega}_t \Delta u_t,
\end{equation}
where $\Delta u_t:=\mathbf{x}'_{t}\bm{\beta} + \varepsilon_t- \mathbf{x}'_{t}\bm{\hat{\beta}}_{t-1}$ and $\bm{\omega}_t:=\frac{1}{1+ \mathbf{x}'_{t}\bm{W}_{t-1} \mathbf{x}_{t}} \bm{W}_{t-1} \mathbf{x}_t$.

Since $\bm{W}_{t-1}$ is positive semi-definite, we know that $1 + \mathbf{x}'_{t}\bm{W}_{t-1} \mathbf{x}_{t} > 0$, ensuring the stability of the update.

We can now expand the term $1 + \mathbf{x}'_{t}\bm{W}_{t-1} \mathbf{x}_{t}$ as:
\begin{align*}
    1 + \sum_{i = 1}^{I} \sum_{j = 1}^{I} w_{ij,t-1} x_{i,t} x_{j,t} &= 1 + \mathbf{x}'_{t}\bm{W}_{t-1} \mathbf{x}_{t} \\
    &= 1 + \frac{1}{\sigma^2}\text{Var}[\mathbf{x}'_{t} \bm{\hat{\beta}}_{t-1}]\\
  %  &= \frac{\sigma^2+ \text{Var}[\mathbf{x}'_{t} \bm{\hat{\beta}}_{t-1}]}{\sigma^2} \\
    &= \frac{\sigma^2+ \text{Var}[\hat{u}_t]}{\sigma^2}.
\end{align*}
Thus, $\bm{\omega}_t$ becomes:
\[
\bm{\omega}_t=\frac{\sigma^2}{\sigma^2+ \text{Var}[\hat{u}_t]} \bm{W}_{t-1} \mathbf{x}_t,
\]
and the update rule for each component $\hat{\beta}_{i,t}$ is:
\[
\hat{\beta}_{i,t} = \hat{\beta}_{i,t-1} + \frac{\sigma^2 }{\sigma^2+  \text{Var}[\hat{u}_t] }\sum_{j=1 }^{n} w_{ij,t-1} x_{j,t} \Delta u_{t}.
\]

\hfill$\blacksquare$

\subsection*{Proof of Proposition \ref{cor:1}}
Fix time $t$ and consider two goods $i, j$ with estimation errors 
$\Delta\beta_{i,t} > 0$ and $\Delta\beta_{j,t} < 0$. Let the bundle be 
$\mathbf{x}_{t+1} = x_{i,t+1} \cdot \mathbf{e}_i + x_{j,t+1} \cdot \mathbf{e}_j$ 
with $x_{i,t+1} > 0$ and $x_{j,t+1} > 0$.
From Proposition~\ref{prop:1}, the expected consumption surprise is:
$
\mathbb{E}[\Delta u_{t+1}] = -\mathbf{x}'_{t+1}\Delta\bm{\beta}_t = 
-x_{i,t+1}\Delta\beta_{i,t} - x_{j,t+1}\Delta\beta_{j,t}$,
and the expected update for each good $k \in \{i,j\}$ is:
\begin{equation}
\mathbb{E}[\hat{\beta}_{k,t+1} - \hat{\beta}_{k,t}] = 
\frac{\sum_{\ell} w_{k\ell,t}\, x_{\ell,t+1}}
{1 + \mathbf{x}'_{t+1}\bm{W}_t \mathbf{x}_{t+1}} 
\cdot \mathbb{E}[\Delta u_{t+1}].
\end{equation}
Since the bundle has nonzero components only on $i$ and $j$, the relevant 
entries of $\bm{W}_t$ are $w_{ii,t}$, $w_{jj,t}$, and $w_{ij,t} = w_{ji,t}$, 
so:
\begin{align}
\sum_\ell w_{i\ell,t}\, x_{\ell,t+1} 
&= w_{ii,t}\, x_{i,t+1} + w_{ij,t}\, x_{j,t+1}, \\
\sum_\ell w_{j\ell,t}\, x_{\ell,t+1} 
&= w_{ji,t}\, x_{i,t+1} + w_{jj,t}\, x_{j,t+1}.
\end{align}
Let $D := 1 + w_{ii,t}x_{i,t+1}^2 + 2w_{ij,t}x_{i,t+1}x_{j,t+1} + 
w_{jj,t}x_{j,t+1}^2 > 0$, where positivity follows from positive 
definiteness of $\bm{W}_t$.

\begin{enumerate}
    \item 
The condition $\frac{x_{j,t+1}}{x_{i,t+1}} = 
-\frac{\Delta\beta_{i,t}}{\Delta\beta_{j,t}}$ is well-defined and positive 
since the two estimation errors have opposite signs. Substituting into the 
expected surprise:
\[
\mathbb{E}[\Delta u_{t+1}] 
= -x_{i,t+1}\!\left(\Delta\beta_{i,t} + 
\frac{x_{j,t+1}}{x_{i,t+1}}\Delta\beta_{j,t}\right)
= -x_{i,t+1}\!\left(\Delta\beta_{i,t} - 
\frac{\Delta\beta_{i,t}}{\Delta\beta_{j,t}}\Delta\beta_{j,t}\right)
= 0.
\]
Since $D > 0$, the entire update vector is zero, so 
$\mathbb{E}[\hat{\bm{\beta}}_{t+1}] = \hat{\bm{\beta}}_t$.

\item
The condition $\frac{x_{j,t+1}}{x_{i,t+1}} > 
-\frac{\Delta\beta_{i,t}}{\Delta\beta_{j,t}}$ implies 
$\mathbb{E}[\Delta u_{t+1}] > 0$. To see this, note that 
$\Delta\beta_{j,t} < 0$ so 
$-\frac{\Delta\beta_{i,t}}{\Delta\beta_{j,t}} > 0$, and:
\[
\mathbb{E}[\Delta u_{t+1}] 
= -x_{i,t+1}\Delta\beta_{i,t} - x_{j,t+1}\Delta\beta_{j,t}.
\]
The first term is negative (since $\Delta\beta_{i,t}>0$) and the second 
is positive (since $\Delta\beta_{j,t}<0$). The stated condition ensures 
the bundle places sufficient relative weight on the underestimated good 
$j$, so the positive term dominates and $\mathbb{E}[\Delta u_{t+1}] > 0$.

Since $D>0$ and $\mathbb{E}[\Delta u_{t+1}]>0$, the sign of 
$\mathbb{E}[\hat{\beta}_{i,t+1} - \hat{\beta}_{i,t}]$ is determined 
entirely by $w_{ii,t}\,x_{i,t+1} + w_{ij,t}\,x_{j,t+1}$, and the sign 
of $\mathbb{E}[\hat{\beta}_{j,t+1} - \hat{\beta}_{j,t}]$ by 
$w_{ji,t}\,x_{i,t+1} + w_{jj,t}\,x_{j,t+1}$.

\begin{itemize}
    \item[(a)]
Let us consider the case in which $w_{ij,t} > 0$.
Since $w_{ii,t} > 0$, $w_{ij,t} > 0$, $x_{i,t+1} > 0$, and 
$x_{j,t+1} > 0$, we have 
$w_{ii,t}\,x_{i,t+1} + w_{ij,t}\,x_{j,t+1} > 0$. By a symmetric 
argument, $w_{ji,t}\,x_{i,t+1} + w_{jj,t}\,x_{j,t+1} > 0$. Hence 
both updates are strictly positive.

\item[(b)] Let us consider the case in which $w_{ij,t} < 0$.
The update for good $i$ is positive if and only if:
\[
w_{ii,t}\, x_{i,t+1} > |w_{ij,t}|\, x_{j,t+1} 
\quad\Longleftrightarrow\quad 
\frac{x_{j,t+1}}{x_{i,t+1}} < \frac{w_{ii,t}}{|w_{ij,t}|}.
\]
The update for good $j$ is positive if and only if:
\[
w_{jj,t}\, x_{j,t+1} > |w_{ji,t}|\, x_{i,t+1} 
\quad\Longleftrightarrow\quad 
\frac{x_{j,t+1}}{x_{i,t+1}} > \frac{|w_{ij,t}|}{w_{jj,t}}.
\]
Both updates are simultaneously positive if and only if:
\[
\frac{|w_{ij,t}|}{w_{jj,t}} < \frac{x_{j,t+1}}{x_{i,t+1}} < 
\frac{w_{ii,t}}{|w_{ij,t}|}.
\]
This interval is non-empty if and only if 
$w_{ij,t}^2 < w_{ii,t}\,w_{jj,t}$, which holds strictly since 
$\bm{W}_t$ is positive definite: for any positive definite matrix, 
the determinant of every principal submatrix is positive, so in 
particular $w_{ii,t}\,w_{jj,t} - w_{ij,t}^2 > 0$. \hfill$\blacksquare$
\end{itemize}
\end{enumerate}

\subsection*{Proof of Proposition \ref{prop:LT}}

Recall that the estimation error is
$\Delta\bm\beta_t = \bm{W}_{t}(\mathbf{X}_t)' \bm{\varepsilon}_t$.
Since $\mathbb{E}[\varepsilon_s]=0$, taking expectations we get
$\mathbb{E}[\Delta\bm\beta_t]=\bm{0}$.

We now find the conditions under which the variance of the estimation
error shrinks to zero. Recall that
\[
\operatorname{Var}\big[\Delta\bm{\beta}_t\big]
= \operatorname{Var}\big[\hat{\bm{\beta}}_t\big]
= \sigma^2\bm{W}_{t} = \sigma^2\bm{Z}_{t}^{-1}.
\]
Convergence in probability, $\hat{\bm\beta}_t \xrightarrow{p} \bm\beta$,
therefore requires $\bm{W}_t\to 0$, i.e., that the uncertainty associated
with each estimated coefficient eventually disappears. Since $\bm{Z}_t$ is
symmetric and positive definite, this happens if and only if all its
eigenvalues grow without bound. The smallest eigenvalue
$\lambda^{\min}(\bm{Z}_t)$ represents the direction in which information
accumulates most slowly; convergence thus requires that even this weakest
direction be infinitely explored, so that
$\lambda^{\min}(\bm{Z}_t)\to\infty$.

Each eigenvalue of $\bm{Z}_t$ measures the cumulative information
accumulated along an independent linear combination of the parameters.
The smallest eigenvalue grows without bound only if every component of
$\mathbf{x}_t$ contributes information infinitely often and no linear
combination of parameters remains unobserved or perfectly collinear with
others. This requires that each coordinate $i$
appears in infinitely many periods and that the bundles $\mathbf{x}_t$ are
not repeatedly orthogonal to the current bias $\Delta\bm{\beta}_t$,
ensuring that every dimension of $\bm{\beta}$ is eventually explored. {Conditions~$(i)$--$(ii)$ are therefore necessary for convergence, and jointly sufficient since together they ensure that no direction in $\mathbb{R}^n$ is permanently unexplored.}
 
 %Hence, convergence in probability, $\hat{\bm{\beta}}_t \xrightarrow{p} \bm{\beta}$, occurs if and only if conditions~$(i)$--$(ii)$ hold.

We now characterize which bundle sequences minimize the total mean
squared error at every $t$. Since
$\hat{\bm{\beta}}_t$ is unbiased:
\[
\mathbb{E} \left[||\Delta\bm{\beta}_t||_2^2\right]
= \operatorname{tr}\!\left(\operatorname{Var}
  \big[\hat{\bm{\beta}}_t\big]\right)
= \sigma^2\operatorname{tr}(\bm{W}_t)
= \sigma^2\sum_{i=1}^{n}\frac{1}{\lambda_i(\bm{Z}_t)}.
\]
Since $\operatorname{tr}(\bm{Z}_t)=\sum_{s=1}^{t}
\|\mathbf{x}_s\|_2^2=t$, the AM-HM inequality gives:
$$
\sum_{i=1}^{n}\frac{1}{\lambda_i(\bm{Z}_t)}
\geq \frac{n^2}{\sum_{i=1}^{n}\lambda_i(\bm{Z}_t)}
= \frac{n^2}{t},
$$
with equality if and only if all eigenvalues equal $t/n$,
i.e., $\bm{Z}_t=\frac{t}{n}\bm{I}$. Therefore:
\[
\mathbb{E}\!\left[||\Delta\bm{\beta}_t||_2^2\right]
\geq \frac{\sigma^2 n^2}{t},
\]
with equality if and only if $\bm{Z}_t=\frac{t}{n}\bm{I}$.
It remains to show this holds if %and only if 
bundles are
equal-frequency singletons.

\medskip
\noindent\textit{Sufficiency.}
Suppose that at each period $s$ the bundle is $\mathbf{x}_s=\pm\mathbf{e}_i$
for some $i\in I$, and that each good appears in exactly $t/n$ periods. Then:
$\bm{Z}_t
= \sum_{s=1}^{t}\mathbf{x}_s\mathbf{x}_s'
= \sum_{i=1}^{n}\frac{t}{n}\mathbf{e}_i\mathbf{e}_i'
= \frac{t}{n}\bm{I}$,
so equality holds.

\medskip
\noindent\textit{Necessity and the role of the sign restriction.}
Equal-frequency singletons are not the only sequences achieving
$\bm{Z}_t=\frac{t}{n}\bm{I}$ in general. Non-singleton bundles with
mixed signs can also attain equality, provided their negative components
allow off-diagonal contributions to cancel across observations.
A simple example is $\mathbf{x}_1=(1/\sqrt{2}, 1/\sqrt{2})$ and
$\mathbf{x}_2=(1/\sqrt{2},-1/\sqrt{2})$ for $n=t=2$, which gives
$\bm{Z}_2=\bm{I}=\frac{t}{n}\bm{I}$.
When bundles are restricted to $\mathbf{x}_t\in\mathbb{R}_+^n$, however,
this cancellation is impossible. If $x_{i,s}>0$ and $x_{j,s}>0$
for some $i\neq j$, then $\mathbf{x}_s\mathbf{x}_s'$ contributes
$x_{i,s}x_{j,s}>0$ to off-diagonal entry $(i,j)$ of $\bm{Z}_t$,
and since all remaining observations also have non-negative components,
every other observation contributes $x_{i,r}x_{j,r}\geq 0$ to the
same entry. Hence $Z_{ij}>0$, contradicting $\bm{Z}_t=\frac{t}{n}\bm{I}$.
Under $\mathbf{x}_t\in\mathbb{R}_+^n$, equal-frequency singletons are
therefore the unique minimizers of $\mathbb{E}[\|\Delta\bm{\beta}_t\|^2]$.
\hfill$\blacksquare$

\subsection*{Proof of Proposition \ref{prop:robust}}

For any symmetric positive definite matrix, its normalized eigenvectors 
$\{\mathbf{v}^i_t\}_{i=1,\dots,n}$ form an orthonormal basis of $\mathbb{R}^n$. 
For a generic $\mathbf{x}\in\mathbb{R}^n$, the rank-one matrix $\mathbf{x}\mathbf{x}'$ 
has $n-1$ zero eigenvalues and one nonzero eigenvalue $\|\mathbf{x}\|_2^2$ with 
eigenvector $\mathbf{x}/\|\mathbf{x}\|_2$. In our model, adding observation 
$\mathbf{x}_{t+1}$ updates the precision matrix as
$
\bm{Z}_{t+1} = \bm{Z}_t + \mathbf{x}_{t+1}\mathbf{x}'_{t+1}$.

We denoted by $\lambda^{min}_t$ and $\lambda^{max}_t$ the minimum and maximum 
eigenvalues of $\bm{Z}_t$, with corresponding normalized eigenvectors $\mathbf{v}^C_t$ 
and $\mathbf{v}^N_t$ respectively, and recall $\kappa_t = \lambda^{max}_t/\lambda^{min}_t$.

\begin{lemma}\label{lemma:add}
If we add observation $\mathbf{x}_{t+1} = \mathbf{v}^i_t$ to $\bm{Z}_t$, 
then $\bm{Z}_{t+1}$ retains the same eigenvectors as $\bm{Z}_t$; the $n-1$ 
eigenvalues $\{\lambda^j_t\}_{j\neq i}$ remain unchanged, while the eigenvalue 
corresponding to $\mathbf{v}^i_t$ becomes $\lambda^i_{t+1} = \lambda^i_t + 1$.\footnote{Lemma~\ref{lemma:add} follows from the fact that adding $\mathbf{x}_{t+1}=\mathbf{v}^i_t$ gives 
$\bm{Z}_{t+1} = \bm{Z}_t + \mathbf{v}^i_t(\mathbf{v}^i_t)'$. 
For any eigenvector $\mathbf{v}^j_t$ with $j\neq i$, 
$\bm{Z}_{t+1}\mathbf{v}^j_t = \lambda^j_t\mathbf{v}^j_t + 
\mathbf{v}^i_t(\mathbf{v}^i_t)'\mathbf{v}^j_t = 
\lambda^j_t\mathbf{v}^j_t$, and for $\mathbf{v}^i_t$, 
$\bm{Z}_{t+1}\mathbf{v}^i_t = \lambda^i_t\mathbf{v}^i_t + 
\mathbf{v}^i_t(\mathbf{v}^i_t)'\mathbf{v}^i_t = 
(\lambda^i_t+1)\mathbf{v}^i_t$, where both steps use 
orthonormality of eigenvectors. Hence all eigenvectors are 
preserved and only $\lambda^i_t$ increases by one. }
\end{lemma}
Let us now prove the two points of Proposition \ref{prop:robust}.

\begin{itemize}
    \item 
By Lemma~\ref{lemma:add}, adding $\mathbf{x}_{t+1}=\mathbf{v}^N_t$ 
yields $\lambda^{\max}_{t+1} = \lambda^{\max}_t + 1$, while all other 
eigenvalues are unchanged. Hence
\[
\kappa_{t+1} 
= \frac{\lambda^{\max}_t+1}{\lambda^{\min}_t} 
> \frac{\lambda^{\max}_t}{\lambda^{\min}_t} 
= \kappa_t.
\]
This is the largest possible increase in $\kappa_t$ achievable with a 
unit-norm bundle. To see this, write any unit-norm bundle as 
$\mathbf{x}_{t+1}=\sum_i a_i\mathbf{v}^i_t$ with $\sum_i a_i^2=1$. 
% By Lemma~\ref{lemma:add}, the new largest eigenvalue satisfies
% \[
% \lambda^{\max}_{t+1}\le \lambda^{\max}_t+\sum_i a_i^2
% =\lambda^{\max}_t+1,
% \]
% with equality if and only if all weight is placed on an eigendirection 
% associated with $\lambda_t^{\max}$. In particular, this is attained by 
% $\mathbf{x}_{t+1}=\pm \mathbf{v}^N_t$.
Since $x_{t+1}x_{t+1}'$ is positive semidefinite and has largest eigenvalue
$\|x_{t+1}\|_2^2=1$, Weyl's inequality implies
\[
\lambda^{\max}_{t+1}
=\lambda^{\max}(Z_t+x_{t+1}x_{t+1}')
\le \lambda^{\max}_t+1.
\]
Equality holds if and only if $x_{t+1}$ is aligned with an eigendirection associated with
$\lambda^{\max}_t$. In particular, this is attained by $x_{t+1}=\pm v_t^N$.

Moreover, Lemma~\ref{lemma:add} implies that adding $\mathbf{v}^N_t$ leaves 
all eigendirections unchanged and only shifts upward the eigenvalue associated 
with $\mathbf{v}^N_t$. Therefore, $\mathbf{v}^N_t$ is again a leading 
eigendirection of $\mathbf{Z}_{t+1}$, and if $\mathbf{x}_s=\mathbf{v}^N_t$ 
for every period $s\ge t$, the same bundle remains optimal at every subsequent 
period. In that case,
\[
\kappa_s = \frac{\lambda^{\max}_t+(s-t)}{\lambda^{\min}_t}\to\infty,
\]
so repeated exposure along $\mathbf{v}^N_t$ asymptotically maximizes learning 
time. This is the \emph{global} optimality of $\mathbf{v}^N_t$: it remains the 
most effective robust choice to slow down learning at every future period, 
independently of $\Delta\bm{\beta}_t$.

\item
By Lemma~\ref{lemma:add}, adding $\mathbf{x}_{t+1}=\mathbf{v}^C_t$ 
shifts only the smallest eigenvalue: $\lambda^{min}_t \mapsto 
\lambda^{min}_t+1$, while all other eigenvalues and eigenvectors 
are unchanged. The new minimum eigenvalue of $\bm{Z}_{t+1}$ is therefore:
\[
\lambda^{min}_{t+1} 
= \min\{\lambda^{min}_t + 1,\; \lambda_{n-1,t}\},
\]
where $\lambda_{n-1,t}$ is the second-smallest eigenvalue of $\bm{Z}_t$, 
which is unaffected by the update. The condition number strictly decreases:
\[
\kappa_{t+1} 
= \frac{\lambda^{max}_t}{\lambda^{min}_{t+1}} 
< \frac{\lambda^{max}_t}{\lambda^{min}_t} 
= \kappa_t.
\]
By the same decomposition argument, no unit-norm bundle can decrease 
$\kappa_t$ by more: any $\mathbf{x}_{t+1}=\sum_i a_i\mathbf{v}^i_t$ 
with $\sum_i a_i^2=1$ satisfies 
$\lambda^{min}_{t+1}\leq \lambda^{min}_t + a_n^2 \leq \lambda^{min}_t+1$, 
with equality if and only if $\mathbf{x}_{t+1}=\pm\mathbf{v}^C_t$.

However, after this update the smallest eigenvector of $\bm{Z}_{t+1}$ 
need not remain $\mathbf{v}^C_t$. If $\lambda^{min}_t+1 > \lambda_{n-1,t}$, 
the new minimum eigenvalue direction becomes $\mathbf{v}^{n-1}_t$, so 
$\mathbf{v}^C_{t+1}\neq\mathbf{v}^C_t$ in general. The correlation-breaking 
bundle must therefore be recomputed at each period as $\mathbf{v}^C_{t+1}$, 
making this a \emph{locally} optimal choice: it delivers the largest possible one-step weak reduction
in $\kappa_t$ at each $t$, though the reduction may be zero when the smallest eigenvalue is not simple.
\end{itemize}

\medskip
\noindent Both results hold independently of the current estimation error 
$\Delta\bm\beta_t$: only the spectral decomposition of $\bm{Z}_t$ is 
required, not knowledge of true or estimated preferences. This is the 
sense in which both strategies are \emph{robust}.

By Sherman--Morrison, the one-step reduction in 
MSE 
for any unit-norm bundle $\mathbf{x}_{t+1}$ is
$$\mathbb{E}\left[\|\Delta\bm{\beta}_t\|_2^2\right]
-\mathbb{E}\left[\|\Delta\bm{\beta}_{t+1}\|_2^2\right]
= \sigma^2\cdot\frac{\mathbf{x}_{t+1}'\bm{W}_t^2\mathbf{x}_{t+1}}
{1+\mathbf{x}_{t+1}'\bm{W}_t\mathbf{x}_{t+1}}.$$
Since $\{\mathbf{v}^i_t\}$ is an orthonormal basis, write 
$\mathbf{x}_{t+1}=\sum_i c_i\mathbf{v}^i_t$ with 
$\sum_i c_i^2=1$. Using $\bm{W}_t\mathbf{v}^i_t=
(1/\lambda^i_t)\mathbf{v}^i_t$ and orthonormality, 
this becomes
$\sigma^2\cdot\frac{\sum_i c_i^2/(\lambda^i_t)^2}
{1+\sum_i c_i^2/\lambda^i_t}$,
which depends only on $(c_1^2,\ldots,c_n^2)$ lying in the 
simplex $\{c_i^2\geq 0,\,\sum_i c_i^2=1\}$. The upper and 
lower level sets of this expression are both halfspaces in 
$(c_1^2,\ldots,c_n^2)$, so it is quasilinear on the simplex 
and its maximum and minimum are both attained at vertices. 
At vertex $c_k^2=1$ and $c_i^2=0$ for $i\neq k$, the 
expression equals $\frac{\sigma^2}{\lambda^k_t(\lambda^k_t+1)}$, 
strictly decreasing in $\lambda^k_t$. Hence the one-step 
MSE reduction is maximized at $c_n^2=1$, 
i.e.\ $\mathbf{x}_{t+1}=\mathbf{v}^C_t$, 
and minimized at $c_1^2=1$, 
i.e.\ $\mathbf{x}_{t+1}=\mathbf{v}^N_t$.
\hfill $\blacksquare$

%Finally, since adding $\mathbf{x}_{t+1}=\mathbf{v}^i_t$ increases only $\lambda^i_t$ by one, the one-step reduction in MSE is $\sigma^2\left(\frac{1}{\lambda^i_t}-\frac{1}{\lambda^i_t+1}\right) = \frac{\sigma^2}{\lambda^i_t(\lambda^i_t+1)}$, which is strictly decreasing in $\lambda^i_t$. Therefore  $\mathbf{v}^C_t$ achieves the largest one-step reduction in $\mathbb{E}[\|\Delta\bm{\beta}_t\|^2]$ among all unit-norm bundles, while $\mathbf{v}^N_t$ achieves the smallest.

%\bigskip

%{\color{red} \bf [the proofs below are for the provider's section -- we have also to decide if we want to put here below the proofs from appendix B]}

\subsection*{Proof of Proposition \ref{prop:monCI}}

At $t=2$, profits $\Pi_2 = \mathbf{x}_2'(\hat{\bm{\beta}}_1 - \bm{\gamma})$ are linear in $\mathbf{x}_2$, so on $\{\mathbf{x} \in \mathbb{R}^n : \|\mathbf{x}\|_1 = 1\}$ the optimum is a signed vertex $\mathbf{x}_2^* = \mathrm{sign}(\hat{\beta}_{i^*,1} - \gamma_{i^*}) \cdot \mathbf{e}_{i^*}$ with $i^* \in \arg\max_{i} |\hat{\beta}_{i,1} - \gamma_i|$. When perceived margins are positive, this reduces to $\mathbf{x}_2^* = \mathbf{e}_{i^*}$ with $i^* \in \arg\max_{i}\{\hat{\beta}_{i,1} - \gamma_i\}$, and second-period profits are $\Pi_2^* = \max_{i}\{\hat{\beta}_{i,1} - \gamma_i\}$. The monopolist therefore solves
\[
\max_{\|\mathbf{x}_1 \|_1=1} \; \mathbf{x}_1'(\hat{\bm{\beta}}_0 - \bm{\gamma}) + \delta\, \mathbb{E}\big[\max_{i}\{\hat{\beta}_{i,1} - \gamma_i\}\big].
\]
From Proposition~\ref{prop:1}, taking expectations:
\[
\mathbb{E}[\hat{\bm{\beta}}_1] = \hat{\bm{\beta}}_0 - \frac{\bm{W}_0\mathbf{x}_1}{1+\mathbf{x}_1'\bm{W}_0\mathbf{x}_1}\mathbf{x}_1'\Delta\bm{\beta}_0.
\]
As $\sigma^2 \to 0$, $\hat{\bm{\beta}}_1 \to \mathbb{E}[\hat{\bm{\beta}}_1]$ in probability, so the monopolist effectively controls $\mathbb{E}[\hat{\bm{\beta}}_1]$ through $\mathbf{x}_1$. We characterize the optimal bundle in each case.

\begin{itemize}
\item If $i = j$, the currently perceived best good coincides with the true best good. When $\hat{\beta}_{i,0} \leq \beta_i$, offering $\mathbf{x}_1 = \mathbf{e}_i$ maximizes period-1 profit and generates non-negative expected surprise, so $\mathbb{E}[\hat{\beta}_{i,1}] \geq \hat{\beta}_{i,0}$ and good $i$ remains the perceived best in period 2; for $\delta$ sufficiently large, $\mathbf{x}_1^* = \mathbf{x}_2^* = \mathbf{e}_i$. When instead $\hat{\beta}_{i,0} > \beta_i$, selling $\mathbf{e}_i$ in period 1 generates negative expected surprise that lowers $\mathbb{E}[\hat{\beta}_{i,1}]$ and erodes the period-2 perceived margin. For $\delta$ sufficiently large, the monopolist prefers a period-1 manipulation bundle on $\mathcal{B}$ that preserves $\mathbb{E}[\hat{\beta}_{i,1}]$, and sells $i$ in period 2.

\item If $i \neq j$ and $\hat{\beta}_{i,0} - \gamma_i > \beta_j - \gamma_j$, good $i$ is currently overestimated and the monopolist wishes to maintain its perceived advantage into period 2. From Proposition~\ref{prop:1}, increasing $\mathbb{E}[\hat{\beta}_{i,1}]$ requires $\sum_k w_{ik,0} x_{k,1}$ and $\mathbf{x}_1'\Delta\bm{\beta}_0$ to have opposite signs. Since $\Delta\beta_{i,0} > 0$, a bundle with $\mathbf{x}_1'\Delta\bm{\beta}_0 < 0$ places weight on underestimated goods and generates positive expected surprise. Under $\|\mathbf{x}_1\|_1=1$, signed bundles guarantee the existence of such a configuration regardless of the structure of $\bm{W}_0$ and $\Delta\bm{\beta}_0$. For $\delta$ sufficiently large, the period-2 gain from this manipulation bundle dominates any period-1 cost of deviating from $\mathbf{e}_i$.

\item If $i \neq j$ and $\beta_j - \gamma_j > \hat{\beta}_{i,0} - \gamma_i$, good $j$ has the highest true margin but is currently underestimated. Selling $j$ in period 2 is optimal once beliefs are corrected, so the monopolist chooses $\mathbf{x}_1$ to maximize $\mathbb{E}[\hat{\beta}_{j,1}]$. Since $\Delta\beta_{j,0} < 0$, a bundle that places weight on the underestimated good $j$ gives $\mathbf{x}_1'\Delta\bm{\beta}_0 < 0$, generating positive expected surprise $\mathbb{E}[\Delta u_1] = -\mathbf{x}_1'\Delta\bm{\beta}_0 > 0$ and an upward revision of $\hat{\beta}_{j,1}$. Under $\|\mathbf{x}_1\|_1=1$, such bundles always exist. For $\delta$ sufficiently large, selling $j$ at its true margin in period 2 dominates any myopic strategy, so this discovery bundle is globally optimal.
\hfill$\blacksquare$
\end{itemize}

\subsection*{Proof of Proposition \ref{prop:MonII}}

Under the stationary strategy $\mathbf{x}_1=\mathbf{x}_2=\mathbf{x}$ with $\|\mathbf{x}\|_2=1$, the monopolist solves
\[
\max_{\|\mathbf{x}\|_2=1}\;
\mathbb{E}\!\left[
\mathbf{x}'(\hat{\bm{\beta}}_0-\bm{\gamma})
+
\delta\,\mathbf{x}'(\hat{\bm{\beta}}_1-\bm{\gamma})
\right].
\]

By the zero-mean assumption $\mathbb{E}[\hat{\beta}_{i,0}-\gamma_i]=0$ for every $i$,
\[
\mathbb{E}\!\left[\mathbf{x}'(\hat{\bm{\beta}}_0-\bm{\gamma})\right]
=
\mathbf{x}'\,\mathbb{E}[\hat{\bm{\beta}}_0-\bm{\gamma}]
=0
\]
for every feasible $\mathbf{x}$, so the first-period term vanishes from the maximization.

From Proposition~\ref{prop:1},
\[
\hat{\bm{\beta}}_1
=
\hat{\bm{\beta}}_0
+
\frac{\bm{W}_0\mathbf{x}}{1+\mathbf{x}'\bm{W}_0\mathbf{x}}\Delta u_1 .
\]
Hence
\[
\mathbf{x}'(\hat{\bm{\beta}}_1-\bm{\gamma})
=
\mathbf{x}'(\hat{\bm{\beta}}_0-\bm{\gamma})
+
\frac{\mathbf{x}'\bm{W}_0\mathbf{x}}{1+\mathbf{x}'\bm{W}_0\mathbf{x}}\Delta u_1 .
\]
Taking expectations,
\[
\mathbb{E}\!\left[\mathbf{x}'(\hat{\bm{\beta}}_1-\bm{\gamma})\right]
=
\mathbb{E}\!\left[\mathbf{x}'(\hat{\bm{\beta}}_0-\bm{\gamma})\right]
+
\frac{\mathbf{x}'\bm{W}_0\mathbf{x}}{1+\mathbf{x}'\bm{W}_0\mathbf{x}}
\mathbb{E}[\Delta u_1].
\]
Using again $\mathbb{E}[\mathbf{x}'(\hat{\bm{\beta}}_0-\bm{\gamma})]=0$, the monopolist's objective reduces to
\[
\max_{\|\mathbf{x}\|_2=1}\;
\delta\,
\frac{\mathbf{x}'\bm{W}_0\mathbf{x}}{1+\mathbf{x}'\bm{W}_0\mathbf{x}}
\mathbb{E}[\Delta u_1].
\]

Since $r/(1+r)$ is strictly increasing in $r\geq 0$, maximizing this objective is equivalent to maximizing or minimizing the Rayleigh quotient $\mathbf{x}'\bm{W}_0\mathbf{x}$ over the unit sphere, depending on the sign of $\mathbb{E}[\Delta u_1]$.

By the Rayleigh--Ritz theorem, and using $\bm{W}_0=\bm{Z}_0^{-1}$, the minimum of $\mathbf{x}'\bm{W}_0\mathbf{x}$ on the unit sphere is attained at $\mathbf{v}_0^N$, the leading eigenvector of $\bm{Z}_0$, while the maximum is attained at $\mathbf{v}_0^C$, the eigenvector of $\bm{Z}_0$ associated with its smallest eigenvalue.

Therefore:
\begin{itemize}
\item if priors are pessimistic, so that $\mathbb{E}[\Delta u_1]<0$, the monopolist maximizes the objective by minimizing $\mathbf{x}'\bm{W}_0\mathbf{x}$, and therefore chooses $\mathbf{x}^*=\mathbf{v}_0^N$;
\item if priors are optimistic, so that $\mathbb{E}[\Delta u_1]>0$, the monopolist maximizes the objective by maximizing $\mathbf{x}'\bm{W}_0\mathbf{x}$, and therefore chooses $\mathbf{x}^*=\mathbf{v}_0^C$. \hfill$\blacksquare$
\end{itemize}

\subsection*{Proof of Proposition \ref{app:ortho_compl}}
With interaction $\Delta\beta_{ij,t}$, the no-surprise (orthogonality) condition for a two-good bundle is
\[
\Delta\beta_{i,t}x_{i,t+1}+\Delta\beta_{j,t}x_{j,t+1}+\Delta\beta_{ij,t}x_{i,t+1}x_{j,t+1}=0.
\]
Imposing $x_{j,t+1}=1-x_{i,t+1}$ yields the quadratic equation
\[
-\Delta\beta_{ij,t}x_{i,t+1}^2
+(\Delta\beta_{i,t}+\Delta\beta_{ij,t}-\Delta\beta_{j,t})x_{i,t+1}
+\Delta\beta_{j,t}=0,
\]
whose two roots (distinct when $\Delta\beta_{ij,t}\neq 0$) are given by the expression stated in the proposition. Hence, the bundle is orthogonal.

To show that one and only one of them is in the open interval $(0,1)$, define $h(x):=\Delta\beta_{i,t}x+\Delta\beta_{j,t}(1-x)+\Delta\beta_{ij,t}x(1-x)$.
Then $h(0)=\Delta\beta_{j,t}<0$ and $h(1)=\Delta\beta_{i,t}>0$, so by continuity, and by the fact that $h(\cdot)$ is a second order polynomial, there exists a unique
$x^\perp\in(0,1)$ with $h(x^\perp)=0$. \hfill $\blacksquare$

\subsection*{Proof of Proposition \ref{prop:Wzeros_interactions}}

When there are complementarities and goods are proposed in dummies, $\bm{Z} $ has dimension $m+\frac{m(m-1)}{2}$: the $m$ goods and all their possible complementarities.
Consider the case in which a consumer has only tried single goods or single couples.
Call $s_i$ and $c_{ij}$ the number of times it happened. 

We can think of $\bm{Z} $ as block matrix. 
The top--left $n \times n$ submatrix records occurrences of the goods:
\begin{itemize}
    \item in the diagonal it has the sum $\bm{Z}_{ii} = s_i +\sum_{j \ne i} c_{ij}$ 
    \item off diagonally it has the occurrences of the couple $\bm{Z}_{ij} =  c_{ij}$ 
\end{itemize}
The bottom right corner is related to co--occurrence of couples:
\begin{itemize}
    \item it is just diagonal, with  $\bm{Z}_{ij,ij} =  c_{ij}$
\end{itemize}
The remaining two rectangular submatrices are symmetric, and depend on whether the single element is part of the couple or not:
\begin{itemize}
    \item $\bm{Z}_{ij,i}=\bm{Z}_{ij,j}=c_{ij}$
    \item $\bm{Z}_{ij,h}=0$ if $h \ne i,j$
\end{itemize}
We want to show that in the inverse $\bm{W}=\bm{Z}^{-1}$, making the same decomposition in submatrices, the two symmetric rectangular matrices are made of $0$'s for all $\bm{W}_{ij,h}$ such that $h \not \in \{i,j\}$.

The formula for the inverse says that 
$\bm{W}_{ij,h}$ is null if and only if the determinant of the corresponding \emph{cofactor} matrix $\bm{C}^{ij,h}$, obtained deleting from $\bm{Z}$ row $ij$ and column $h$, is null.

\bigskip
Since  $h \not \in \{i,j\}$, then:
\begin{itemize}
    \item  row $h$ of $\bm{C}^{ij,h}$ is equal to $c_{hk}$, for $k \ne h$, in the first $m-1$ elements, and $c_{hk}$, for $k \ne h$, in the remaining elements, for the couples where $h$ belongs - so, each $c_{hk}$ appears twice, once in the first $n-1$ elements and once in the remaining elements;
    \item each row $k h$, for $k\ne h$ of $\bm{C}^{ij,h}$ has only element $c_{hk}$ once in the first $m-1$ elements and once in the remaining elements;
    \item so, row $h$ of $\bm{C}^{ij,h}$ is equal to the sum of  rows $k h$, for $k\ne h$ of $\bm{C}^{ij,h}$ (the point is that we can exclude any singular occurrence $s_k$, and column--positions are respected);
    \item hence, $\bm{C}^{ij,h}$ has determinant $0$ and $\bm{W}_{ij,h}=0$. 
\end{itemize}

This proves the result. \hfill $\blacksquare$ 

\subsection*{Proof of Proposition \ref{prop:nolearn}}
    Recall that the consumption surprise at time $t+1$ is defined as
$
\Delta u_{t+1}
:= 
u_{t+1}-\mathbb{E}_{\hat{\bm{\beta}}_{t}}[u_{t+1}]
=
\alpha+\mathbf{x}_{t+1}'\bm{\beta}+\varepsilon_{t+1}
-\hat{\alpha}-\mathbf{x}_{t+1}'\hat{\bm{\beta}}_{t},
$
so that
\begin{equation}
\mathbb{E}[\Delta u_{t+1}]
=
-(\hat{\alpha}-\alpha)
-
\mathbf{x}_{t+1}'\Delta\bm{\beta}_t.
   \label{eq:delta_u_expand}
\end{equation}
Hence, the consumer experiences no surprise in expectation, $\mathbb{E}[\Delta u_{t+1}]=0$, if and only if
\begin{equation}
\mathbf{x}_{t+1}'\Delta\bm{\beta}_t
=
-(\hat{\alpha}-\alpha).
\label{eq:zero_surprise_condition}
\end{equation}
This condition defines a hyperplane of bundles that, on expectation, generate zero prediction error.

To characterize the set of bundles that satisfy condition~(\ref{eq:zero_surprise_condition}), 
note that any vector $\mathbf{x}_{t+1}\in\mathbb{R}^I$ can be uniquely decomposed into a component parallel to the bias vector $\Delta\bm{\beta}_t$ and a component orthogonal to it. 
Formally, by the projection theorem in Euclidean space, there exist scalars $ \xi \in\mathbb{R}$ and a vector $\mathbf{z}$ such that
$$
\mathbf{x}_{t+1} =  \xi \cdot   \Delta\bm{\beta}_t + \mathbf{z},
\qquad 
\text{with } \mathbf{z} \perp\Delta\bm{\beta}_t = 0.
$$
Substituting this orthogonal decomposition into~(\ref{eq:zero_surprise_condition}) 
and using the property $\Delta\bm{\beta}_t'\Delta\bm{\beta}_t = \|\Delta\bm{\beta}_t\|^2$, we obtain
$$
\xi \cdot  \|\Delta\bm{\beta}_t\|^2 = -(\hat{\alpha}-\alpha),
$$
from which we get
\begin{equation}
\mathbf{x}_{t+1}
=
-\frac{\hat{\alpha}-\alpha}{\|\Delta\bm{\beta}_t\|^2}\,\Delta\bm{\beta}_t
+
\mathbf{z},
\qquad
\mathbf{z}' \Delta\bm{\beta}_t = 0.
\label{eq:x_no_learning}
\end{equation}
Therefore, any bundle of this form yields zero expected surprise.
\\
\\
Also recall that
$$\hat{\bm{\beta}}_{t+1} = \hat{\bm{\beta}}_t + \bm{\omega}_{t+1}\Delta u_{t+1}$$
Taking expectations and using 
(\ref{eq:delta_u_expand})--(\ref{eq:x_no_learning}), we get
\[
\mathbb{E}[\hat{\bm{\beta}}_{t+1}]
=
\hat{\bm{\beta}}_t + 
\bm{\omega}_{t+1}\mathbb{E}[\Delta u_{t+1}]
=
\hat{\bm{\beta}}_t.
\]
Thus, whenever condition (\ref{eq:x_no_learning}) holds, the consumer’s expected beliefs remain unchanged. Conversely, if
\[
\mathbb{E}[\hat{\bm{\beta}}_{t+1}] = \hat{\bm{\beta}}_{t},
\]
then
\[
\mathbb{E}[\Delta u_{t+1}] = 0,
\]
implying (\ref{eq:zero_surprise_condition}) and therefore (\ref{eq:x_no_learning}).
It follows that the consumer’s expected beliefs remain unchanged if and only if
the bundle $\mathbf{x}_{t+1}$ satisfies (\ref{eq:x_no_learning}).
\hfill $\blacksquare$

%\end{document}

%\newpage
\section{Additional analysis on the movies' database} 
\label{sec:appendix_movies}

This appendix gathers additional results that complement the analysis of Section~\ref{sec:movies} and could not be included in the main text.
Table \ref{table:top_beta} reports the actors/actresses with the highest estimated $\hat\beta$, computed at the end of time/movie $t=680$. To compute these, we perform an ordinary least square regression with all the 680 movies in the subsample:
$$
u_t = \alpha + \mathbf{x}_t' \bm{\beta} + \varepsilon_t,
$$
where $\mathbf{x}_t \in \mathbb R^{121}$ is a vector of 0's and 1's, with 1 in position $i$ if the $i$-th actor appears in movie $t = 1, ..., 680$.

\begin{table}[h!]
    \centering
    \begin{tabular}{|c|c|c|}
    \toprule
    \textbf{Rank} & \textbf{Actor/Actress} & {$\hat\beta$ (million \$)} \\ 
    \hline
    1  & Julie Andrews        & 260 \\
    \hline
    2  & Harrison Ford     & 181 \\
    \hline
    3  & Orlando Bloom        & 172 \\
    \hline
    4  & Richard Dreyfuss    & 168 \\
    \hline
    5  & Robert Downey Jr.        & 140 \\
    \hline
    6  & Ewan McGregor    & 125 \\
    \hline
    7  & Tobey Maguire        & 112 \\
    \hline
    8  & Mike Myers   & 99 \\
    \hline
    9  & Christian Bale         & 97 \\
    \hline
    10 & Michael Keaton      & 93 \\
    \hline
    11 & Tim Allen         & 91 \\
    \hline
    12 & Dustin Hoffman           & 84 \\
    \hline
    13 & Michelle Rodriguez       & 82 \\
    \hline
    14 & Willem Dafoe       & 80 \\
    \hline
    15 & Emma Watson        & 77 \\
    \hline
    16 & Robert Redford            & 74 \\
    \hline
    17 & Sigourney Weaver     & 73 \\
    \hline
    18 & Daniel Radcliffe       & 70 \\
    \hline
    19 & Ian McKellen & 66 \\
    \hline
    20 & Mark Ruffalo          & 62 \\
    \bottomrule
    \end{tabular}
    \caption{\footnotesize  Top 20 of $\hat\beta$'s when considering all the movies in our dataset.}
    \label{table:top_beta}
\end{table}

In our interpretation, each individual's $\hat\beta$ is the estimated contribution of that actor to the movie, which would on average obtain the box office earning given by the intercept value $\alpha_{680} = 279$ M\$ (which, thus, indicates the baseline box office result for a movie). 
The network measures, $v^N$ and $v^C$, are computed using the co-starring network at the final date, that is, considering the network whose adjacency matrix is $\mathbf Z_{680}$. 

\begin{figure}[h]
    % \centering
    \vspace{-30mm}
    \hspace{-20mm}
    \setlength{\abovecaptionskip}{-20mm}
    % \hspace*{-30mm}
    \includegraphics[width=1.24\textwidth]{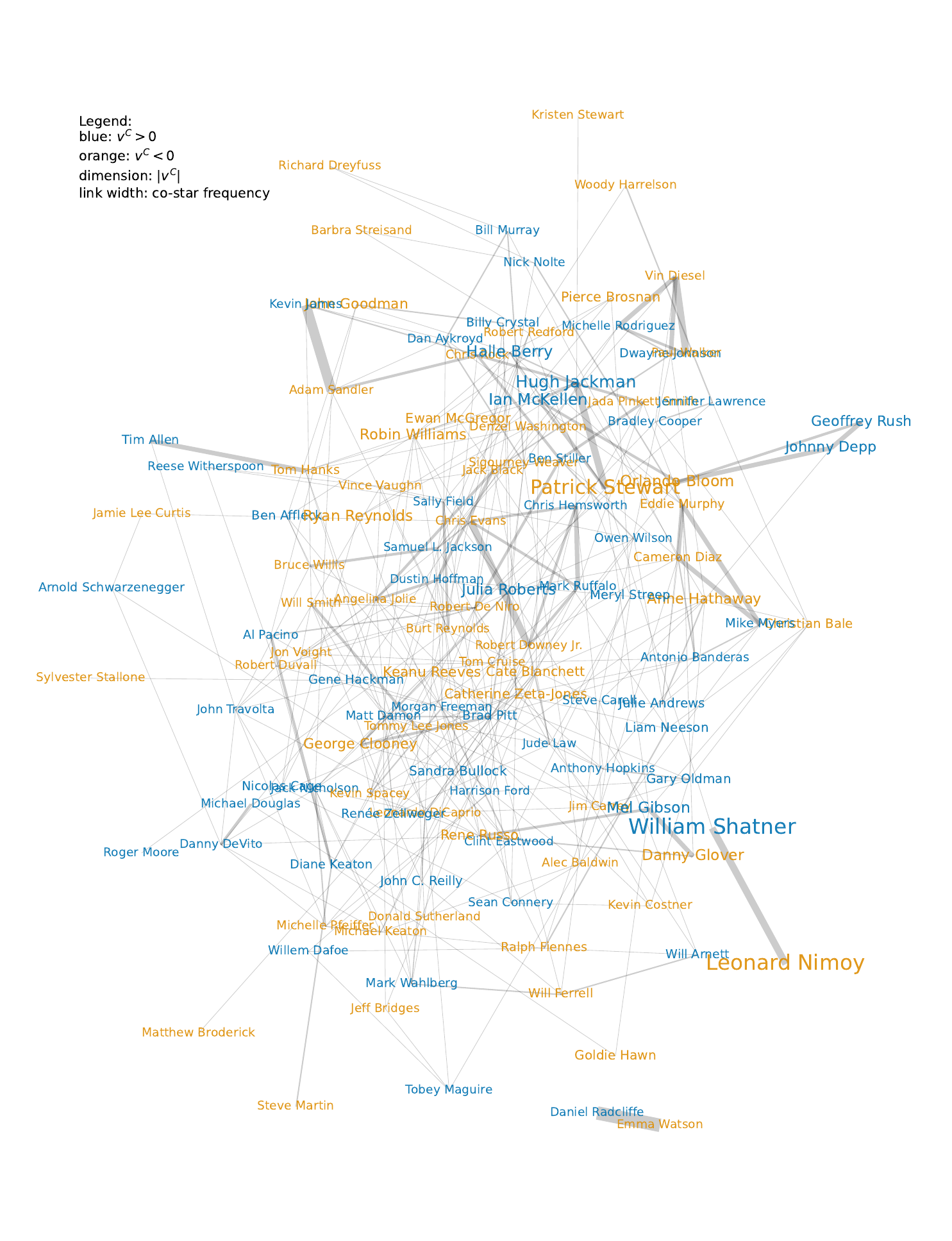}
    \caption{\footnotesize  Co--starring network of the 121 actors in our database, with larger font size for more central nodes according to correlation centrality $\mathbf v^C$. The two groups induced by this eigenvector are in different colors. As in Figure \ref{fig:net1}, the names' position is the same and link width is proportional to the number of movies two actors co-star in. Here, however, the name's dimension is different, because it is proportional to $|v^C|$ (rather than to $v^N$).}
    \label{fig:net2}
\end{figure}

The co-starring network in Figure \ref{fig:net2} shows the 121 actors in the dataset with name's size proportional to $|v^C|$ and color given by the sign of $v^C$.

\clearpage
\subsection{Considering complementarities between actors}

This subsection extends the empirical exercise to the case of complementarities between actors, as introduced in Appendix~\ref{subsec:complementarity}. As is standard in econometric specifications with interactions, the vector $\mathbf{x}_t$ includes not only individual actor indicators, but also indicators for selected pairs of actors. Thus, the empirical exercise no longer evaluates only the nodes of the co-starring network, corresponding to individual actors, but also some of its links, corresponding to pairs of actors who appear jointly in movie $t$.

\begin{figure}[htb]
    \centering
    \includegraphics[width=.8\linewidth]{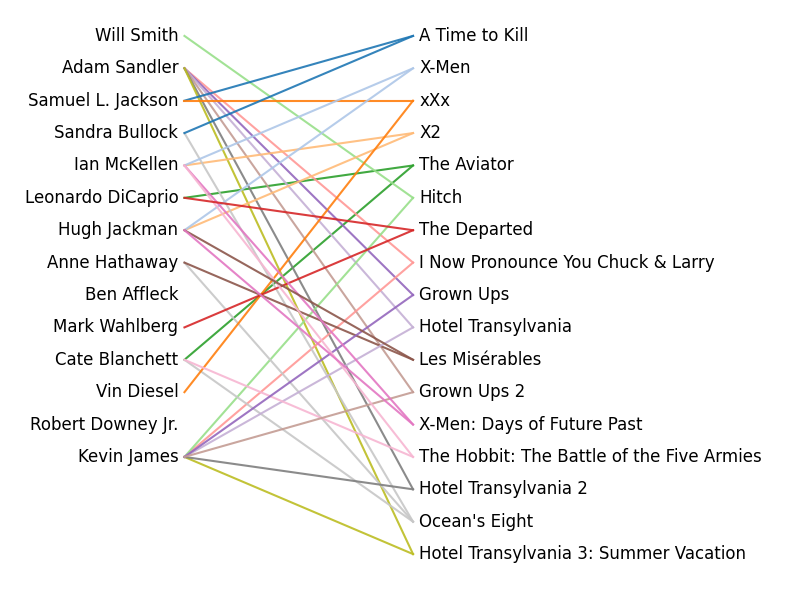}
    \caption{\small On the left, the 14 actors among the 121 in the dataset who appear in at least 4 movies both before and after the adjacency matrix $\bm Z_t$ reaches full rank. On the right, the 17 movies in which they co-star together (except Robert Downey Jr. who does not co-appear with others).}
    \label{fig:network_complementarities}
\end{figure}

To avoid an excessive number of pairs, given the number of variables relative to the total number of available observations, we restrict attention to the actors who appear most frequently in the dataset. Applying the same actor-selection criterion discussed in Section~\ref{subsec:moviebeta}, based on appearances before and after the full-rank threshold at $t=554$, we retain 14 actors, who co-star together in 17 movies and generate 12 distinct pairs. Figure~\ref{fig:network_complementarities} displays this restricted set.

\begin{figure}[htb]
    \centering
    \includegraphics[width=0.8\linewidth]{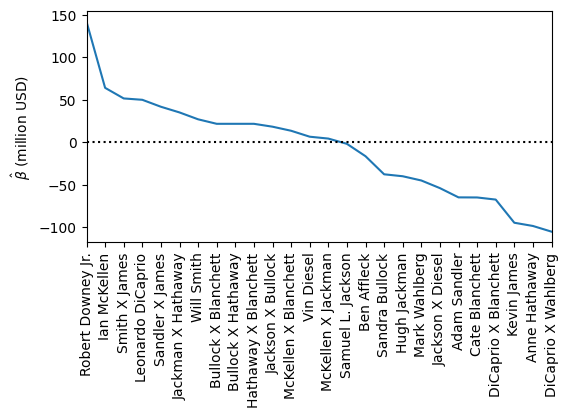}
    \caption{\small Estimated $\hat\beta$'s for 14 selected actors and for their complementarities, indicated by an ``X'' between actors' last names.}
    \label{fig:beta_hat_complementarities}
\end{figure}

For example, since \emph{A Time to Kill} features both Samuel L. Jackson and Sandra Bullock, we include the interaction between these two actors among the regressors. The total number of variables in the estimation is therefore given by 121 actor indicators plus 12 pair indicators. The corresponding estimated $\hat\beta$'s are shown in Figure~\ref{fig:beta_hat_complementarities}. Interestingly, the figure shows that actors such as Leonardo DiCaprio can have a positive estimated $\hat\beta$ as individual actors and, at the same time, negative estimated interaction coefficients when paired with others (e.g., with Cate Blanchett and with Mark Wahlberg).

\end{document}